\documentclass[letter,11pt]{article}
\usepackage{latexsym}
\usepackage{amsmath,amssymb,bm,ascmac,color,calc}
\usepackage{slashed}
\usepackage[mathscr]{eucal}
\usepackage{graphicx}
\usepackage{cite}
\usepackage{lineno}
\usepackage{jheppub}
\usepackage{multirow}
\usepackage{subfigure}
\usepackage[normalem]{ulem}
\usepackage[T1]{fontenc}

\usepackage{color}

\usepackage{makecell}

\allowdisplaybreaks

\newcommand{\mpt}{{\;/\!\!\!\! {P}_T}} 
\newcommand{\mptvec}{{\;/\!\!\!\! \vec{P}_T}} 
\newcommand{\ignore}[1]{}
\renewcommand{\bf}{\textbf}

\newcommand{\gev}{\textrm{ GeV}}

\newcommand{\amc}{{\sc MadGraph5\textunderscore}a{\sc MC@NLO}}

\def\beq{\begin{equation}}
\def\eeq{\end{equation}}
\newcommand{\ba}{\begin{array}}
\newcommand{\ea}{\end{array}}
\newcommand{\bea}{\begin{eqnarray}}
\newcommand{\eea}{\end{eqnarray} }
\newcommand{\bal}{\begin{align}}
\newcommand{\eal}{\end{align}}
\def\bi{\begin{itemize}}
\def\ei{\end{itemize}}
\def\ben{\begin{enumerate}}
\def\een{\end{enumerate}}
\def\beq{\begin{equation}}
\def\eeq{\end{equation}}
\def\bc{\begin{center}}
\def\ec{\end{center}}
\def\bt{\begin{table}}
\def\et{\end{table}}
\def\btb{\begin{tabular}}
\def\etb{\end{tabular}}

\title{Portraying Double Higgs at the Large Hadron Collider}

\author[a]{Jeong Han Kim,}
\author[b,c]{Minho Kim,}
\author[a]{Kyoungchul Kong,}
\author[d]{Konstantin T. Matchev,}
\author[c]{Myeonghun Park} 
 
\affiliation[a]{Department of Physics and Astronomy, University of Kansas, Lawrence, KS 66045, USA}
\affiliation[b]{Department of Physics, POSTECH, 77 Cheongam-ro, Nam-gu, Pohang, 37673, Korea}
\affiliation[c]{Institute of Convergence Fundamental Studies and School of Liberal Arts, Seoultech, 232 Gongneung-ro, Nowon-gu, Seoul, 01811, Korea}
\affiliation[d]{Institute for Fundamental Theory, Physics Department, University of Florida, Gainesville, FL 32611, USA}

\emailAdd{jeonghan.kim@ku.edu} 
\emailAdd{kmhmon@postech.ac.kr}
\emailAdd{kckong@ku.edu}
\emailAdd{matchev@phys.ufl.edu}
\emailAdd{parc.seoultech@seoultech.ac.kr} 
 
\abstract{
We examine the discovery potential for double Higgs production at the high luminosity LHC in the final state with two $b$-tagged jets, two leptons and missing transverse momentum. Although this dilepton final state has been considered a difficult channel due to the large backgrounds, we argue that it is possible to obtain sizable signal significance, by adopting a deep learning framework making full use of the relevant kinematics along with the jet images from the Higgs decay. 
For the relevant number of signal events we obtain a substantial increase in signal sensitivity over existing analyses. We discuss relative improvements at each stage and the correlations among the different input variables for the neutral network. The proposed method can be easily generalized to the semi-leptonic channel of double Higgs production, as well as to other processes with similar final states.
  }

\begin{document} 
\maketitle

\section{Introduction} \label{sec:intro}

The discovery of the Higgs boson \cite{Aad:2012tfa,Chatrchyan:2012xdj} jumpstarted the comprehensive program of precision measurements of all Higgs couplings. While the Higgs boson couplings to fermions and gauge bosons are in good agreement with the Standard Model (SM) predictions \cite{Khachatryan:2016vau}, the Higgs self-couplings are difficult to measure experimentally \cite{ATL-PHYS-PUB-2017-001,ATL-PHYS-PUB-2016-024,Kim:2018uty,Sirunyan:2018two,CMS:2015nat,CMS:2017cwx,Baglio:2012np,Sirunyan:2017guj,Cepeda:2019klc}. Yet, the knowledge of those couplings is crucial for understanding the exact mechanism of electroweak symmetry breaking and the origin of mass in our universe. It is also a guaranteed physics target which can be probed at the upgraded Large Hadron Collider (LHC) or at future colliders. The resulting experimental constraints on the Higgs self-couplings will have an immediate and long-lasting impact on model-building efforts beyond the SM.

We parameterize the Higgs self-interaction as follows:
\begin{equation}
V = \frac{m_h^2}{2} h^2 + \kappa_3 \lambda_{3}^{\rm SM} v h^3 + \frac{1}{4}\kappa_4  \lambda_{4}^{\rm SM} h^4 \label{eq:Vh} \, ,
\end{equation}
where $m_h$ is the mass of the SM Higgs boson ($h$), $v \approx 256$ GeV is the Higgs vacuum expectation value,
 $$\lambda_{3}^{\rm SM} = \lambda_{4}^{\rm SM} = \frac{m_h^2}{2 v^2}$$ are the SM values for the Higgs self-couplings, 
 while $\kappa_3$ and $\kappa_4$ parametrize the corresponding deviations from them. In order to access $\kappa_3$ ($\kappa_4$), one has to measure the process of
double (triple) Higgs boson production at the LHC, possibly with high luminosity (HL), or at future colliders.

Double Higgs ($hh$) production has been studied in many channels, including $b \bar b b \bar b$ \cite{ATLAS:2018combi, Aaboud:2018knk, CMS:2018smw, deLima:2014dta, Wardrope:2014kya, Behr:2015oqq}, $b \bar b \gamma \gamma$ \cite{Sirunyan:2018iwt, Aaboud:2018ftw, CMS-PAS-FTR-15-002, ATL-PHYS-PUB-2014-019, Kim:2018uty, Kling:2016lay, Baur:2003gp, Baglio:2012np, Huang:2015tdv, Azatov:2015oxa,Cao:2015oaa,Cao:2016zob,Alves:2017ued, Barger:2013jfa, Chang:2018uwu}, $b \bar b \tau\tau$ \cite{CMS-PAS-FTR-15-002, Aaboud:2018sfw, Sirunyan:2017djm, Kim:2018uty, Baur:2003gpa, Goertz:2014qta, Dolan:2012rv}, $b \bar b W^+ W^- /ZZ$ \cite{Aaboud:2018zhh, CMS:2017ums, CMS:2015nat, CMS:2017cwx, Kim:2018cxf,  Papaefstathiou:2012qe, Huang:2017jws}, $W^+ W^- W^+ W^-$ \cite{Aaboud:2018ksn}, etc. Among the different possible final states, here we focus on $hh$ production at the HL-LHC in the final state with two $b$-tagged jets, two leptons and missing transverse momentum. The signal process is $(h\to b{\bar b})(h\to W^\pm W^{*\mp}\to \ell^+\nu_\ell {\ell'}^-\bar{\nu}_{\ell'})$ and it suffers from large SM backgrounds, primarily due to top quark pair production ($t\bar{t}$). 
The few existing studies in this channel therefore employ sophisticated algorithms (neutral network (NN) \cite{CMS:2015nat}, deep neutral network (DNN) \cite{Sirunyan:2017guj}, boosted decision tree (BDT) \cite{Adhikary:2017jtu,CMS:2017cwx}, etc.) to increase the signal sensitivity, but show somewhat pessimistic results, with a significance no better than $1\sigma$ at the HL-LHC with 3 ab$^{-1}$ luminosity. 

The recent study in Ref.~\cite{Kim:2018cxf} introduced some new ideas for reducing the SM backgrounds in this channel.
For example, the new variables {\it Topness} and {\it Higgsness} were designed to test whether the event kinematics is consistent with $t\bar{t}$ or $hh$, respectively.  
The use of Topness and Higgsness already effectively reduced the $t\bar t$ background to a manageable level, and additional variables were then
employed to handle the remaining SM background processes --- e.g.,
the subsystem variable $M_{T2}^{(\ell)}$ is effective in eliminating background arising from $\tau$ decays. 
In this paper, we supplement the novel kinematic method from Ref.~\cite{Kim:2018cxf} with the analysis of the jet image in the $h\to b \bar b$ decay, where the basic idea is to treat the detector as a camera and the streams of jets as an image \cite{Bhattacherjee:2019fpt,Gallicchio:2010sw,Gallicchio:2010dq,Hook:2011cq,Cogan:2014oua,deOliveira:2015xxd,Lin:2018cin,deOliveira:2017pjk}.
In our case, the collimated nature of the Higgs decay will hopefully differ from the patterns obtained in SM production processes. In addition, 
we adopt a deep learning framework in our main analysis, since it is known that modern deep learning algorithms trained on jet images provide improved signal-to-background discrimination \cite{Gallicchio:2010sw,Gallicchio:2010dq,Hook:2011cq,Baldi:2014kfa,Cogan:2014oua,deOliveira:2015xxd,Komiske:2016rsd,Kasieczka:2017nvn,Lin:2018cin}. 

The analysis presented in this paper contains a number of improvements in comparison to previous studies:
\begin{itemize}
\item Unlike the customized detector simulation performed in Ref.~\cite{Kim:2018cxf}, here we employ {\sc Delphes}  \cite{deFavereau:2013fsa}  to simulate detector effects such as detector resolution, reconstruction efficiency, etc., and {\sc Fastjet} \cite{Cacciari:2011ma} for jet-reconstruction.
\item We use deep learning framework to optimize the cuts, which further increases the significance compared to the conventional cut-and-count as performed in Ref.~\cite{Kim:2018cxf}. 
\item We exploit an enlarged set of relevant variables which consists of the 10 variables originally considered in Ref.~\cite{Adhikary:2017jtu}:
$p_{T\ell_1}$, $p_{T\ell_2}$, $\mpt$, $m_{\ell\ell}$, $m_{bb}$, $\Delta R_{\ell\ell}$, $\Delta R _{bb}$, $p_{Tbb}$, $p_{T\ell\ell}$, and $\Delta\phi_{bb,\ell\ell}$,
supplemented with the six recent variables from Ref.~\cite{Kim:2018cxf}:
Topness, Higgsness, $M_{T2}^{(b)}$, $M_{T2}^{(\ell)}$, $\hat s_{min}^{(\ell\ell)}$ and $\hat s_{min}^{(bb\ell\ell)}$.
\item We include a SM background process, $tW$ production, which was missing from all previous discussions of this channel, yet
it turns out to be the next dominant background once the $t\bar t$ background is under control. 
\item The fact that the Higgs boson $h$ is a color-singlet allows us to use the jet image of the $h \to b\bar b$ decay for further background suppression \cite{Cogan:2014oua,Gallicchio:2010sw,Gallicchio:2010dq,Hook:2011cq,Lin:2018cin}.
\item We examine the effect of pile-up, which was missing from previous studies. The expected average number of pile-up $\left<\mu\right>$ at the HL-LHC is $\mathcal{O}(200)$ collisions per bunch crossing\,\cite{ATL-PHYS-PUB-2019-005}. Thus for any precision measurements, it is crucial to have a strategy in place to ensure that pile-up effects do not jeopardize the analysis. Here we choose to apply the Soft Drop algorithm\,\cite{Larkoski:2014wba} for QCD analyses, which is a powerful pile-up mitigation technique. In order to reduce pile-up effects on the relevant kinematic variables, we adopt the definition for a missing transverse momentum from ATLAS, which excludes contributions from soft neutral particles\,\cite{Aaboud:2018tkc}.
\end{itemize}

Our results show that the dominant $t\bar t$ background can be significantly reduced until it is comparable to the other subdominant backgrounds,
i.e., after all cuts, we find that all SM backgrounds contribute at similar levels. 
This reduction can be accomplished without sacrificing too much of the signal rate, which leads to an improved signal significance. 
Our study indicates that the dilepton channel from $hh\to b\bar{b} W^+W^-$ could contribute to the combined significance for $hh$ discovery
on par with the other final states, making double Higgs production sooner accessible at the HL-LHC.

This paper is structured as follows.
We begin our discussion of the SM backgrounds and present the details of our simulation in section \ref{sec:analysis}.
In the following two sections \ref{sec:method1} and \ref{sec:method2}, we provide some basic information on the kinematic variables
used later in the analysis and on jet images, respectively. 
Then in section \ref{sec:dnn} we discuss how we set up our analysis in a deep learning framework. 
Section \ref{sec:results} presents our results, while section \ref{sec:discussion} is reserved for the discussion and conclusions. 
We include a brief review on deep neural networks in Appendix \ref{sec:appendix}.

\section{Event generation and detector simulation} \label{sec:analysis}

Parton-level signal and background events were generated using \amc{} v2.6 \cite{Alwall:2014hca} with the default NNPDF2.3QED parton distribution functions \cite{Ball:2013hta} at leading order QCD accuracy at the $\sqrt{s} = 14$ TeV LHC. The default dynamical renormalization and factorization scales were used. We assume 3ab$^{-1}$ of luminosity throughout this paper. 
Parton-level events were generated with the following cuts: 
$p_{Tj} > 20$ GeV, 
$p_{Tb} > 20$ GeV, 
$p_{T\gamma} > 10$ GeV, 
$p_{T\ell} > 10$ GeV, 
$\eta_{j}$ < 5, 
$\eta_b$ < 5, 
$\eta_{\gamma}$ < 2.5, 
$\eta_\ell$ < 2.5, 
$\Delta R_{bb} < $ 1.8, 
$\Delta R_{\ell\ell} < $ 1.3,
70 GeV $< m_{jj}, m_{bb}< $ 160 GeV and  
$m_{\ell\ell} < $ 75 GeV. %
For $jj\ell\ell\nu\bar\nu$, $\ell\ell b j$ and $tW+j$ backgrounds, we impose 5 GeV $< m_{\ell\ell} < 75$ GeV additionally.
Here the angular distance $\Delta R_{ij}$ is defined by
\begin{eqnarray}
\Delta R_{ij} = \sqrt{(\Delta\phi_{ij})^2+(\Delta \eta_{ij})^2},
\label{deltaR}
\end{eqnarray}
where $\Delta\phi_{ij} = \phi_i-\phi_j$ and $\Delta\eta_{ij}=\eta_i-\eta_j$ are respectively the differences of the azimuthal angles and rapidities between particles $i$ and $j$.

The double Higgs production cross-section is normalized to $\sigma_{hh} = 40.7$ fb, the next-to-next-to-leading order (NNLO) accuracy in QCD \cite{Grigo:2014jma}. 
Considering all relevant branching fractions, we obtain signal cross section $\sigma_{hh} \cdot 2 \cdot \text{BR}(h \rightarrow b \overline{b}) \cdot \text{BR}(h \rightarrow W W^{*} \rightarrow \ell^+ \ell^- \nu \bar \nu)  =0.648$ fb, where $\ell$ denotes an electron or a muon, including leptons from tau decays. 
The major background is $t \overline{t}$ production, whose cross section is normalized to the NNLO QCD cross-section 953.6~pb~\cite{Czakon:2013goa}. 
Another important background is $t \overline{t} h$, which is normalized to the next-to-leading order (NLO) QCD cross-section of 611.3~fb~\cite{Dittmaier:2011ti}.
For the $t \overline{t} V$ ($V=W^\pm, Z$) background, we apply an NLO k-factor of 1.54, resulting in a cross-section of 1.71 pb \cite{deFlorian:2016spz}. 
We apply an NLO k-factor of 1.0 for the Drell-Yan type backgrounds $\ell \ell b j$ and $\tau \tau b b$, where $j$ denotes partons in the five-flavor scheme. 
Note that a recent study indicates that ${\rm k}^{NNLO, DY}_{QCD\otimes QED} \approx 1$ \cite{deFlorian:2018wcj}. 
The irreducible $jj\ell\ell\nu\nu$ background from the mixed QCD+EW process is included with ${\rm k}_{NLO}$ = 2. 
Finally, we generate $tW + j$ events with up to one additional matched jet (in the five-flavor scheme), whose cross-section turns out to be 0.51 pb (after the cuts) including all relevant branching fractions. 
As we try to reconstruct events, off-shell effects for the top quark and $W$ boson need to be taken care of properly. We generate parton level events with MadGraph5, which includes the proper treatment of the off-shell effects for the top quark and the $W$ boson for both signal and all backgrounds.

Events are further processed for parton-shower/hadronization using {\sc Pythia8235} \cite{Sjostrand:2014zea}.
We use {\sc Delphes} 3.4.1 \cite{deFavereau:2013fsa} for simulating the detector effects and {\sc Fastjet} 3.3.1 \cite{Cacciari:2011ma} for jet-reconstruction, with modified ATLAS settings 
as follows. 
\begin{itemize}
\item Jets are clustered with the anti-$k_T$ algorithm \cite{Cacciari:2008gp} with cone-size $\Delta R = 0.4$, 
where $\Delta R$ is the distance (\ref{deltaR}) in the ($\phi$, $\eta$) space. For the analysis, we consider jets with $p_{Tj} > 20$ GeV and
$|\eta_j|<2.5$.
 \item We use the a flat $b$-tagging efficiency, $\epsilon_{b\rightarrow b} = 0.75$, and flat mis-tagging rates for non-$b$ jets of
 $\epsilon_{c\rightarrow b} = 0.1$ and $\epsilon_{j\rightarrow b} = 0.01$ \cite{ATL-PHYS-PUB-2019-005}.
\item For lepton isolation, we require $\frac{p_{T \ell}}{ p_{ T \ell} + \sum_i p_{T i} } > 0.7$, where
the sum is taken over the transverse momenta $p_{Ti}$ of all final states particles $i$,  $i\neq \ell$, with 
$p_{Ti} > 0.5$ GeV and within $\Delta R_{i  \ell} < 0.3$ of the lepton candidate $\ell$.  
Leptons are also required to have $p_{T\ell} > 10$ GeV and $|\eta_\ell|<2.5$.
\item For photon isolation, we analogously require $\frac{\sum_i p_{T i} }{p_{ T \gamma}} < 0.12$ for particles within $\Delta R_{i \gamma} < 0.3$
of the photon candidate $\gamma$. 
Photons are also required to have $p_{T\gamma} > 25$ GeV and $|\eta_\gamma|<2.5$.
 \item The missing transverse momentum $\mptvec$ is defined as the negative vector sum of the transverse momenta of the accepted leptons, photons, jets and soft tracks as follows\,\cite{Aaboud:2018tkc};
 \begin{equation}
  \mptvec= -\left(\sum \vec p_{T\ell} + \sum \vec p_{T\gamma} + \sum \vec p_{Tj} + \sum \vec p_{T\textrm{(track)}} \right) .
  \end{equation}
Here the last term is added to consider unused soft tracks. These tracks are required to have $p_{T} > 0.4$ GeV,  $|\eta|<2.5$ and transverse (longitudinal) impact parameter  $|d_0| <1.5\,\textrm{mm}\,(|z_0 \sin\theta| < 1.5\,\textrm{mm})$. To reduce effects from pile-up, we only use particles which have track information. 

 \end{itemize}

After particle reconstruction, we employ the following {\it baseline selection cuts}\footnote{For the motivation behind these cuts, see 
Fig.~\ref{fig:var1} (in which the cut values are indicated with vertical dotted lines) and the related discussion in Sec.~\ref{sec:method1} below.} from Ref.~\cite{Kim:2018cxf}:
\begin{itemize}
\item the two leading jets must be $b$-tagged,  each with $p_{T} > 30$ GeV, 
\item exactly two isolated leptons of opposite sign, each with $p_{T\ell} > 20$ GeV, 
\item $\mpt = | \mptvec | > 20 $ GeV for the reconstructed missing transverse momentum, 
\item proximity cut of $\Delta R_{\ell\ell} < 1.0$ for the two leptons, 
\item proximity cut of $\Delta R_{bb} < 1.3$ for the two $b$-tagged jets, 
\item $m_{\ell\ell} < 65$ GeV for the two leptons,  
\item $95$ GeV $< m_{bb}<  140$ GeV for the two $b$-tagged jets. 
\end{itemize}

For those events which passed the baseline cuts, we form 16 kinematic variables, as well as jet images.
As we will see later, the jet images can capture additional features which are not already contained in the 16 standard kinematic variables. 
Therefore one can obtain better performance by combining kinematics and jet images, which is one of the main ideas of this paper.

\section{Kinematics in signal and backgrounds} \label{sec:method1}

\begin{figure}[t]
\centering
\includegraphics[width=4.05cm]{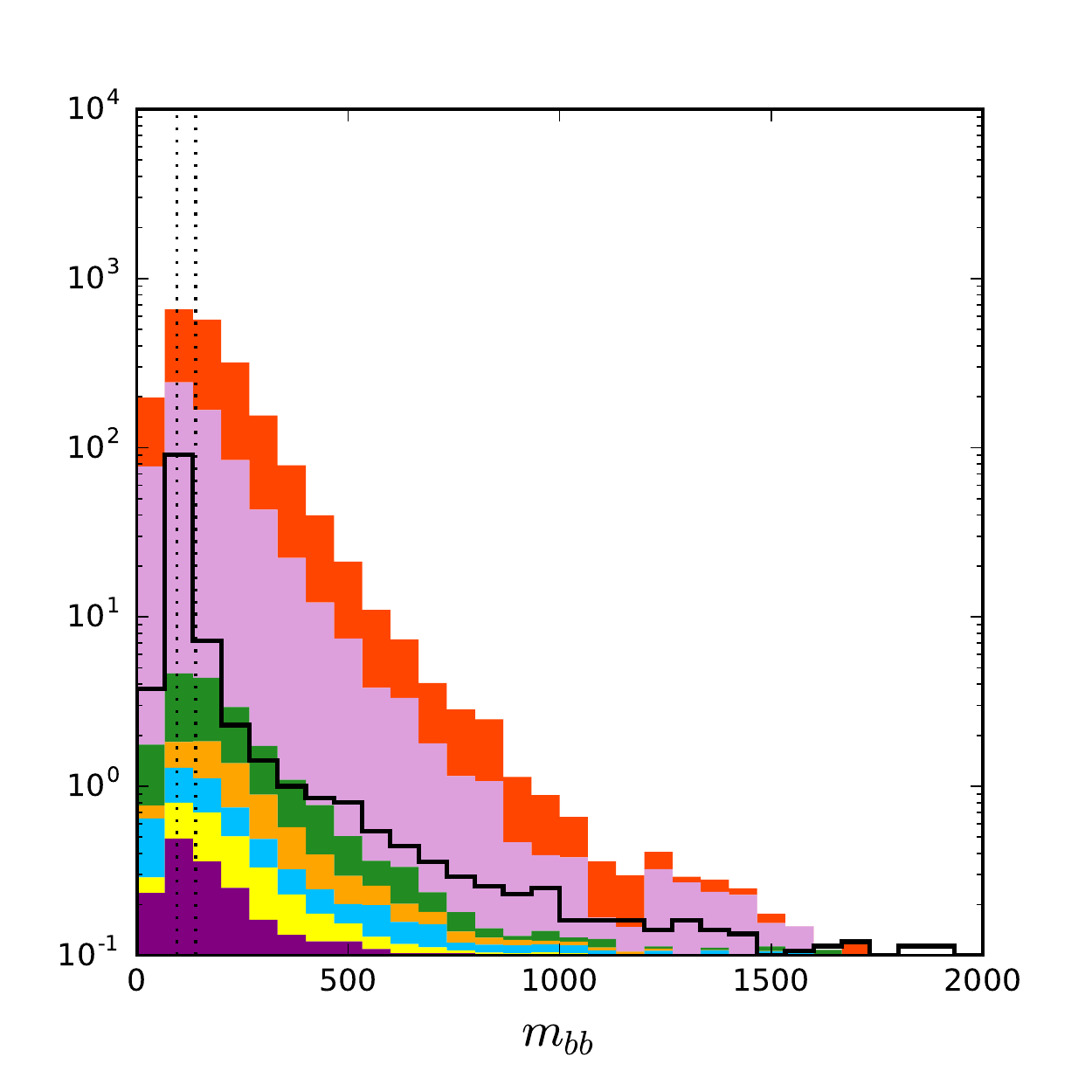} \hspace*{-0.5cm}
\includegraphics[width=4.05cm]{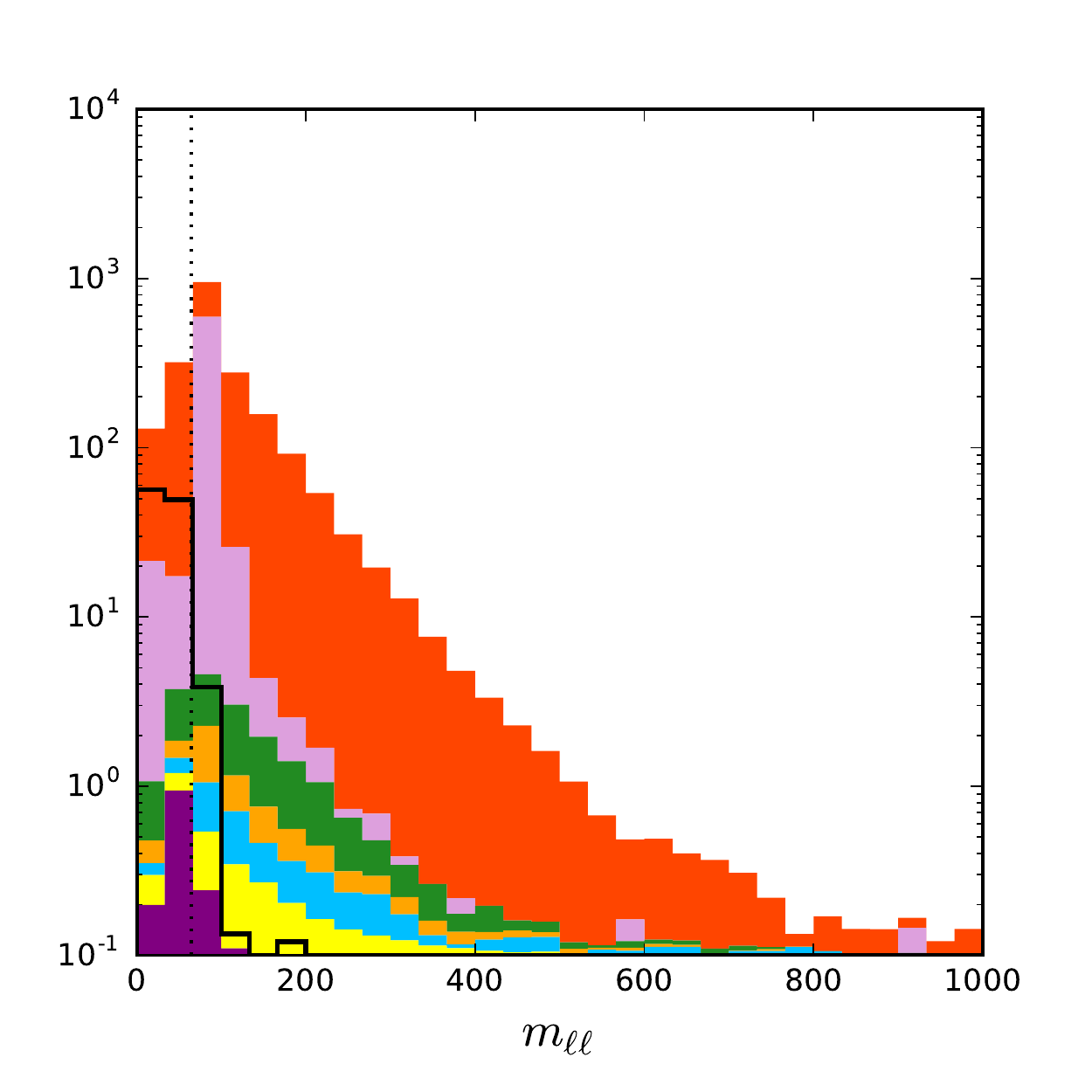} \hspace*{-0.5cm}
\includegraphics[width=4.05cm]{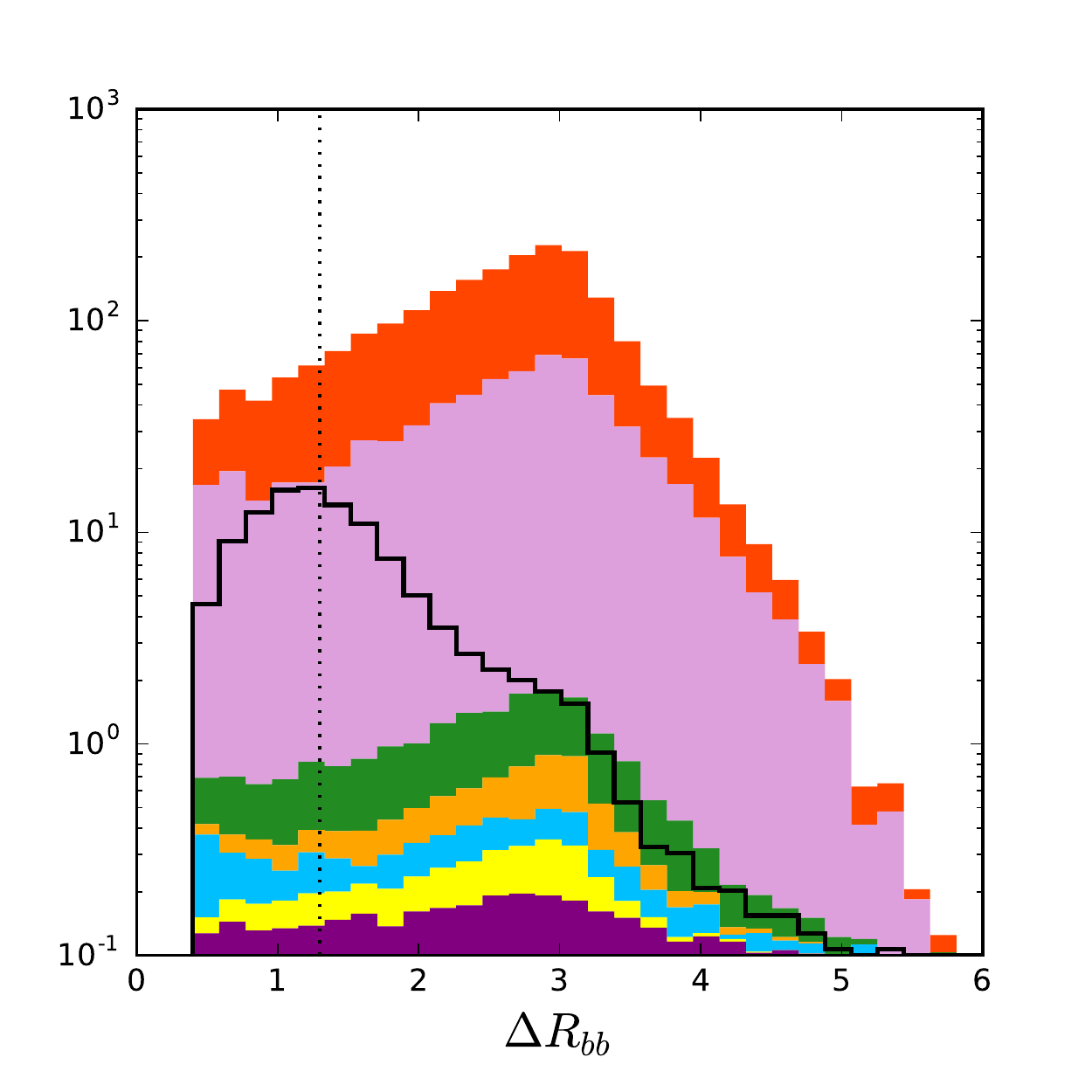} \hspace*{-0.5cm}
\includegraphics[width=4.05cm]{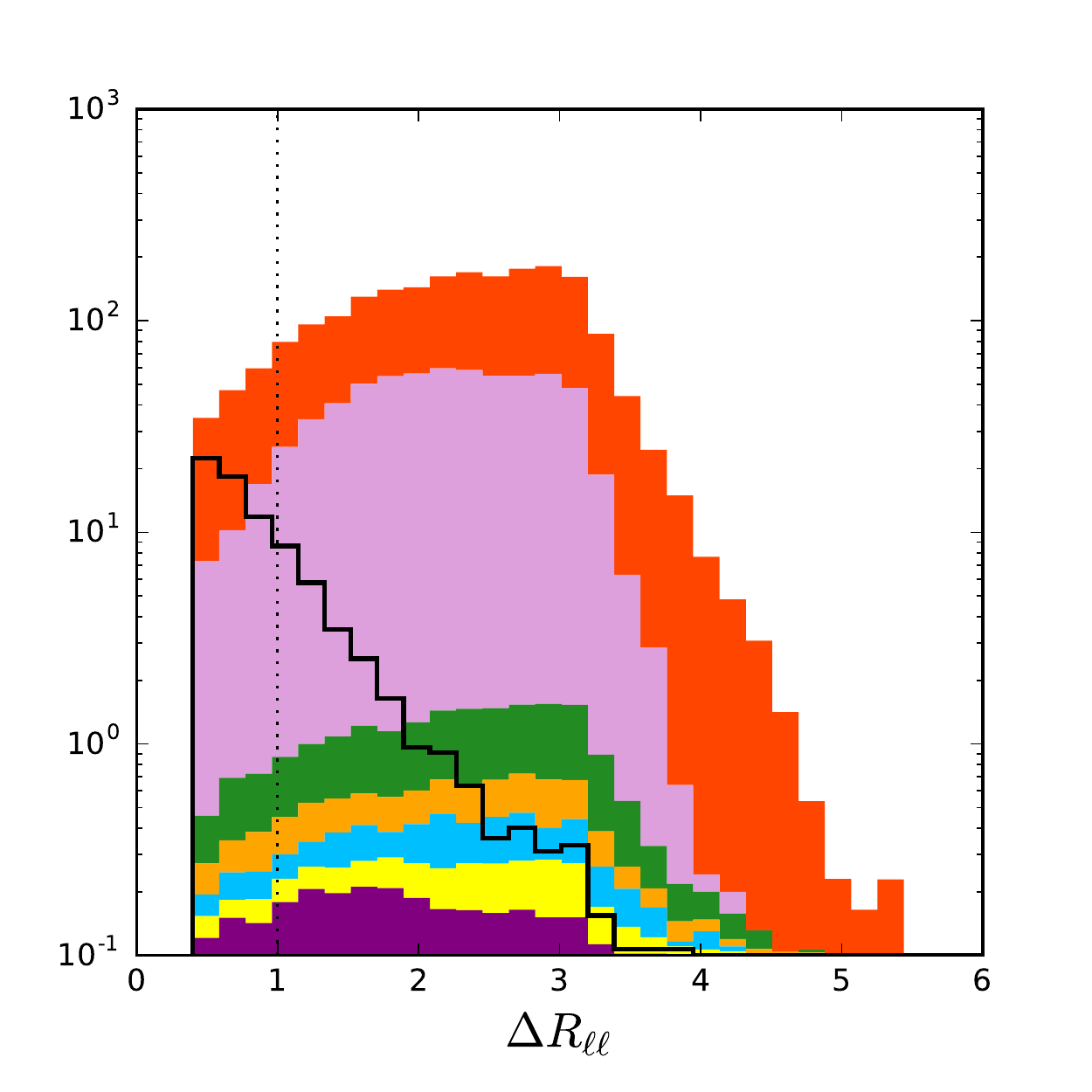} 
\\ \vspace*{-0.1cm}
\includegraphics[width=4.05cm]{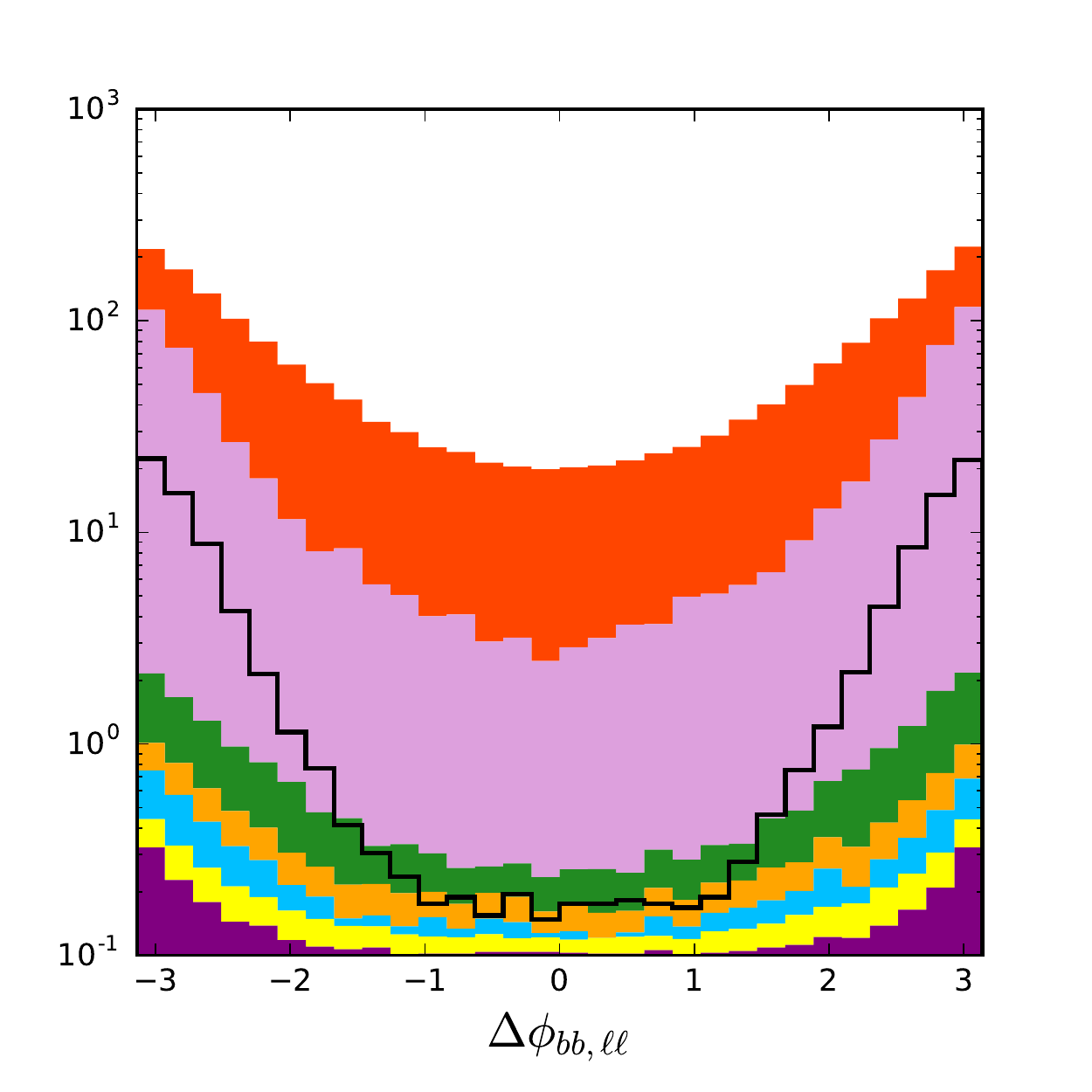} \hspace*{-0.5cm}
\includegraphics[width=4.05cm]{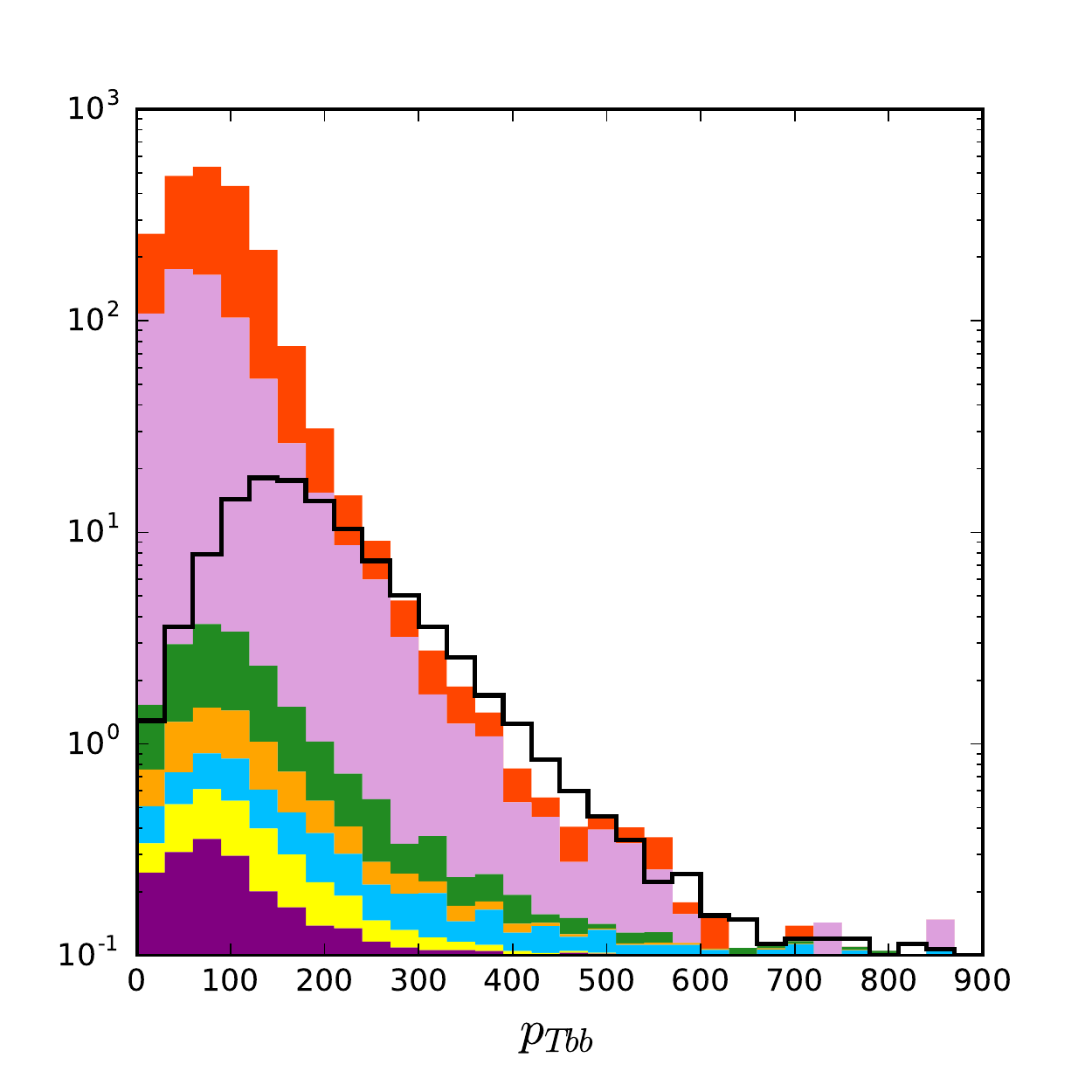} \hspace*{-0.5cm}
\includegraphics[width=4.05cm]{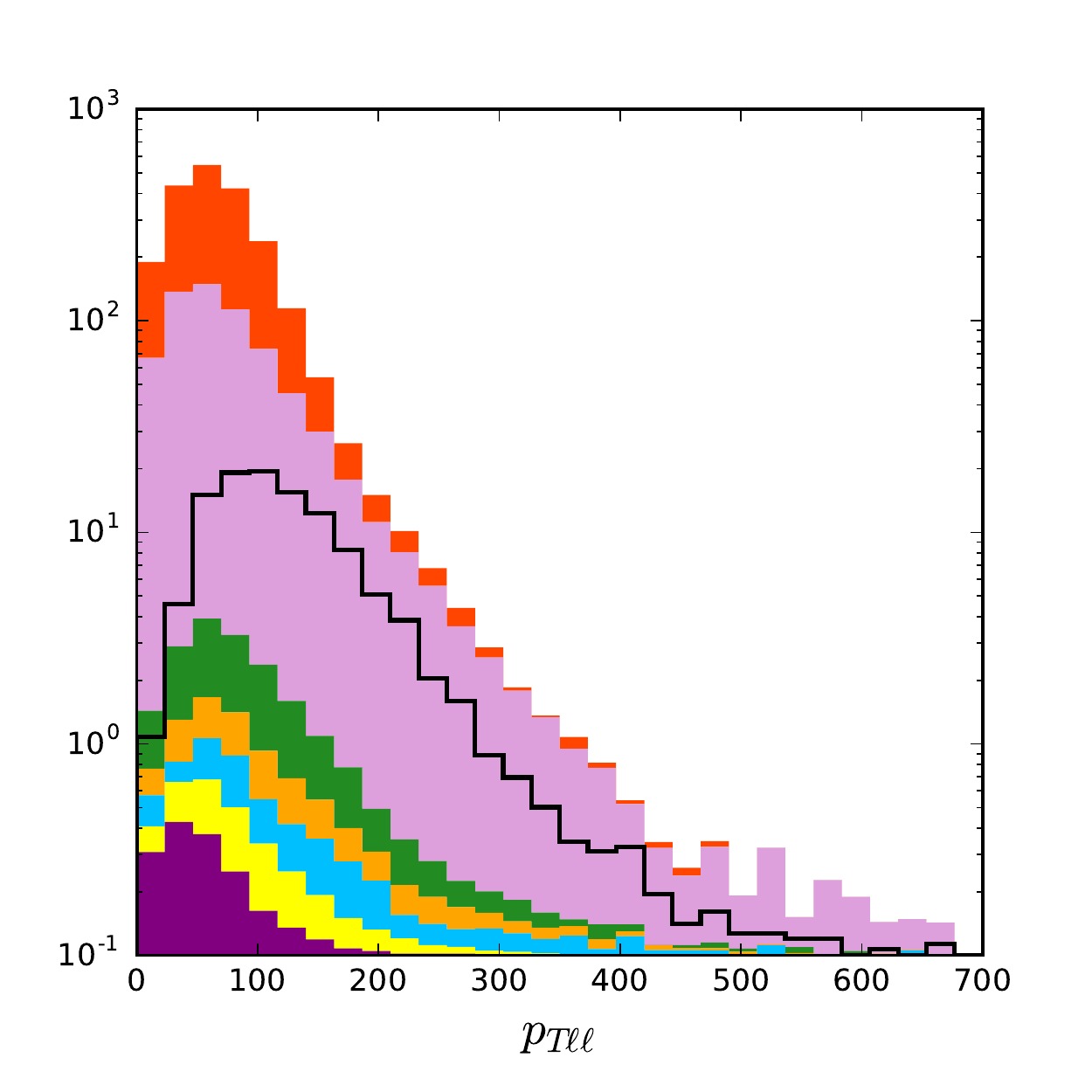} \hspace*{-0.5cm}
\includegraphics[width=4.05cm]{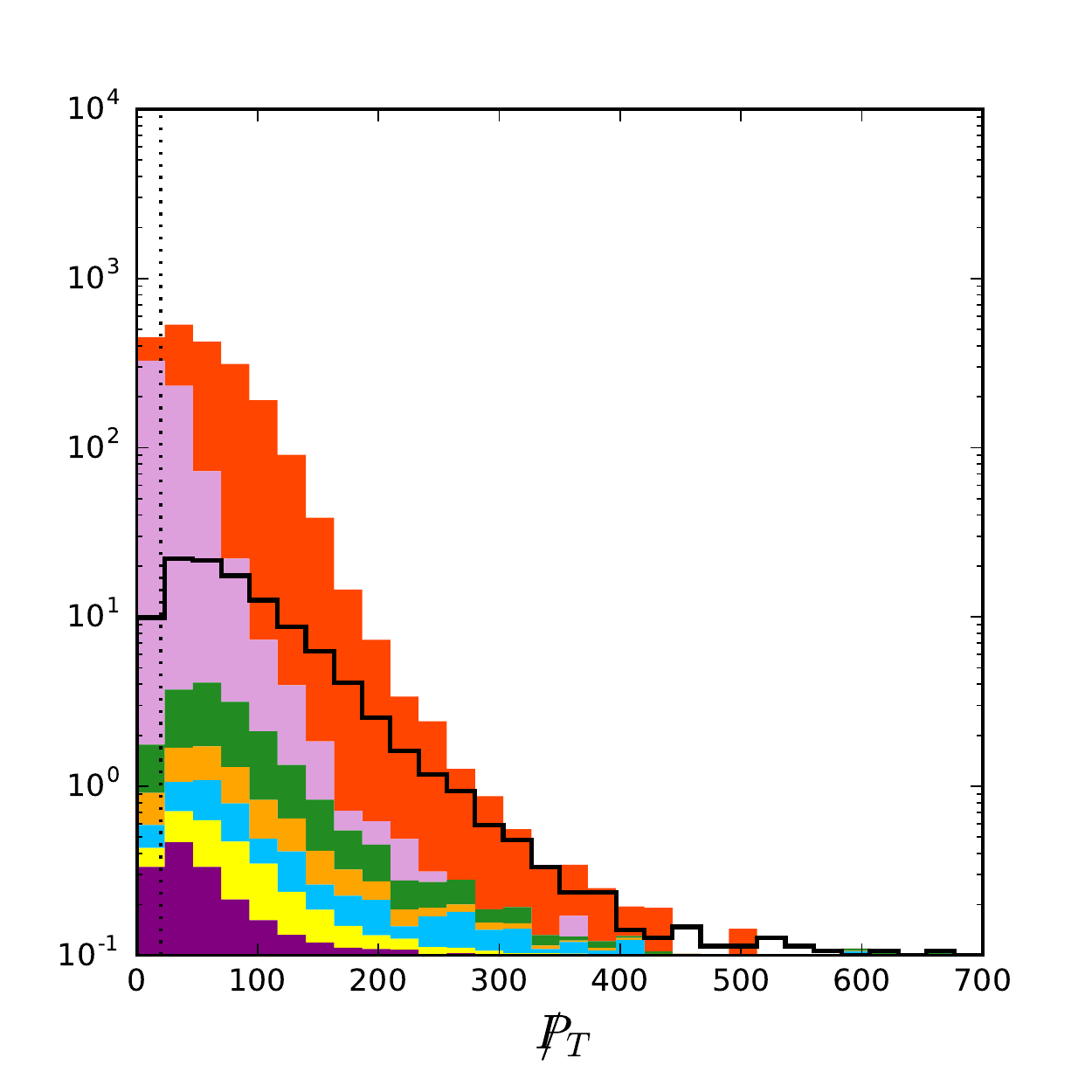} \\ 
\vspace*{-0.1cm}
\includegraphics[width=4.05cm]{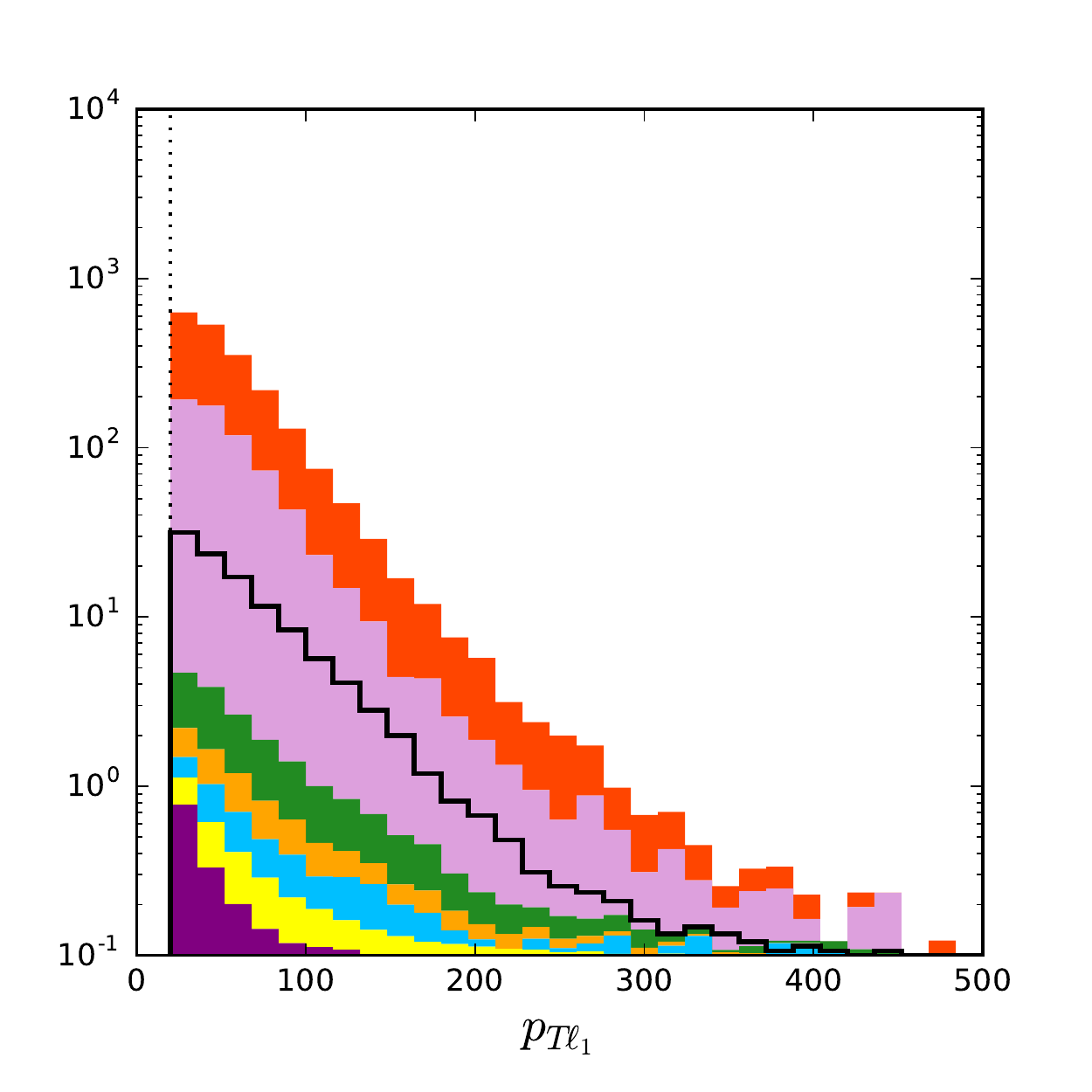} \hspace*{-0.5cm}
\includegraphics[width=4.05cm]{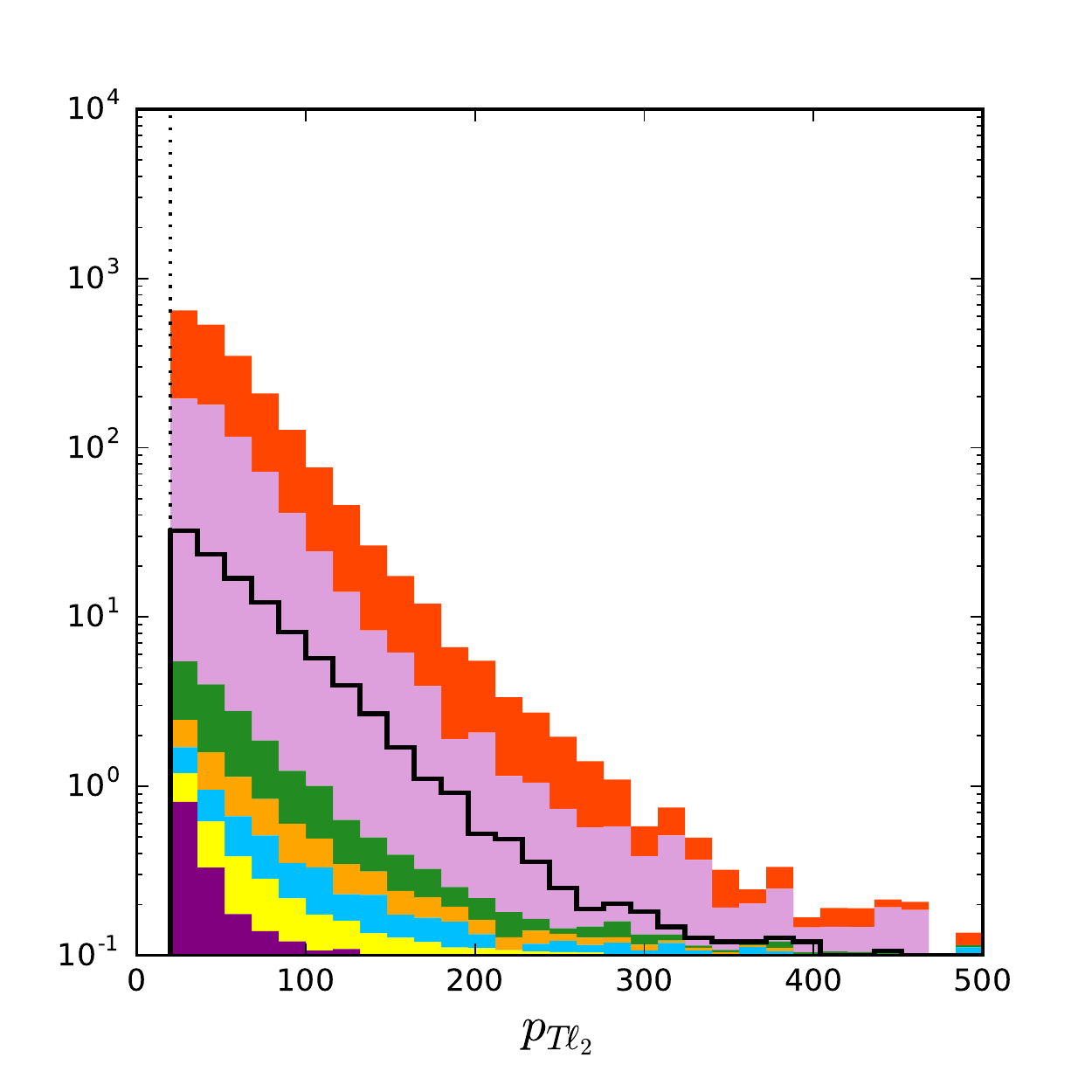} \hspace*{-0.5cm}
\includegraphics[width=4.05cm]{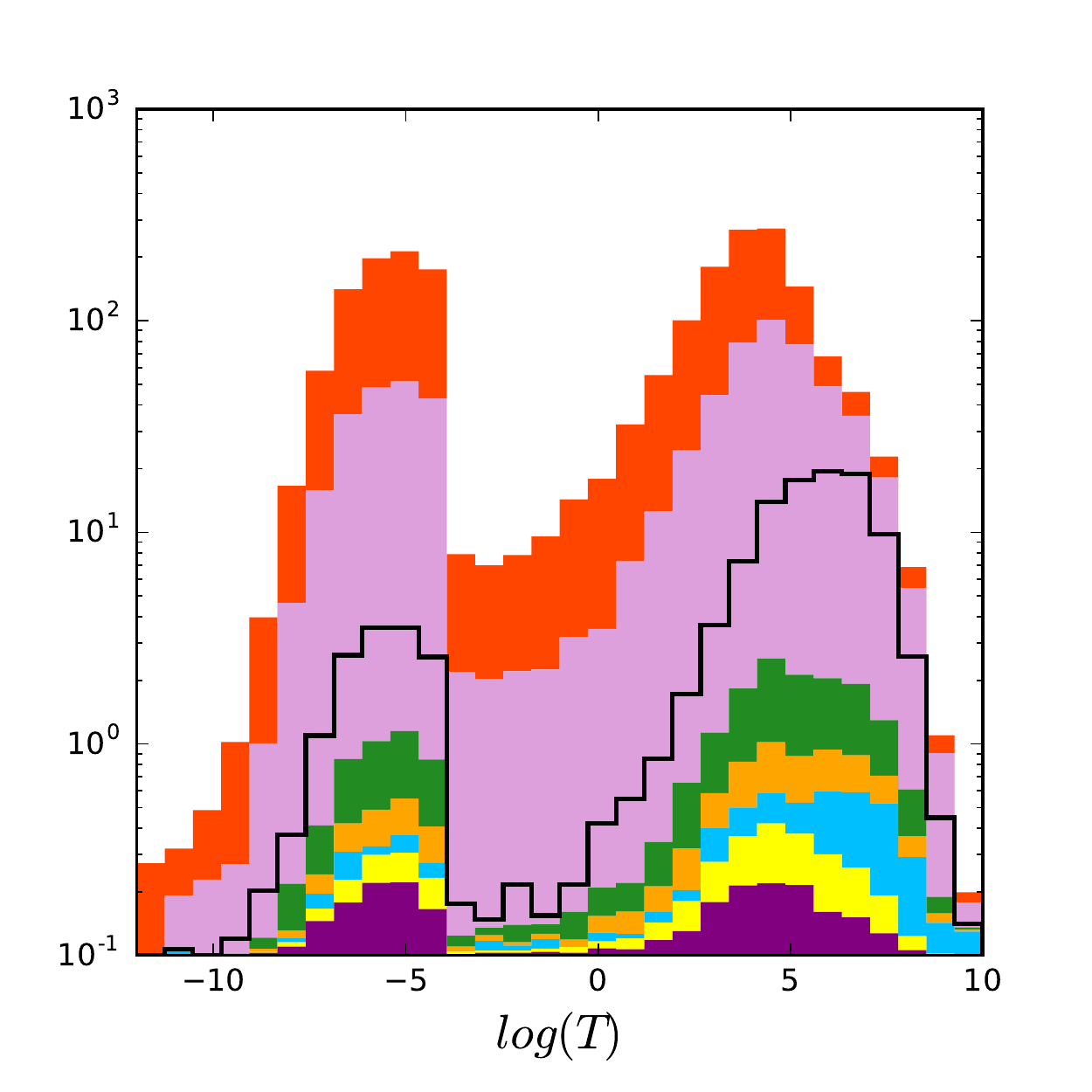} \hspace*{-0.5cm}
\includegraphics[width=4.05cm]{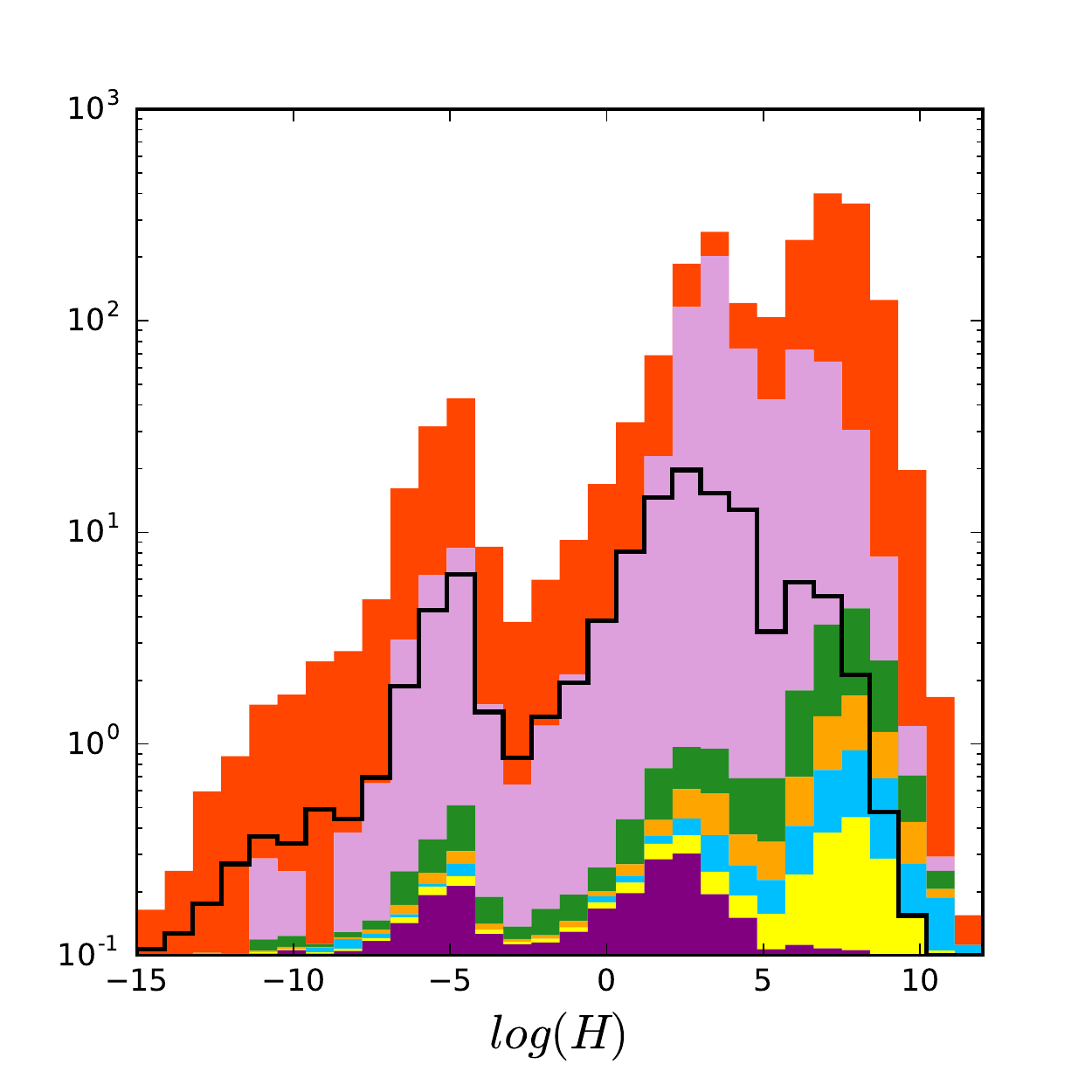}  \\ 
\vspace*{-0.1cm}
\includegraphics[width=4.05cm]{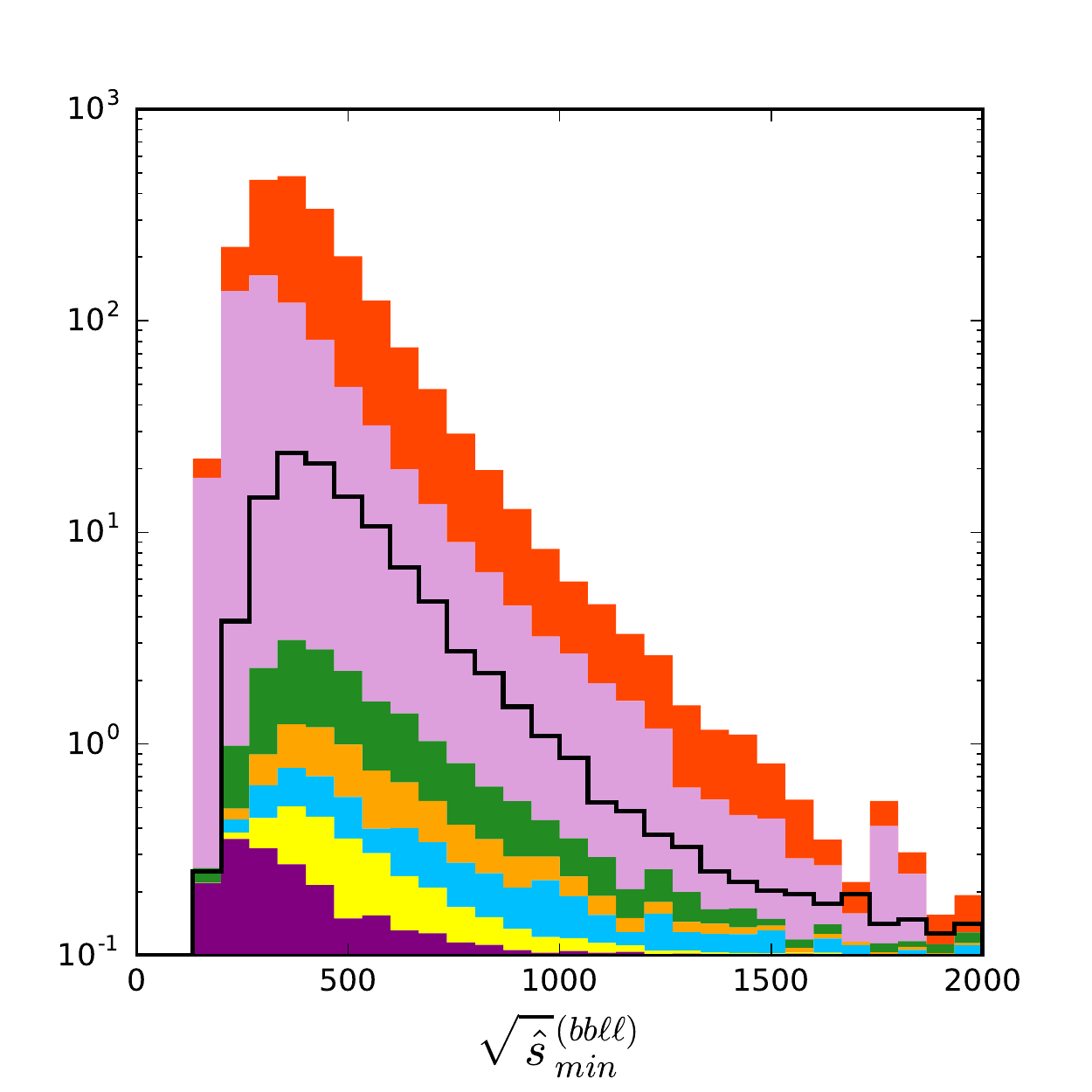} \hspace*{-0.5cm}
\includegraphics[width=4.05cm]{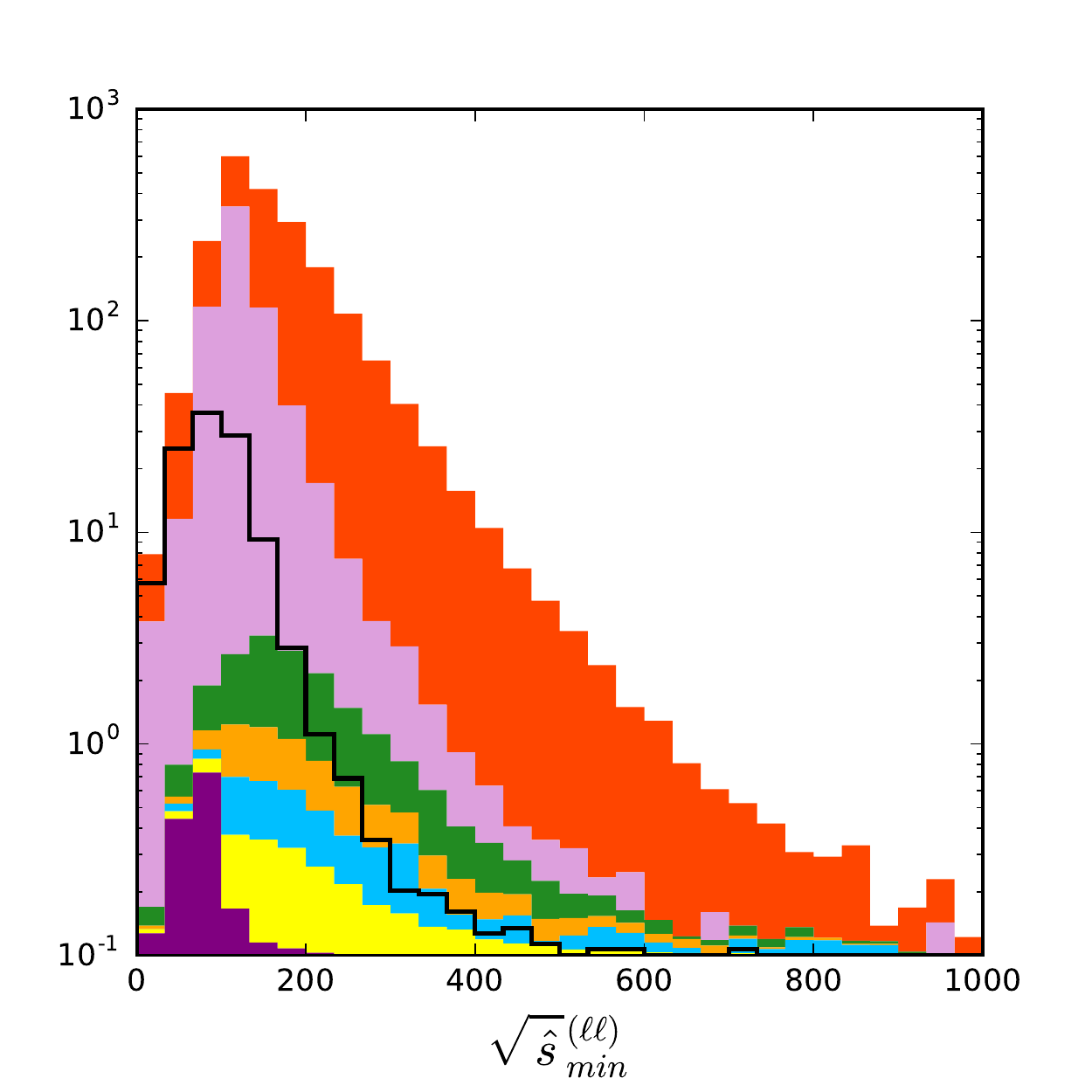} \hspace*{-0.5cm}
\includegraphics[width=4.05cm]{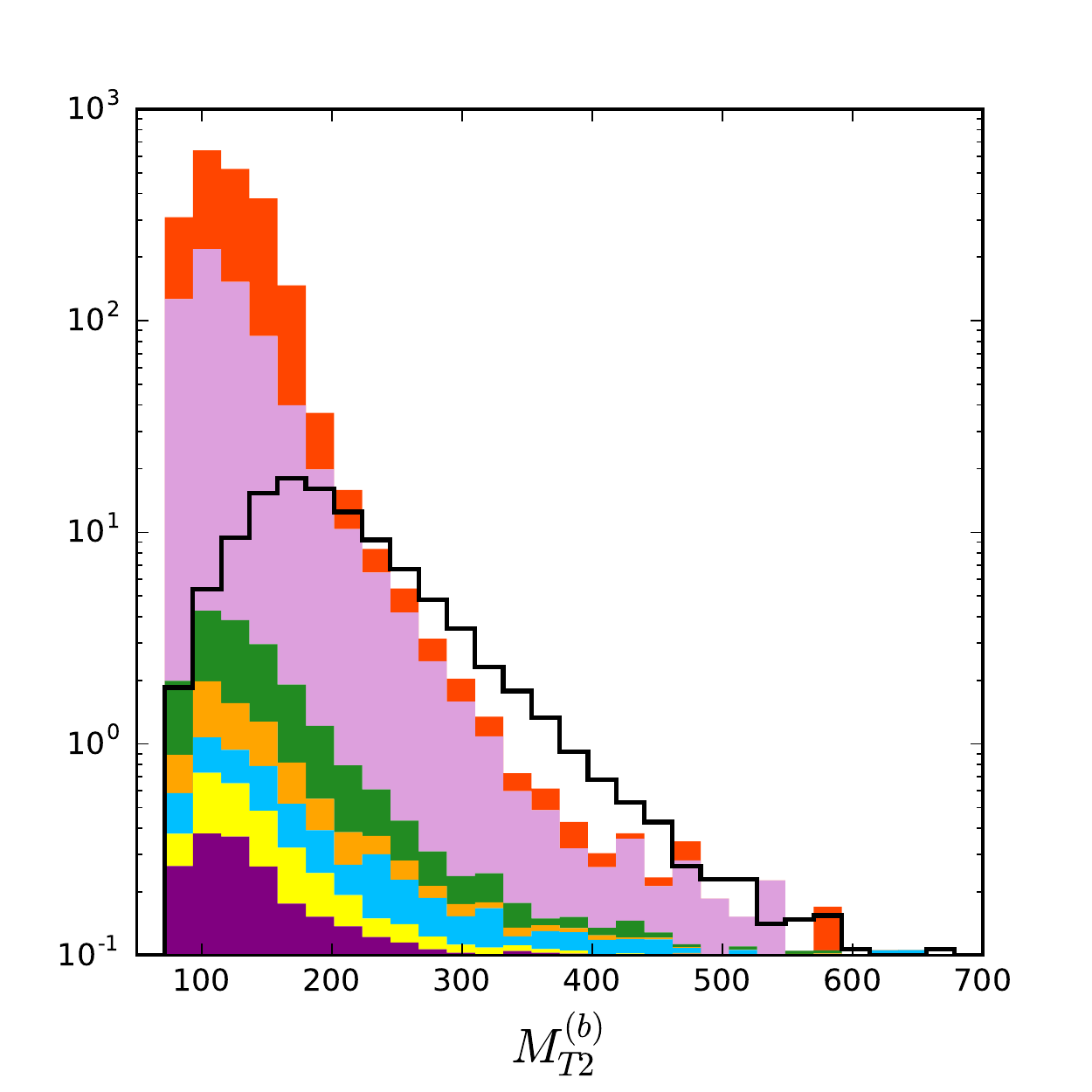} \hspace*{-0.5cm}
\includegraphics[width=4.05cm]{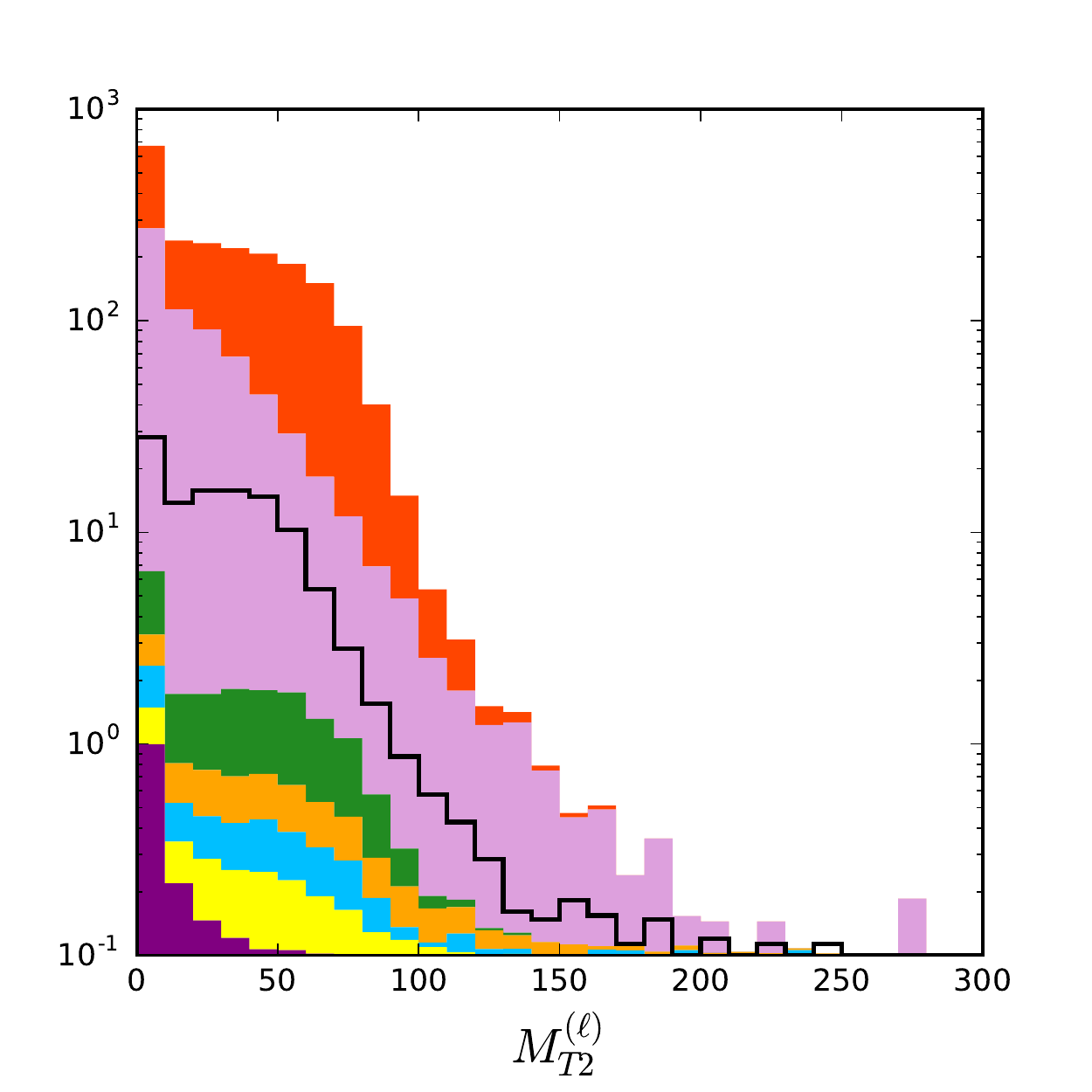}  \\
\includegraphics[width=14cm]{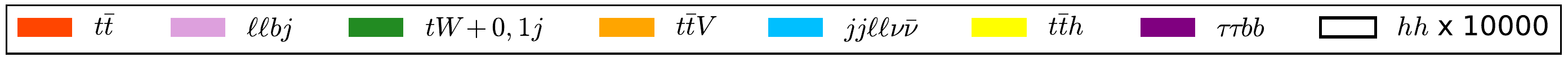} 
\caption{\label{fig:var1} 
Distributions of the 16 kinematic variables for signal ($hh$) and different types of backgrounds ($t\bar t$, $t\bar W$, $t\bar t V$, $t\bar t h$, $\tau\tau bb$, $\ell\ell b j$ and $jj\ell\ell\nu\nu$) before baseline cuts. The $y$-axis represents the number of events for each process and all individual distributions are normalized properly according to their respective cross-sections assuming 3 ab$^{-1}$ at the 14 TeV LHC. The dotted vertical lines indicate the baseline cuts introduced in Section \ref{sec:analysis}. 
}
\end{figure}

In this section we introduce the 16 kinematic variables used in this analysis. Their kinematic distributions (for signal and all relevant backgrounds) are shown in Fig.~\ref{fig:var1} and will be discussed shortly.

We begin with ten standard kinematic variables, which were previously considered in Refs.~\cite{Adhikary:2017jtu,CMS:2017cwx}
(their distributions are shown in the first ten panels of Fig.~\ref{fig:var1}): 
\begin{itemize}
\item $m_{b b}$, the invariant mass of the two $b$-tagged jets (1st plot in the 1st row). This is expected to be a good variable, since for signal events, the 
two $b$-jets originate from the decay of a narrow resonance (the Higgs boson) and would therefore reconstruct to the Higgs mass, up to resolution effects: $m_{bb}\sim m_h$.
This justifies the baseline cut of $95$ GeV $< m_{bb}<  140$ GeV, as indicated with the vertical dotted lines.
In contrast, no such correlations exists for backgrounds events: the two $b$-jets either originate from different decay chains and are uncorrelated (as in the case of $t\bar{t}$, for example),
or they reconstruct to the mass of a $Z$-boson or an off-shell gluon, with a mass lower than $m_h$. The plot in Fig.~\ref{fig:var1}, 
while confirming those expectations, also shows that the total background happens to peak at a value of $m_{bb}$ which, unfortunately, is not too far away from $m_h$, providing
the motivation to explore other variables.
\item $m_{\ell\ell}$, the invariant mass of the two leptons (2nd plot in the 1st row). For the case of the signal, the two leptons ultimately originate from the Higgs boson decay, 
and therefore their invariant mass $m_{\ell\ell}$ is bounded from above, hence the baseline cut of $m_{\ell\ell} < 65$ GeV. Note that the $m_{\ell\ell}$ distribution (which is observable) 
should be the same as the distribution of $m_{\nu\nu}$ (which is unobservable).
\item $\Delta R_{bb}$, the angular separation (\ref{deltaR}) between the two $b$-tagged jets (3rd plot in the 1st row). Given the relatively low Higgs mass, the two Higgs particles in $hh$ production
have sizable transverse momentum and their respective decay products (e.g., the two $b$-quarks) tend to go in the same direction. This and the next four variables try to exploit this kinematic property of the signal.
For example, the Higgs boost implies that $\Delta R_{bb}$ is relatively small for signal events, and this motivates the proximity cut of $\Delta R_{bb} < 1.3$.
\item $\Delta R_{\ell\ell}$, the angular separation (\ref{deltaR}) between the two leptons (4th plot in the 1st row). Here the same arguments apply as in the case of $\Delta R_{bb}$
just discussed. The corresponding plot in Fig.~\ref{fig:var1} confirms that the signal $\Delta R_{\ell\ell}$ distribution peaks well below most of the background processes, prompting the
baseline cut of $\Delta R_{\ell\ell} < 1.0$. 
\item $\Delta \phi_{bb, \ell\ell}$, the azimuthal angle in the transverse plane between the two $b$-jet system and the two lepton system (1st plot in the 2nd row). 
This is yet another way to capture the back-to-back boost of the two Higgs bosons in double Higgs production. 
Fig.~\ref{fig:var1} shows that the signal peaks at $\Delta \phi_{bb , \ell\ell}= \pm \pi$ more sharply than the background, which could be exploited later in the neural network analysis. 
However, no baseline cut was applied in this case, since $\Delta \phi_{bb , \ell\ell}$ is expected to be largely correlated with $\Delta R_{bb}$  and $\Delta R_{\ell\ell} $.
\item $p_{Tbb}$, the transverse momentum of the two $b$-jet system (2nd plot in the 2nd row). 
Like the previous three variables, this variable is motivated by the significant boost of the Higgs bosons in the signal, but no baseline cut was applied. 
\item $p_{T\ell\ell}$, the transverse momentum of the two lepton system (3rd plot in the 2nd row). 
This variable behaves similarly to $p_{Tbb}$, but to a lesser extent, since the two leptons come from separate $W$s, while the two $b$-quarks are direct decay products of the Higgs boson.
\item $\mpt = |\mptvec|$, the magnitude of the missing transverse momentum (4th plot in the 2nd row). 
A $\mpt$ cut is routinely applied in order to fight the QCD backgrounds (not shown in Fig.~\ref{fig:var1}). 
Following Ref. \cite{CMS:2017cwx}, here we use a baseline cut of $\mpt > 20 $ GeV.
\item $p_{T\ell_1}$, the transverse momentum of the hardest lepton (1st plot in the 3rd row).
\item $p_{T\ell_2}$, transverse momentum of the next-hardest lepton (2nd plot in the 3rd row).
As shown in Fig.~\ref{fig:var1}, the individual transverse momenta of the two leptons are similar for both signal and backgrounds.
Therefore, the lepton $p_T$'s may be good for triggering purposes, but not for background rejection.  
\end{itemize}
We note that for the signal, many of these 10 variables are strongly correlated to each other\footnote{The strong correlation arises due to the very nature of double Higgs production --- the two Higgs particles are produced with a sizable 
transverse momentum, which restricts the kinematics of their decay products.}. This implies that cutting on one variable significantly reduces the power of other variables.
At the same time, while these 10 variables are among the most commonly used in high energy physics, it is not guaranteed that they fully capture all 
kinematic differences between signal and background. This is why we introduce six additional variables \cite{Kim:2018cxf}: Topness, Higgsness, 
$\sqrt{\hat s}^{(bb\ell\ell)}_{min}$, $\sqrt{\hat s}^{(\ell\ell)}_{min}$, $M_{T2}^{(b)}$ and $M_{T2}^{(\ell)}$, shown in the last six panels of Fig.~\ref{fig:var1},
which are meant to take full advantage of the kinematic differences between the signal and background event topologies.

The Topness variable measures the degree of consistency of a given event with the kinematics of 
dilepton $t\bar t$ production, where there are 6 unknowns (the three-momenta of the two neutrinos, $\vec p_{\nu}$ and $\vec p_{\bar\nu}$) and four on-shell constraints, 
$m_t$, $m_{\bar t}$, $m_{W^+}$ and $m_{W^-}$. 
Here $m_t = m_{\bar t}$ is the mass of top or antitop quark, and $m_{W^\pm} = m_{W}$ is the mass of the $W$ boson.
Then the neutrino momenta can be fixed by minimizing the following quantity 
\begin{eqnarray}
\chi^2_{ij} &\equiv& \min_{\tiny \mptvec = \vec p_{T \nu } + \vec p_{T \bar\nu }}  \left [ 
\frac{\left ( m^2_{b_i \ell^+ \nu} - m^2_t \right )^2}{\sigma_t^4} \,   +
\frac{\left ( m^2_{\ell^+ \nu} - m^2_W \right )^2}{\sigma_W^4}  \, \right. \nonumber  \\ 
&&\qquad\qquad\qquad \left. +\, \frac{\left ( m^2_{b_j \ell^- \bar \nu} - m^2_t \right )^2}{\sigma_t^4} \, +
\frac{\left ( m^2_{\ell^- \bar\nu} - m^2_W \right )^2}{\sigma_W^4}   \right ]  , 
\label{eq:tt}  
\end{eqnarray}
subject to the missing transverse momentum constraint, $ \mptvec = \vec p_{T \nu } + \vec p_{T \bar\nu }$. 
The parameters $\sigma_t$ and $\sigma_W$ are indicative of the corresponding experimental resolutions and intrinsic particle widths.
In principle, they can be treated as free parameters and one can tune them using NN, BDT, etc. In our numerical
study, we shall use $\sigma_t=5$ GeV and $\sigma_W=5$ GeV.
Since there is a twofold ambiguity in the paring of a $b$-quark and a lepton, Topness is defined as the smaller of the two $\chi^2$s \cite{Kim:2018cxf},
\begin{eqnarray}
T &\equiv&  {\rm min} \left ( \chi^2_{12} \, , \, \chi^2_{21} \right ) \, .
\end{eqnarray}
The Topness distributions for both signal and backgrounds before baseline cuts are shown in Fig.~\ref{fig:var1} (3rd plot in the 3rd row). 
We observe that, as expected, $T$ tends to have smaller values for the main background ($t\bar{t}$) than for signal.

In our signal of $hh$ production, the two $b$-quarks arise from a Higgs decay ($h \to b \bar b$), and therefore their invariant mass $m_{bb}$
can be used as a first cut to enhance the signal sensitivity. 
For the decay of the other Higgs boson, $h \to W^\pm W^{*\mp}$, Higgsness is defined as follows \cite{Kim:2018cxf} 
\begin{eqnarray}
H &\equiv&    {\rm min}_{\tiny \mptvec = \vec p_{T \nu } + \vec p_{T \bar\nu }}   \left [
 \frac{\left ( m^2_{\ell^+\ell^-\nu \bar\nu} - m^2_h \right )^2}{\sigma_{h_\ell}^4}    \right. 
 + \frac{ \left ( m_{\nu  \bar\nu}^2 -  (m^{peak}_{\nu\bar\nu})^2 \right )^2}{ \sigma^4_{\nu}}  \nonumber \\
&&\qquad \qquad \qquad \qquad  +\, {\rm min} \left ( 
\frac{\left ( m^2_{\ell^+ \nu } - m^2_W \right )^2}{\sigma_W^4} + 
\frac{\left ( m^2_{\ell^- \bar \nu} - (m^{peak}_{W^*})^2 \right )^2}{\sigma_{W^*}^4}  \, , \right.  \nonumber \\
&&\qquad \qquad \qquad \qquad \qquad \qquad 
\left. \left.
\frac{\left ( m^2_{\ell^- \bar \nu} - m^2_W \right )^2}{\sigma_W^4} + 
\frac{\left ( m^2_{\ell^+ \nu} - (m^{peak}_{W^*})^2 \right )^2}{\sigma_{W^*}^4}  
\right )  \right ]   \, .  \label{eq:hww}
\end{eqnarray}
It tests whether the neutrino kinematics can be compatible with having the Higgs boson and one of the $W$-bosons on-shell, 
while at the same time being consistent with the invariant mass distributions expected for the off-shell $W$-boson, $W^\ast$, and the 
neutrino pair, $\nu\bar\nu$.
The invariant mass $m_{W^*}$ is bounded by $0 \leq m_{W^*} \leq m_h- m_W$ and the peak of its distribution is at 
\begin{equation}
m_{W^*}^{peak} = \frac{1}{\sqrt{3}} \sqrt{ 2 \left ( m_h^2 + m_W^2 \right ) - \sqrt{m_h^4 + 14 m_h^2 m_W^2 + m_W^4}} \, .
\label{mwastpeak}
\end{equation}

\begin{figure}[t]
\centering
\includegraphics[width=7cm]{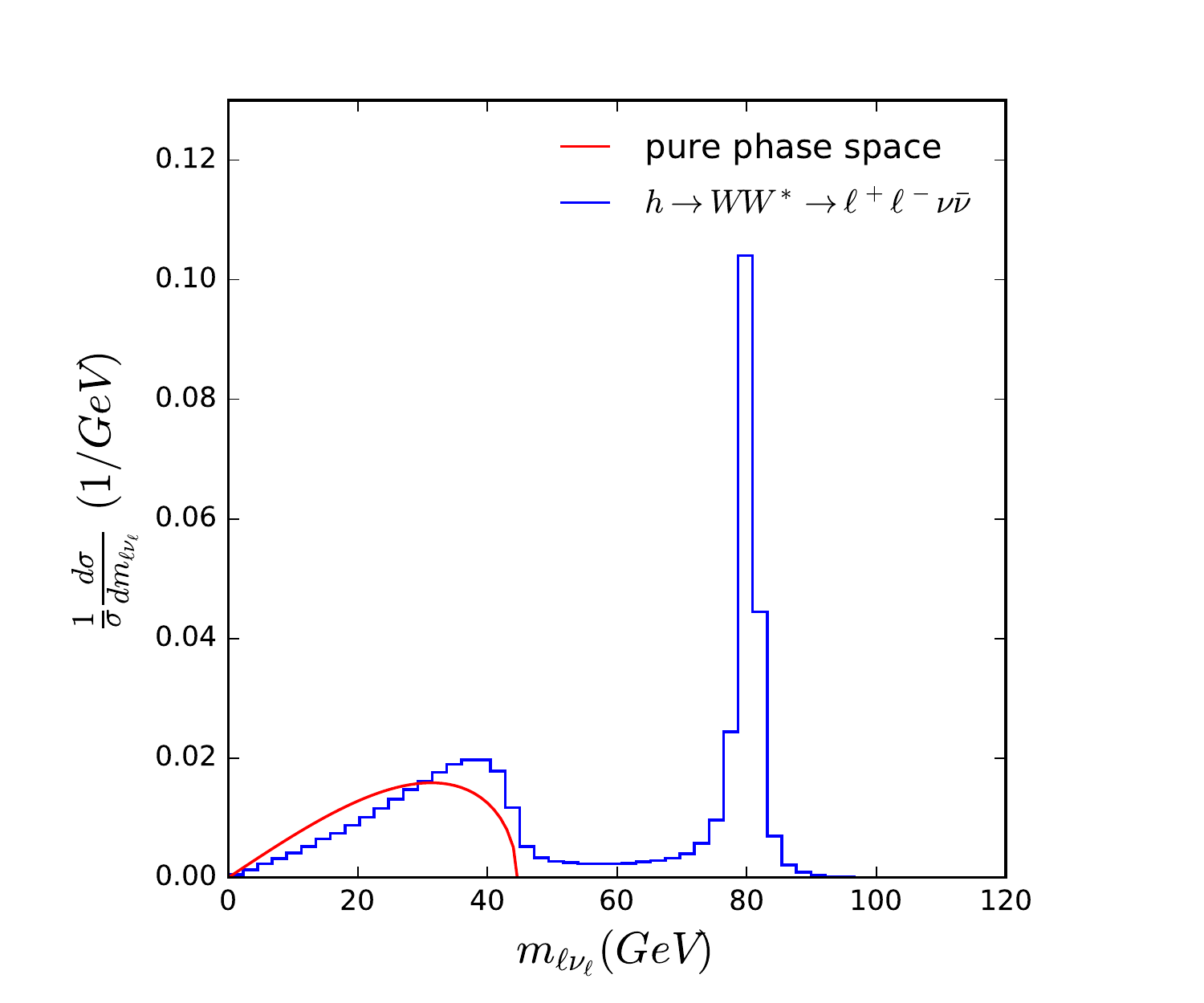} \hspace*{-0.5cm}
\includegraphics[width=7cm]{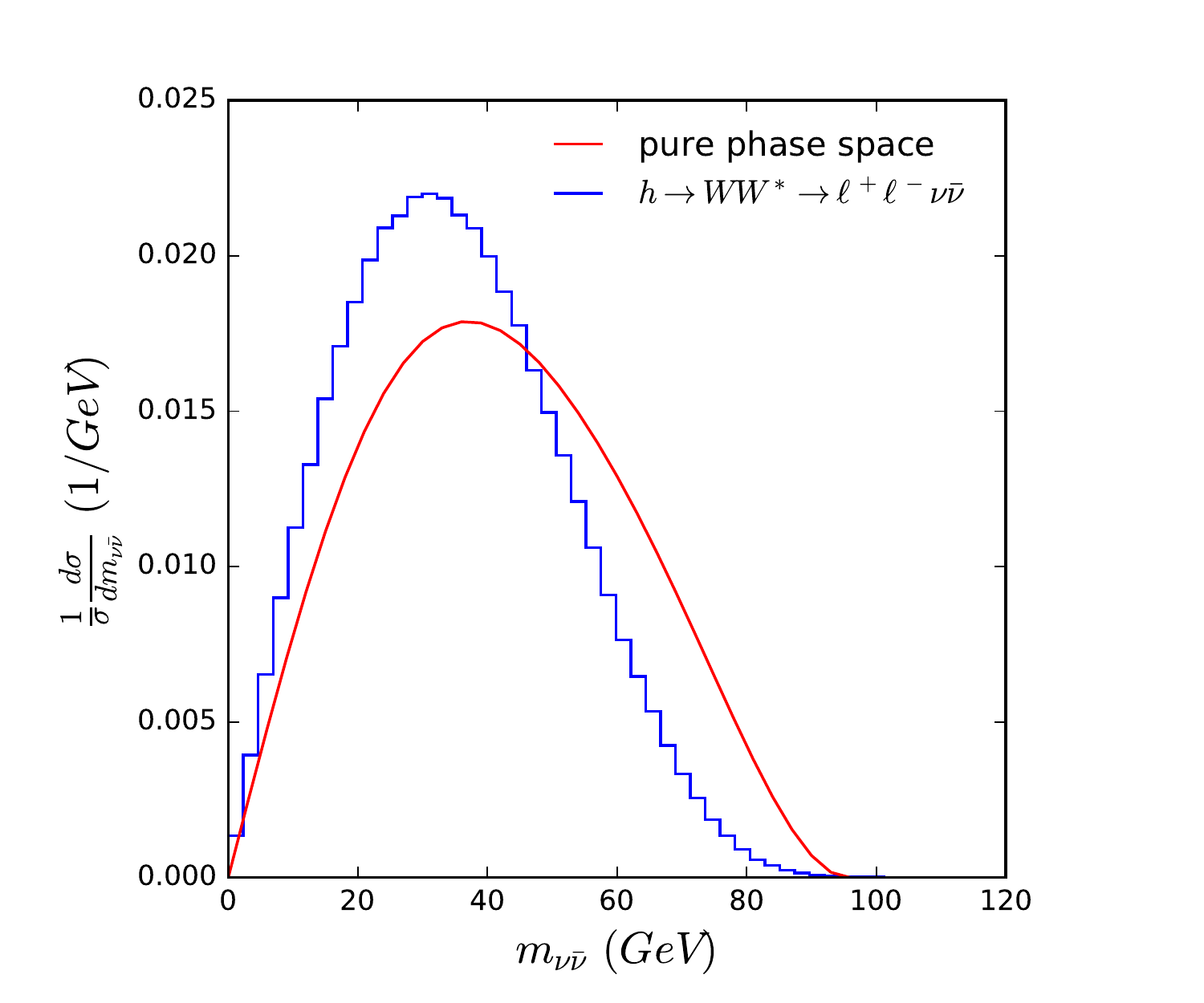}
\caption{\label{fig:mnnmll} 
Unit-normalized invariant mass distribution of the the lepton-neutrino ($m_{\ell\nu}$, left) and the two neutrinos ($m_{\nu\bar\nu}$, right).}
\end{figure}

The left panel of Fig.~\ref{fig:mnnmll} shows the unit-normalized invariant mass distribution of the proper lepton-neutrino system ($m_{\ell\nu}$). 
The distribution has a bimodal shape --- the narrow peak on the right near 80 GeV corresponds to the on-shell $W$-boson resonance, 
while the broader hump to the left is due to the off-shell $W^\ast$, with
a clear end-point at $m_h- m_W=45$ GeV and a maximum near {$m_{W^*}^{peak}=40$ GeV} in accordance with (\ref{mwastpeak}). 

The definition of Higgsness (\ref{eq:hww}) also includes a term which tests for consistency with the expected invariant mass distribution 
$\frac{\textrm{d}\sigma}{\textrm{d} m_{\nu\bar\nu}}$ for the neutrino pair\footnote{In the limit of massless leptons, the distribution 
$\frac{\textrm{d}\sigma}{\textrm{d} m_{\nu\bar\nu}}$ is the same as the dilepton mass distribution 
$\frac{\textrm{d}\sigma}{\textrm{d} m_{\ell^+\ell^-}}$, which is directly observable and therefore more 
commonly discussed in the literature \cite{Han:2009ss,Han:2012nr,Han:2012nm,Cho:2012er}.}, which is shown in the right panel of Fig.~\ref{fig:mnnmll}.
The red solid curve gives the pure phase space prediction 
\begin{equation}
\frac{\textrm{d}\sigma}{\textrm{d}m_{\nu \bar\nu}} \propto \int\textrm{d}m_{W^*}^2 \lambda^{1/2}(m_h^2,m_W^2, m_{W^*}^2) f(m_{\nu \bar\nu}) \, , 
\label{ps}
\end{equation}
where 
$\lambda(x,y,z)=x^2+y^2+z^2-2xy-2yz-2zx$ is the two-body phase space function and $f(m)$ is the invariant mass distribution of the antler topology with $h\to W W^* \to \ell^+ \ell^- \nu \bar \nu$:
\begin{eqnarray}
f(m)\sim \left\{
\begin{array}{l l}
\eta\, m \, , & 0 \leq m \leq e^{-\eta}E, \\ [1mm]
m \ln(E/m)  \, , & e^{-\eta}E \leq m \leq E,
\end{array}\right.  \label{eq:antlerf}
\end{eqnarray}
where the endpoint $E$ and the parameter $\eta$ are defined in terms of the particle masses as
\begin{eqnarray}
E              &=&  \sqrt{ m_W m_{W^*} \, e^{\eta}}  \, ,  \\ [2mm]
\cosh \eta &=&\left ( \frac{m_h^2- m_{W}^2 - m_{W^*}^2}{2 m_{W} m_{W^*} } \right ) \, .
\label{etadef}
\end{eqnarray}
Note that by allowing one of the $W$-bosons to be on-shell, eqs.~(\ref{ps}-\ref{etadef})
generalize the results previously derived in Refs. 
\cite{Han:2009ss,Han:2012nr,Han:2012nm,Cho:2012er} for the purely on-shell case.
The blue histogram in the right panel of Fig.~\ref{fig:mnnmll} shows the actual $m_{\nu\bar\nu}$ distribution, whose shape
is slightly different from the pure phase space result (\ref{ps}), due a helicity suppression in the $W$-$\ell$-$\nu$ vertex. 
In particular, we observe that the actual peak is at $m_{\nu\bar\nu}^{peak} \approx 30$ GeV, 
which is the value that we shall use in the definition of Higgsness (\ref{eq:hww}).\footnote{We note that other variants of Higgsness
are also possible --- for example, instead of penalizing the function $H$ by the distances to the peaks in the corresponding distributions, 
one can introduce penalty terms which take advantage of the knowledge of the exact probability distributions (the blue histograms in Fig.~\ref{fig:mnnmll}). }

The definition of Higgsness (\ref{eq:hww}) contains some additional resolution parameters:
$\sigma_h$ for the reconstructed mass of the Higgs boson, 
$\sigma_{W^\ast}$ for the reconstructed mass of the off-shell $W$ boson,
and $\sigma_{\nu}$ for the $m_{\nu\bar\nu}$ resolution.
In what follows, we shall take $\sigma_{W^\ast}=5$ GeV, $\sigma_{h_\ell}=2$ GeV, and $\sigma_\nu = 10 $ GeV.\footnote{We have checked that our results are not very sensitive to these choices.}

\begin{figure}[t]
\centering
\includegraphics[width=14cm]{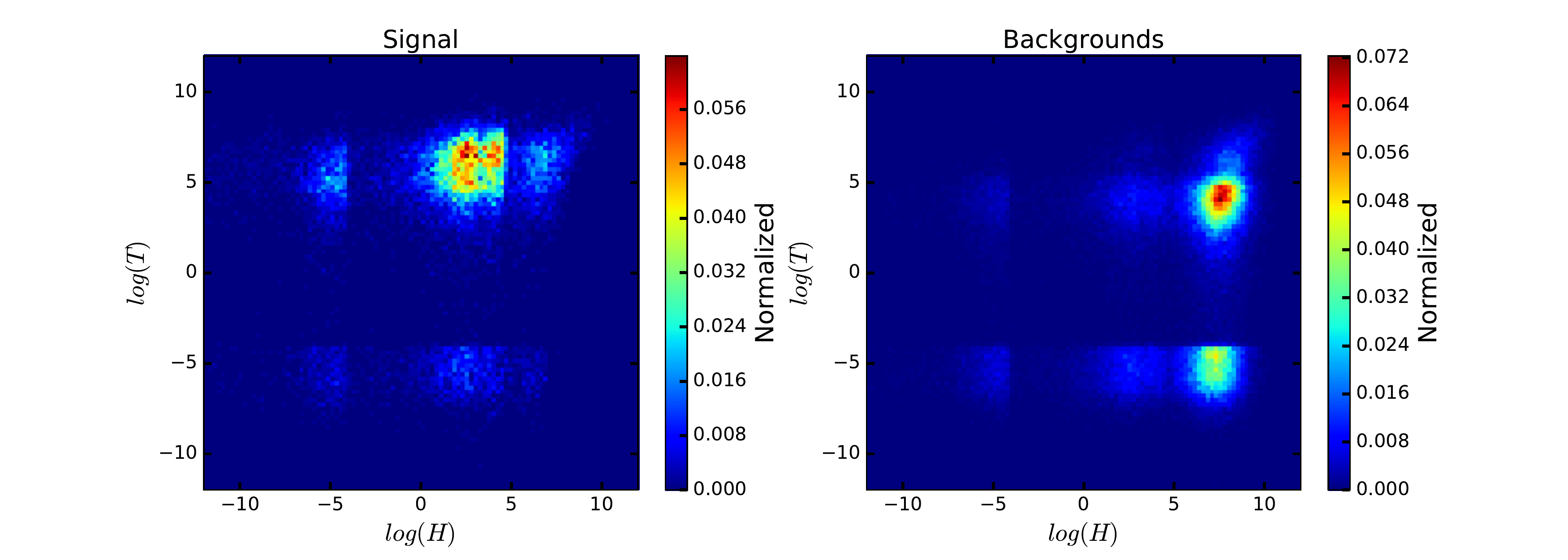}
\includegraphics[width=14cm]{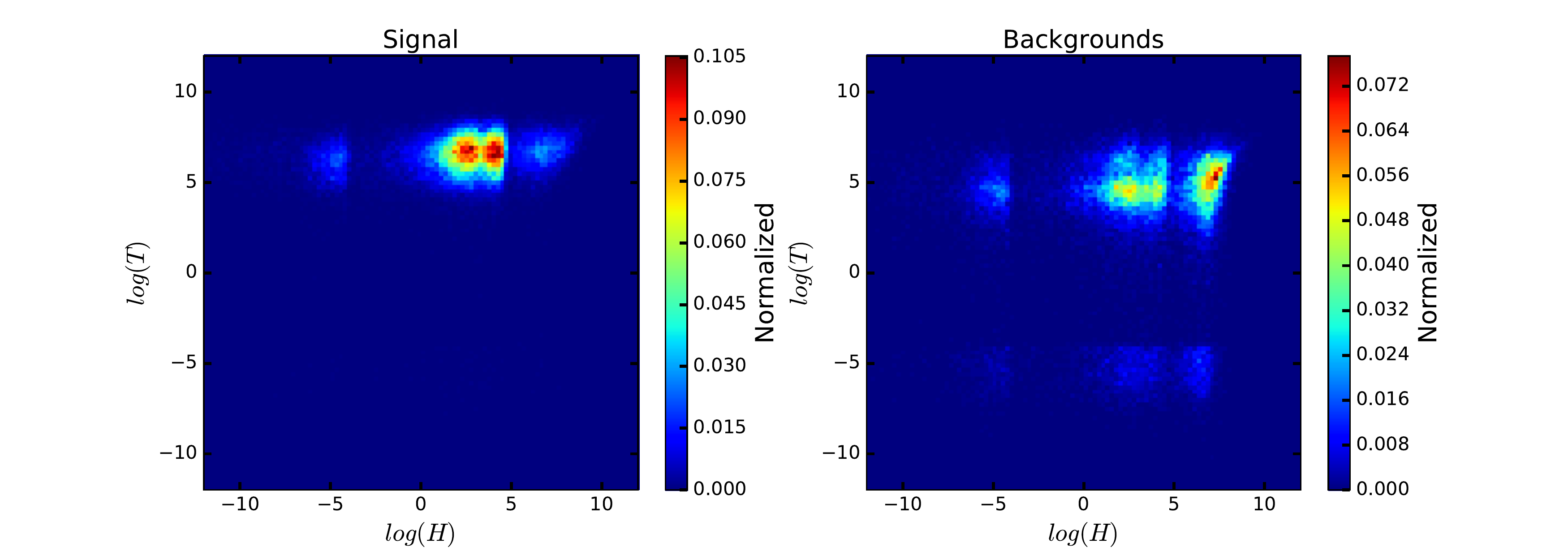}
\caption{\label{fig:THness} Two-dimensional correlation plots for 
Higgsness and Topness for signal (left) and backgrounds (right) before (top) and after (bottom) baseline cuts.}
\end{figure}

The Higgsness distributions for both signal and backgrounds before baseline cuts are shown in Fig. \ref{fig:var1} 
(4th plot in the 3rd line). 
The two dimensional map of (Higgsness, Topness) on a log-log scale is depicted in Fig.~\ref{fig:THness}. 
The Higgsness and Topness distributions in Fig. \ref{fig:var1} are projections of this two dimensional scatter plot onto the $x$-axis and $y$-axis, respectively. 
Although the signal and the backgrounds do not exhibit a very clean separation in the individual one-dimensional projections in Fig. \ref{fig:var1}, 
their two dimensional correlation plots show some visible differences. 
We note that even after employing the baseline cuts, one can still see a difference in the two dimensional correlation of Higgsness and Topness (bottom row plots). 

Along with Higgsness and Topness, we also consider two versions of the $ \hat{s}_{min}$ variable \cite{Konar:2008ei,Konar:2010ma}, which is defined as 
\begin{equation}
 \hat{s}_{min}^{({\rm v})} = m_{{\rm v}}^2 + 2 \left ( \sqrt{ |\vec P_{T}^{\rm v} |^2 + m_{\rm v}^2 }\ |\mptvec| - \vec P_{T}^{\rm v} \cdot \mptvec \right ) \, ,
 \label{smindef}
\end{equation}
where $({\rm v})$ represents a set of visible particles under consideration, while $m_{\rm v}$ and $\vec P_{T}^{\rm v}$ are their invariant mass and transverse momentum, respectively. 
The variable (\ref{smindef}) characterizes the system comprising of the visible particles $({\rm v})$ 
and the invisible particles (here assumed to be massless) which are responsible for the measured missing transverse momentum $\mptvec$. 
It provides the minimum value of the Mandelstam invariant mass $\hat s$ for the system which is consistent with the observed visible 4-momentum vector.
We shall apply (\ref{smindef}) to the whole event, where ${\rm v}=\{bb\ell\ell\}$, or to the 
subsystem resulting from the decay $h \to W^\pm W^{*\mp}\to\ell^+\ell^- \nu \bar \nu$, where ${\rm v}=\{\ell\ell\}$. 
The distributions of the resulting variables $\hat{s}_{min}^{(bb\ell\ell)}$ and $\hat{s}_{min}^{(\ell\ell)}$
are shown in the left two panels on the fourth row of Fig.~\ref{fig:var1}. 
The $\hat{s}_{min}^{(bb\ell\ell)}$ variable represents the minimum energy required to produce the two original parent particles 
(the two Higgs bosons in the case of the signal and the two top quarks in the case of the major $t\bar{t}$ background).
This is why one would expect the distribution to peak around the parent mass threshold, $2m_h$ for the signal and $2m_t$ for the background \cite{Konar:2008ei}.
However, the first panel in the fourth row of Fig.~\ref{fig:var1} shows that while the background $\hat{s}_{min}^{(bb\ell\ell)}$ distribution peaks near $2 m_t$, which is expected,
the signal $\hat{s}_{min}^{(bb\ell\ell)}$ distribution peaks around 400 GeV, which is substantially higher than $2m_h$. 
This implies that the two top quarks are produced more or less at rest, while the two Higgs bosons have a sizable boost. 
Similarly, the variable $\hat{s}_{min}^{(\ell\ell)}$ is the minimum energy required to produce the two $W$ bosons. 
For the $t\bar{t}$ background, where both $W$ bosons are on-shell, the peak is expected to occur around $2 m_W$.
On the other hand, the signal distribution should be softer, since 
one of the $W$ bosons is off-shell, and furthermore, the peak should be located slightly below the Higgs boson mass. 
These kinematic differences are illustrated in the second plot on the fourth row of Fig.~\ref{fig:var1}, and motivate the use of 
$\hat{s}_{min}^{(\ell\ell)}$ as an analysis variable.

The last two panels in the fourth row of Fig.~\ref{fig:var1} show distributions of the subsystem $M_{T2}$ variable \cite{Burns:2008va} --- first when 
it is applied to the $b \bar b$ visible system resulting from the $t\to bW$ decays ($M_{T2}^{(b)}$), and then when it is applied to the 
$\ell^+\ell^-$ visible system resulting from the $W\to \ell\nu$ decays ($M_{T2}^{(\ell)}$). In principle, $M_{T2}$ is defined as \cite{Lester:1999tx}
\begin{equation}
M_{T2} (\tilde m) \equiv \min\left\{\max\left[M_{TP_1}(\vec{p}_{T \nu },\tilde m),\;M_{TP_2} (\vec{p}_{T \bar\nu },\tilde m)\right] \right\} \, , 
\label{MT2def}
\end{equation}
where the minimization over the transverse masses of the parent particles $M_{TP_i}$ ($i=1, 2$) is performed over the transverse neutrino 
momenta $\vec{p}_{\nu T}$ and $\vec{p}_{\bar\nu T}$, subject to the $\mptvec $ 
constraint\footnote{See Refs.~\cite{Barr:2011xt,Kim:2017awi,Cho:2014naa,Konar:2009wn,Konar:2009qr,Baringer:2011nh,Kim:2015uea,Goncalves:2018agy,Debnath:2017ktz} 
for more information and other variants of $M_{T2}$.}. 
The parameter $\tilde m$ in (\ref{MT2def}) is the test mass for the daughter particle:
in the case of $M_{T2}^{(\ell)}$ one should use $\tilde m = m_\nu = 0$, 
while in the case of $M_{T2}^{(b)}$, the daughter particles are the $W$ bosons, and $\tilde m = m_W = 80$ GeV,
which leads to the lower bound $m_W \le M_{T2}^{(b)}$ visible in the plot.
By construction, the $M_{T2}$ variables are bounded by the mass of the corresponding parent particle. Indeed,
the $M_{T2}^{(b)}$ distribution for $t\bar t$ production shows a sharp drop around $M_{T2}^{(b)} = m_t$, 
while the signal distribution extends well above $m_t$. Similarly, the $M_{T2}^{(\ell)}$ distribution for $t\bar{t}$ 
drops around $m_W$, as expected. In addition, it exhibits a peak structure in the first bin, which is due to leptonic tau decays.
This suggests that $M_{T2}^{(\ell)}$ can be effective in eliminating backgrounds with $\tau$s. 

This concludes our discussion of the 16 kinematic variables depicted in Fig.~\ref{fig:var1}.
The newly introduced 6 variables (Topness, Higgsness, $\hat s_{min}^{(bb\ell\ell)}$, $\hat s_{min}^{(\ell\ell)}$, $M_{T2}^{(b)}$ and $M_{T2}^{(\ell)}$) 
typically require a few extra steps to compute them, thus we shall refer to them as high-level kinematic variables, 
while the remaining 10 traditional variables will be called low-level kinematic variables.  
We will perform two independent analyses --- one with and one without the high-level kinematic variables, in order
to estimate the performance benefit from adding the additional 6 variables.

\section{Color flow in signal and backgrounds} \label{sec:method2}

%
\begin{figure}[t]
\centering
\includegraphics[width=0.97\textwidth,trim=0 0.5cm 0  0]{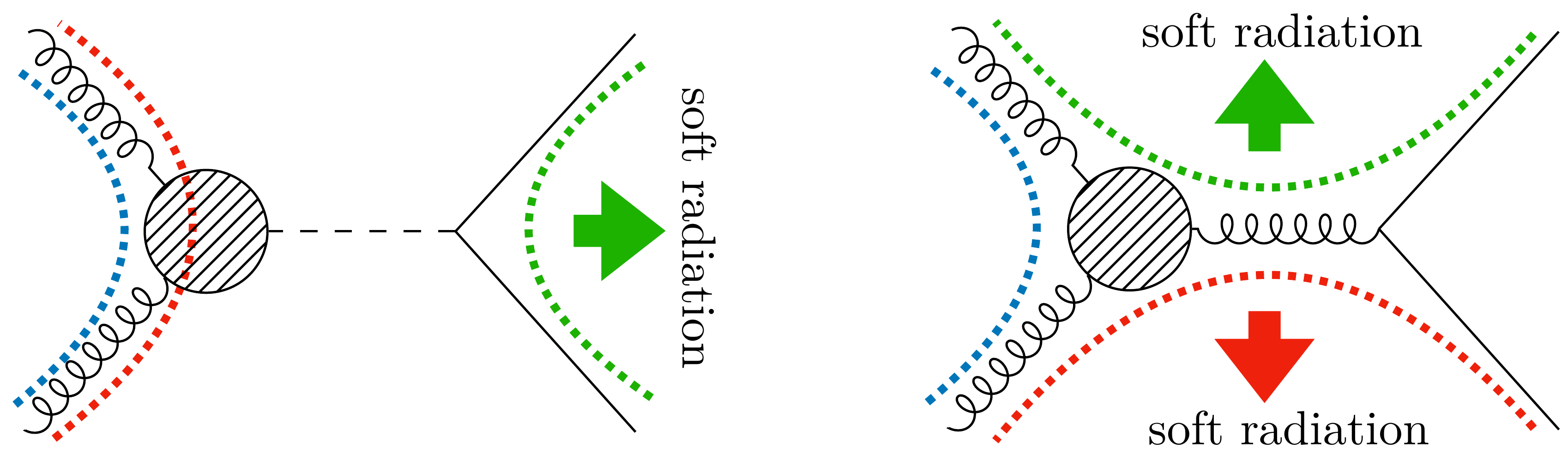}
\caption{\label{fig:colorflow} Color flow diagrams for a color-singlet particle (left) and a color-octet particle (right). 
The colored dotted lines represent QCD color-connection and arrows denote the direction of hadron concentration. 
}
\end{figure}

We note that the two $b$-quarks in the signal result from the decay of a single non-colored object, the Higgs boson.
In contrast, the two $b$-quarks in $t\bar t$ production (which is the dominant background) arise from the decays of top quarks, 
which in turn are produced via the strong interactions from a gluon-gluon initial state. This distinction is pictorially illustrated in
Fig.~\ref{fig:colorflow}. The different color-flow \cite{Maltoni:2002mq} will lead to different hadronization patterns, 
which can be used to discriminate a color singlet particle from a color octet (or triplet) at hadron colliders such as the LHC. 
Since the quarks which originate from a color singlet particle are color-connected to each other, their hadronization will not involve the initial state partons. 
On the contrary, the quarks which originate from a color octet particle are color-connected to the annihilating partons in the initial state, 
and consequently their hadronization is correlated with these initial state partons, see Fig.~\ref{fig:colorflow}.

\begin{figure}[t]
\centering
\includegraphics[width=8cm]{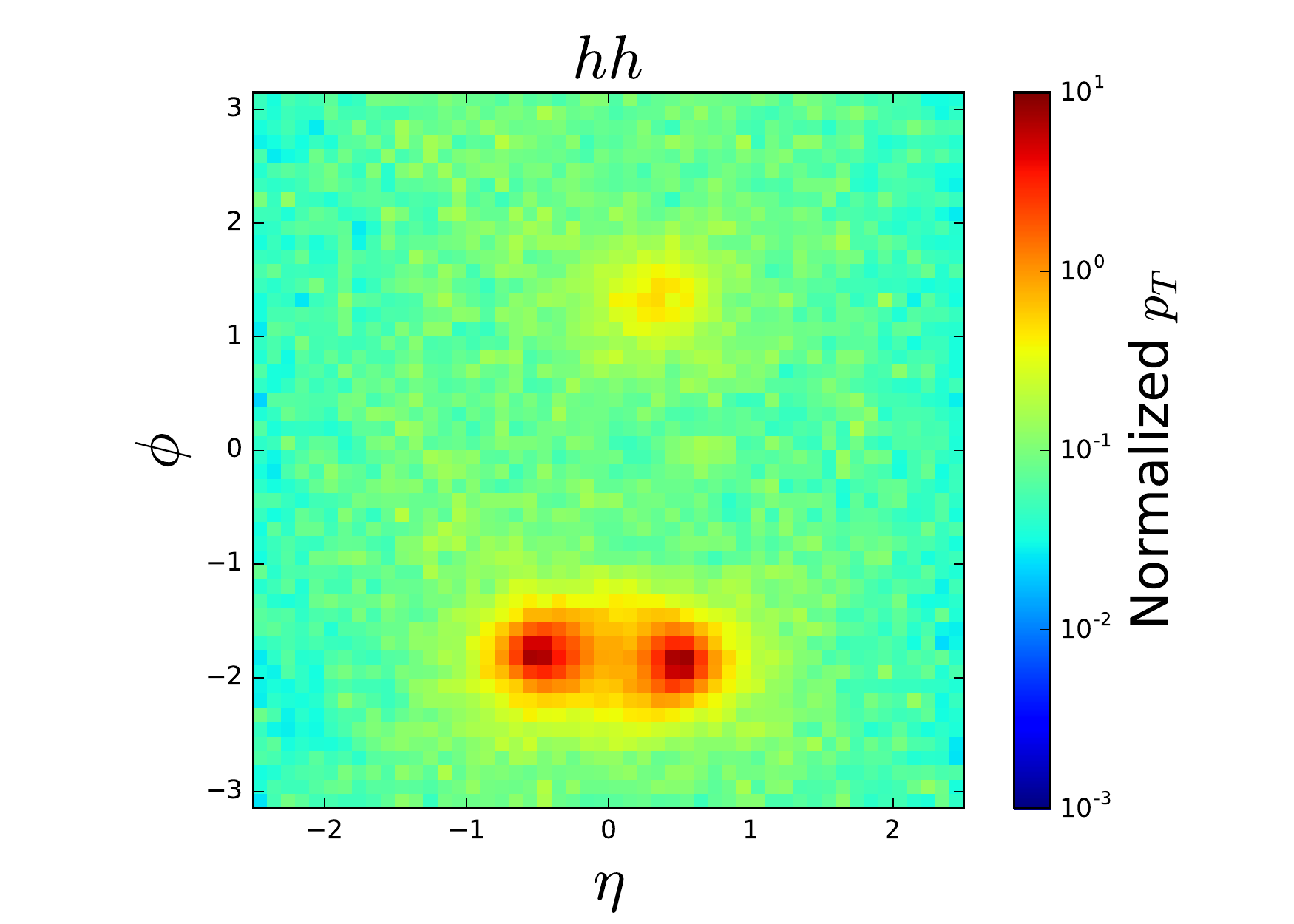}  \hspace*{-1cm}
\includegraphics[width=8cm]{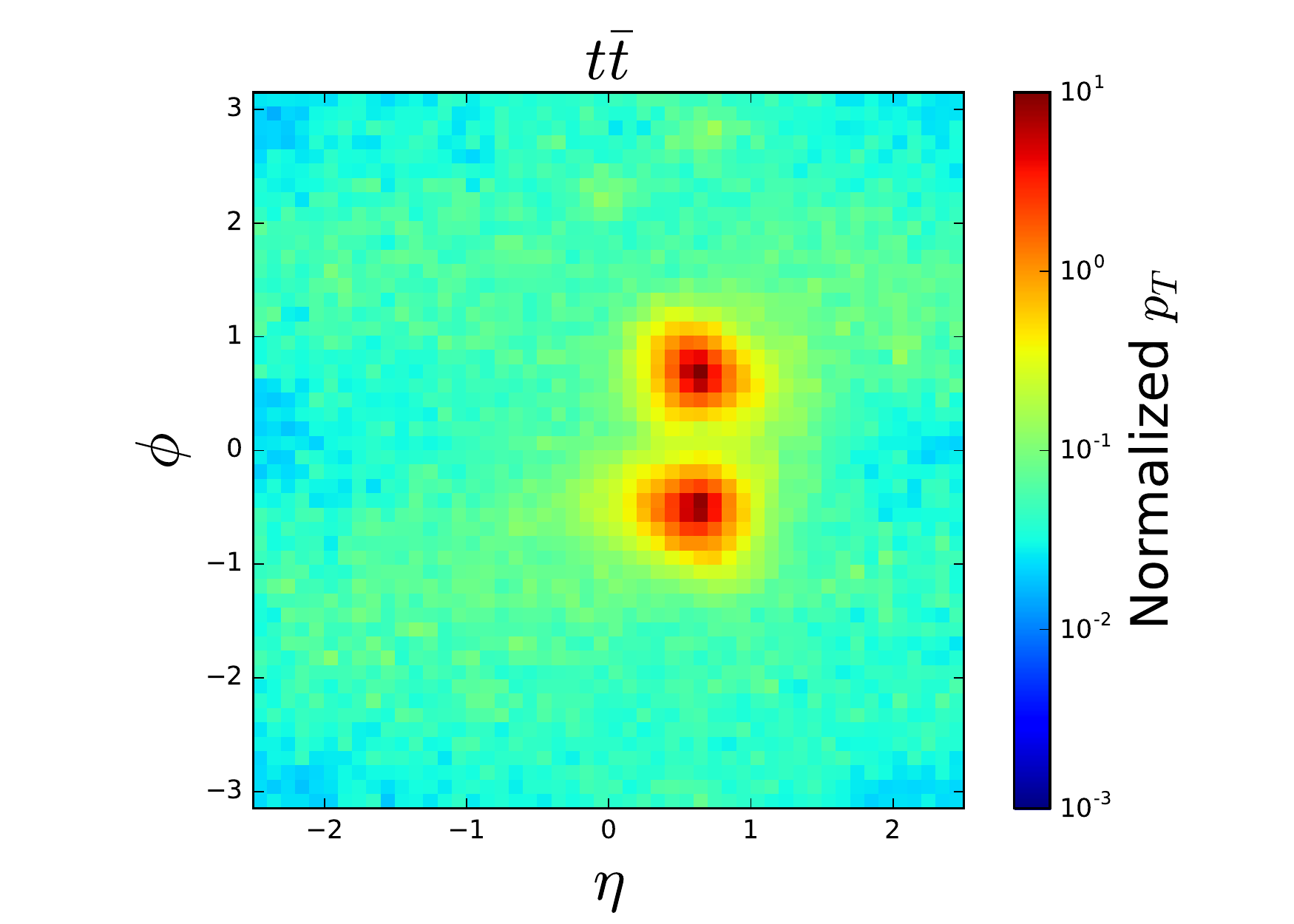} 
\caption{\label{fig:jetimage1} 
Cumulative $p_T$ distributions resulting from showering 10,000 times a single partonic event for the signal (left) and $t\bar t$ production (right). 
The two $b$ quarks from $h \to b \bar b$ are color-connected to each other and the soft radiation tends to fill in the region between them (left panel), 
while the two $b$ quarks from $t \bar t$ production are not color-connected and the two clusters from their hadronization tend to be more isolated (right panel).}
\end{figure}

The difference in color flow will be reflected in the resulting hadron distributions. 
Hadrons coming from a color-singlet object will tend to be closer to the direction of the original mother particle, 
and as a result, the soft radiation will tend to populate the region between the two $b$ quarks.
On the other hand, hadrons from the decay of a color-octet particle will not be so narrowly focused, 
due to the influence of the initial state partons.
These features are illustrated in Fig.~\ref{fig:jetimage1}, where we show the cumulative $p_T$ distributions in the $(\eta,\phi)$ plane 
after showering the same partonic event 10,000 times. In the left panel we used a signal event, while in the right panel 
we used an event from $t\bar t$ production. We see that the b-jet clusters in the right panel tend to be better defined and more isolated,
since they are not color-correlated among themselves. On the other hand, in the left panel we observe quite a bit of soft radiation in the region between the 
two $b$ jets, due to the existing color connection between them.

\begin{figure}[t]
\centering
\hspace*{-0.8cm}
\includegraphics[width=12.3cm]{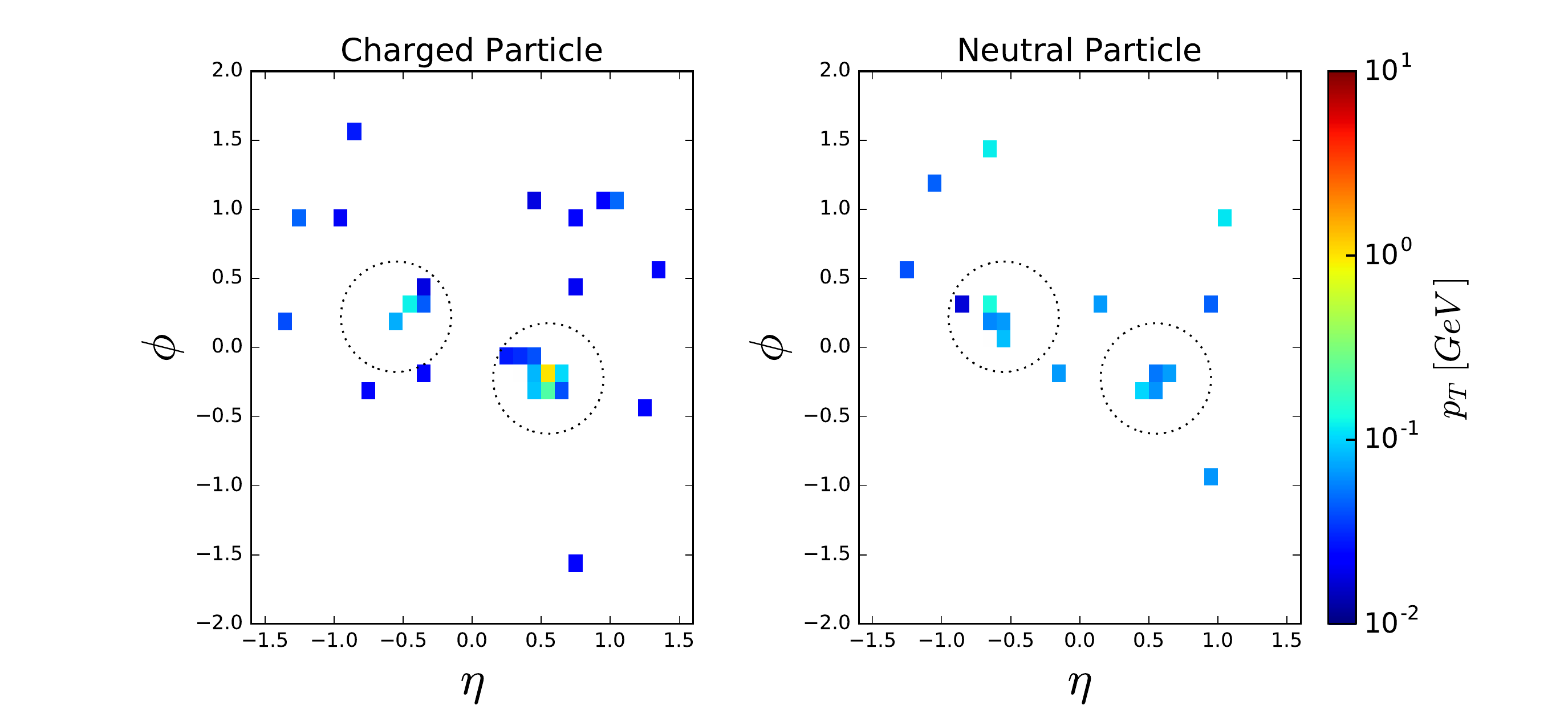} \hspace*{-1cm}
\includegraphics[width=4.7cm]{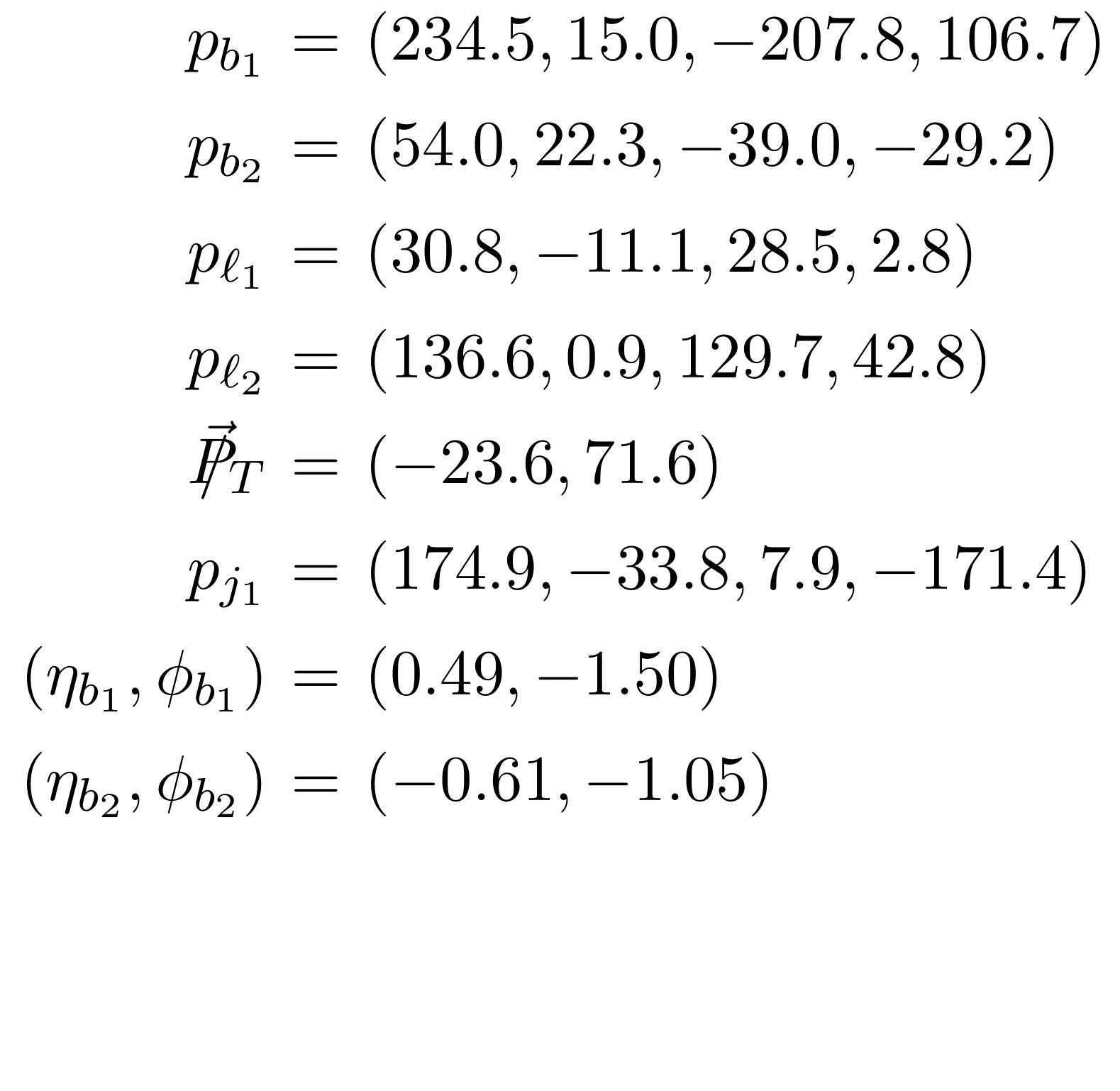}   \\
\hspace*{-0.8cm}
\includegraphics[width=12.3cm]{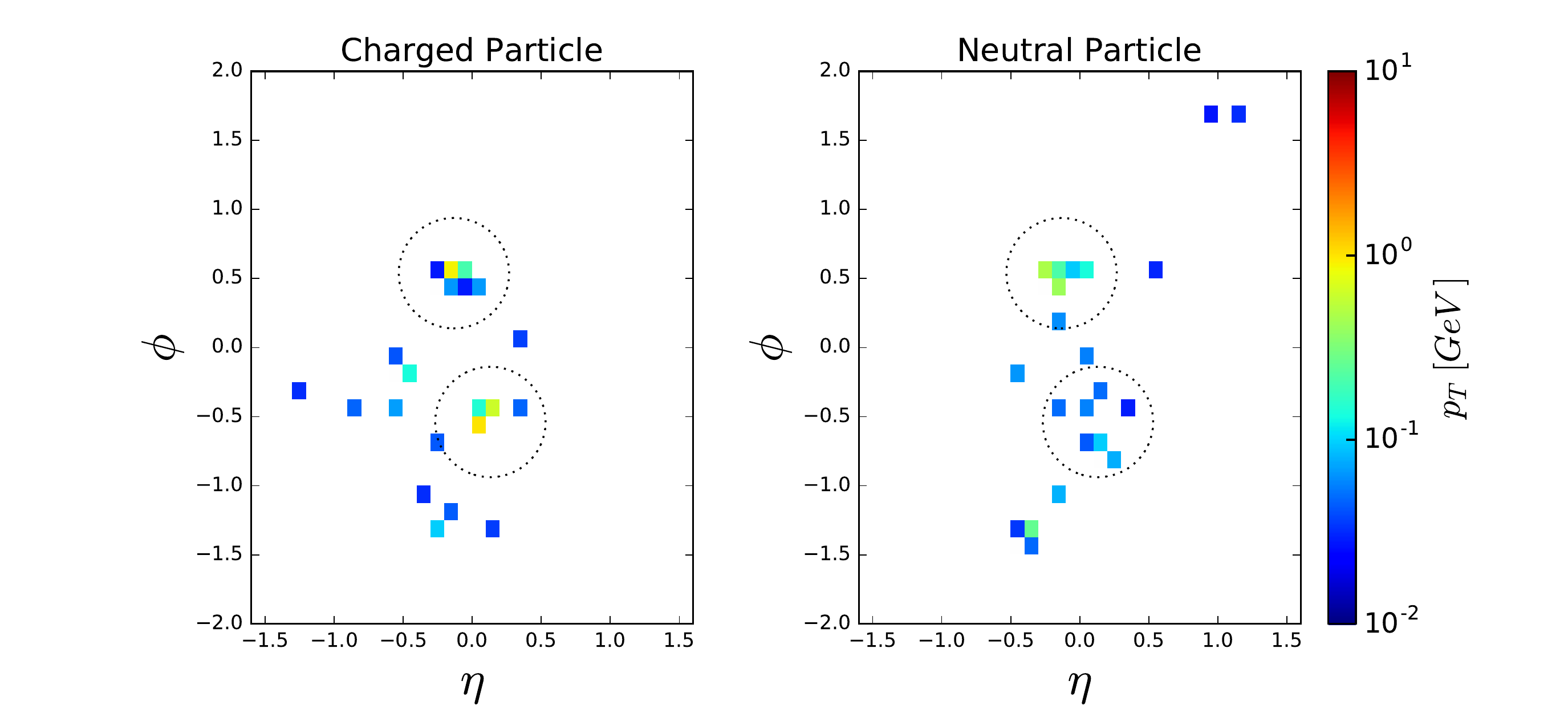} \hspace*{-1cm}
\includegraphics[width=4.7cm]{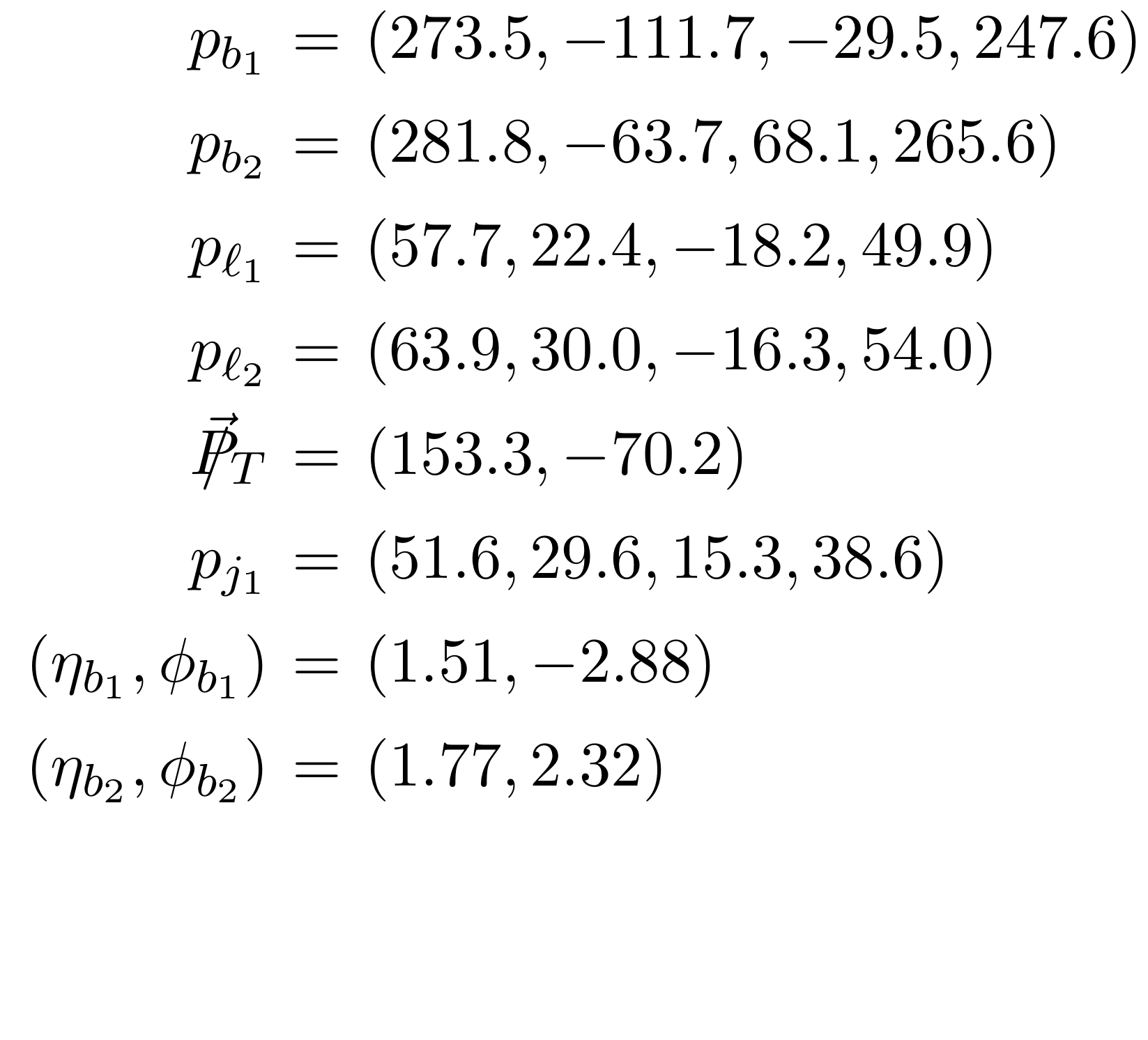}   
\caption{\label{fig:jetimage0} 
Transverse momentum distribution of charged particles (left) and neutral particles (right) 
for one chosen signal event (top row) and one chosen $t\bar t$ event (bottom row), 
where the origin is taken to be the center of the $b$-quark pair. 
The dotted circles represent the $\Delta R$=0.4 cones for reconstructing the corresponding $b$ jets. 
The four-momentum information of each event is given to the right of each panel row.}
\end{figure}

Of course, the results in Fig.~\ref{fig:jetimage1} are only valid in the statistical sense, since we took the same parton-level event and hadronized it multiple times.
In reality, only one instance of this hadronization will be realized, as illustrated in Fig.~\ref{fig:jetimage0}.
The top row of plots shows the hadronization patterns for charged particles (left panel) and neutral particles (right panel)
in the case of one signal event, while the bottom row shows the same, but for one $t\bar{t}$ event.
The parton-level event information is quoted (in GeV) to the right of each row of panels, and then each event is translated in the 
($\eta, \phi$) plane until the origin is aligned with the direction of the  $b$-quark pair.
The color scheme indicates the total $p_T$ in each pixel, while the dotted circles represent the $\Delta R=0.4$ cones for reconstruction of the corresponding $b$ jets. 

\begin{figure}[t]
\centering
\includegraphics[width=4.67cm]{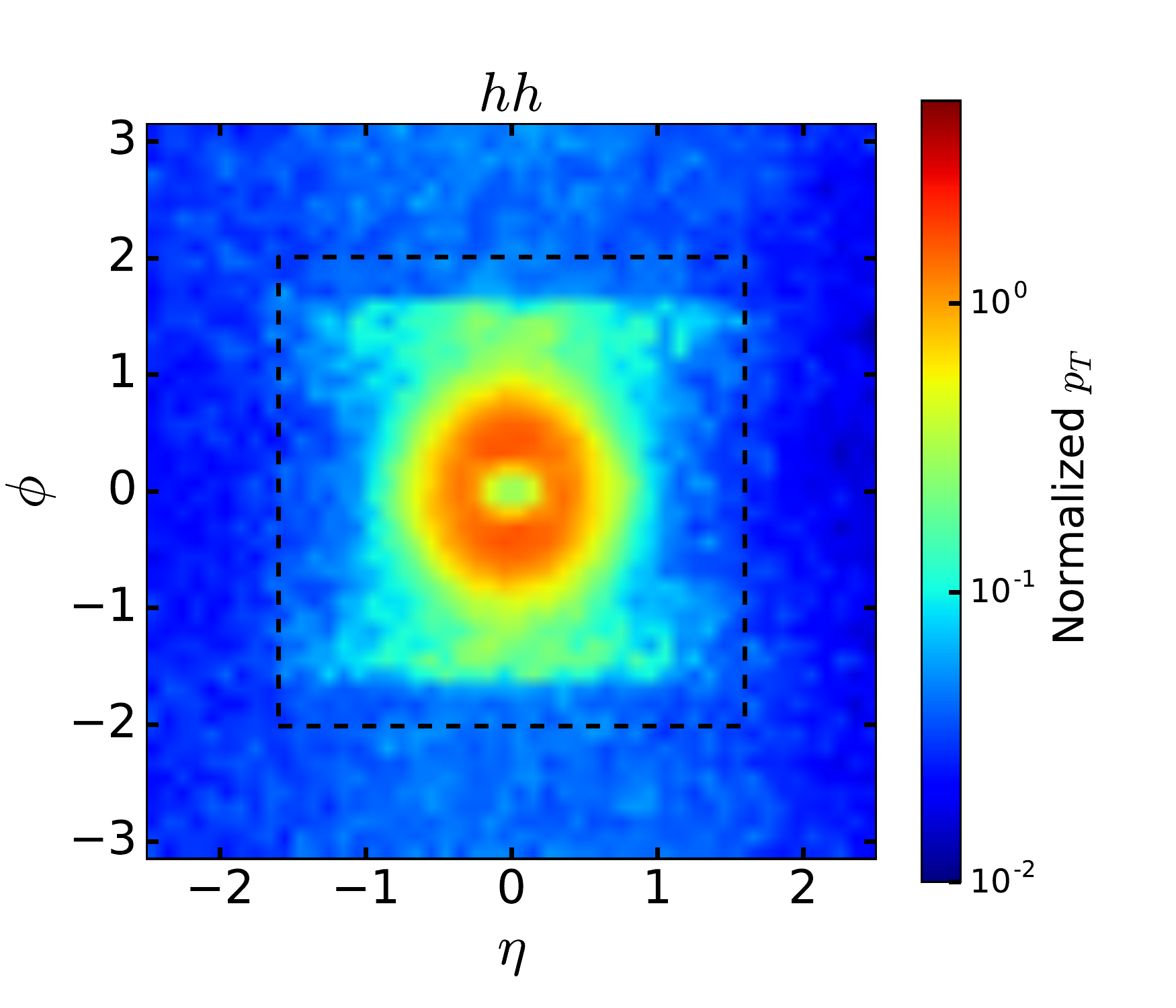} \hspace*{-1.3cm}
\includegraphics[width=4.67cm]{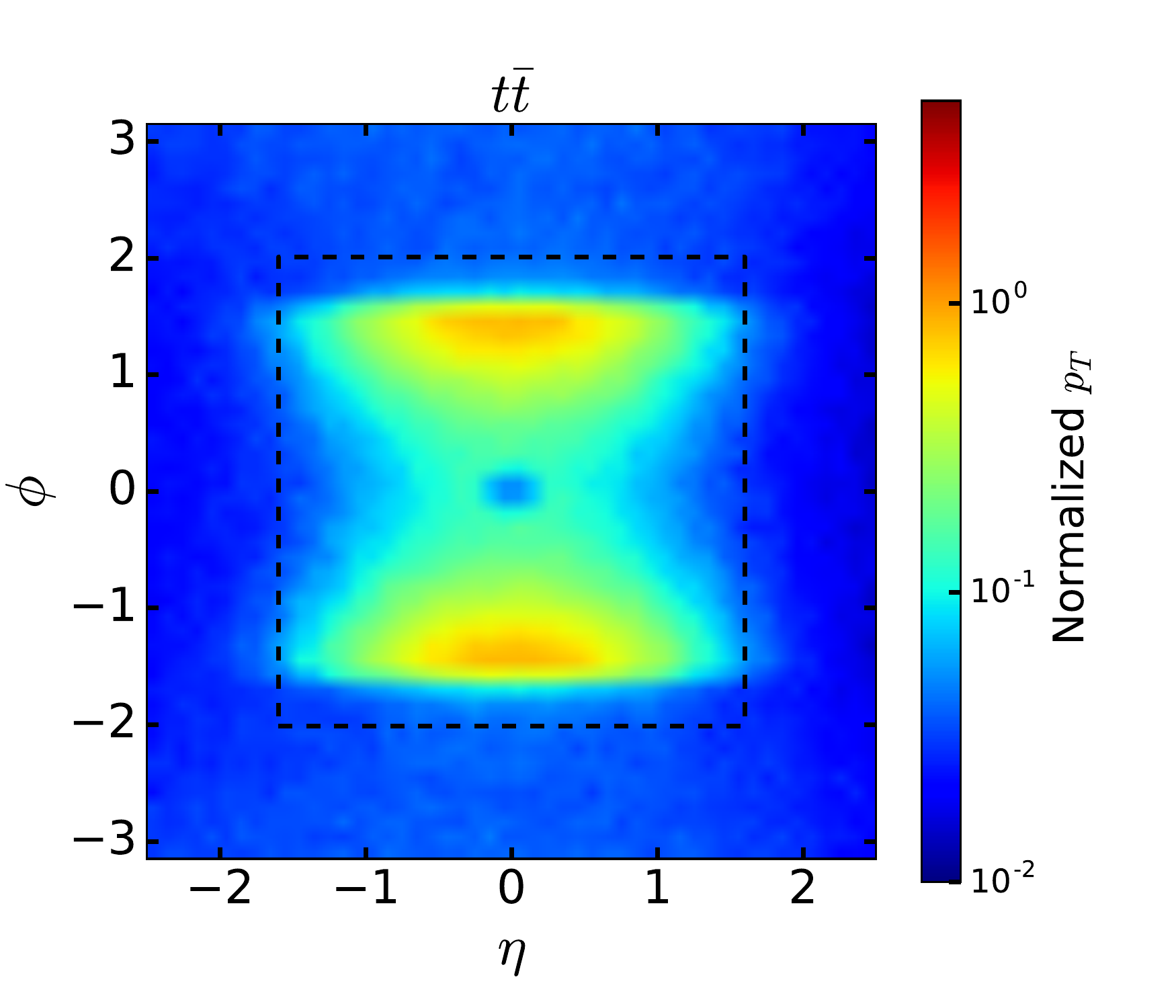} \hspace*{-1.3cm}
\includegraphics[width=4.67cm]{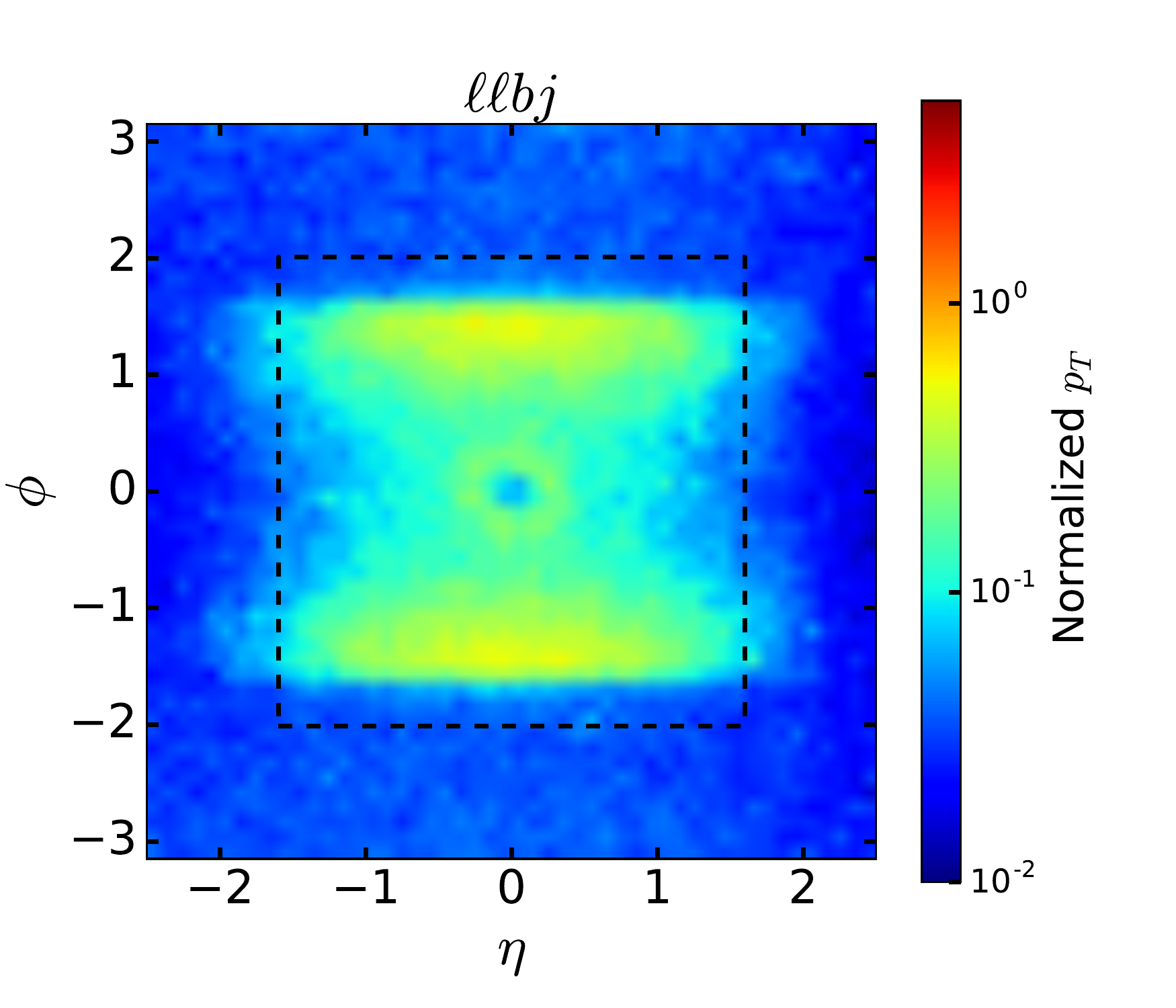} \hspace*{-1.3cm}
\includegraphics[width=4.67cm]{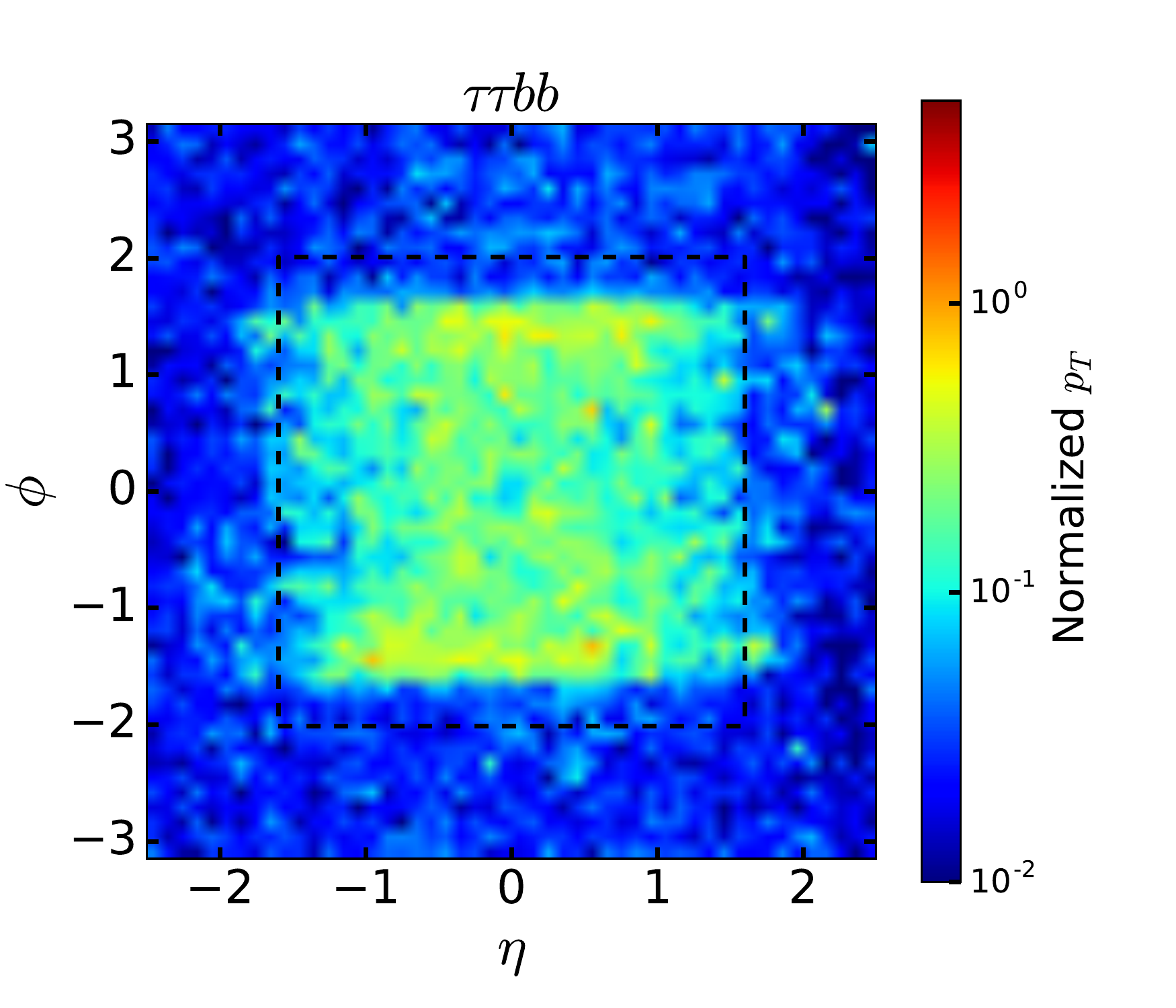}  \\
\includegraphics[width=4.67cm]{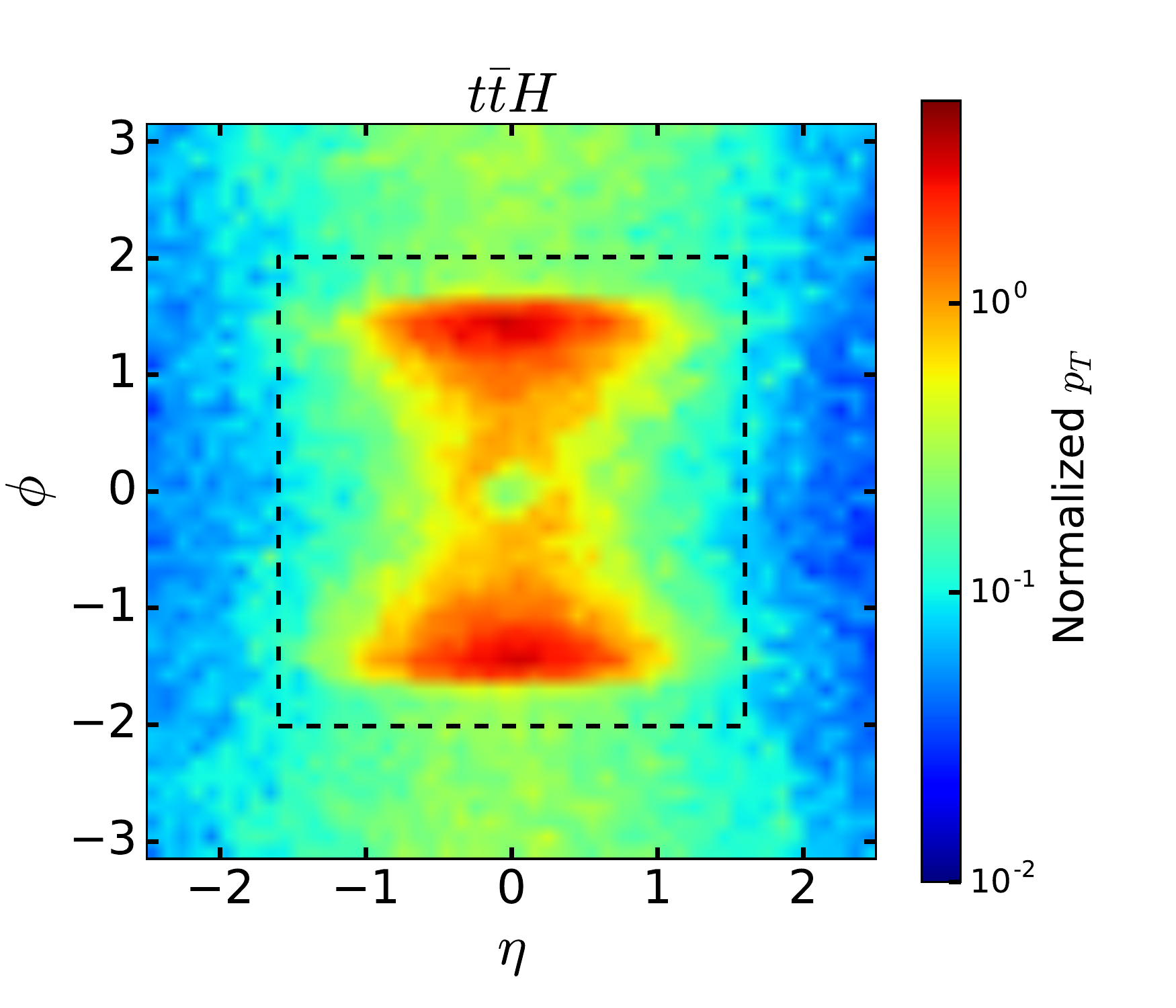} \hspace*{-1.3cm}
\includegraphics[width=4.67cm]{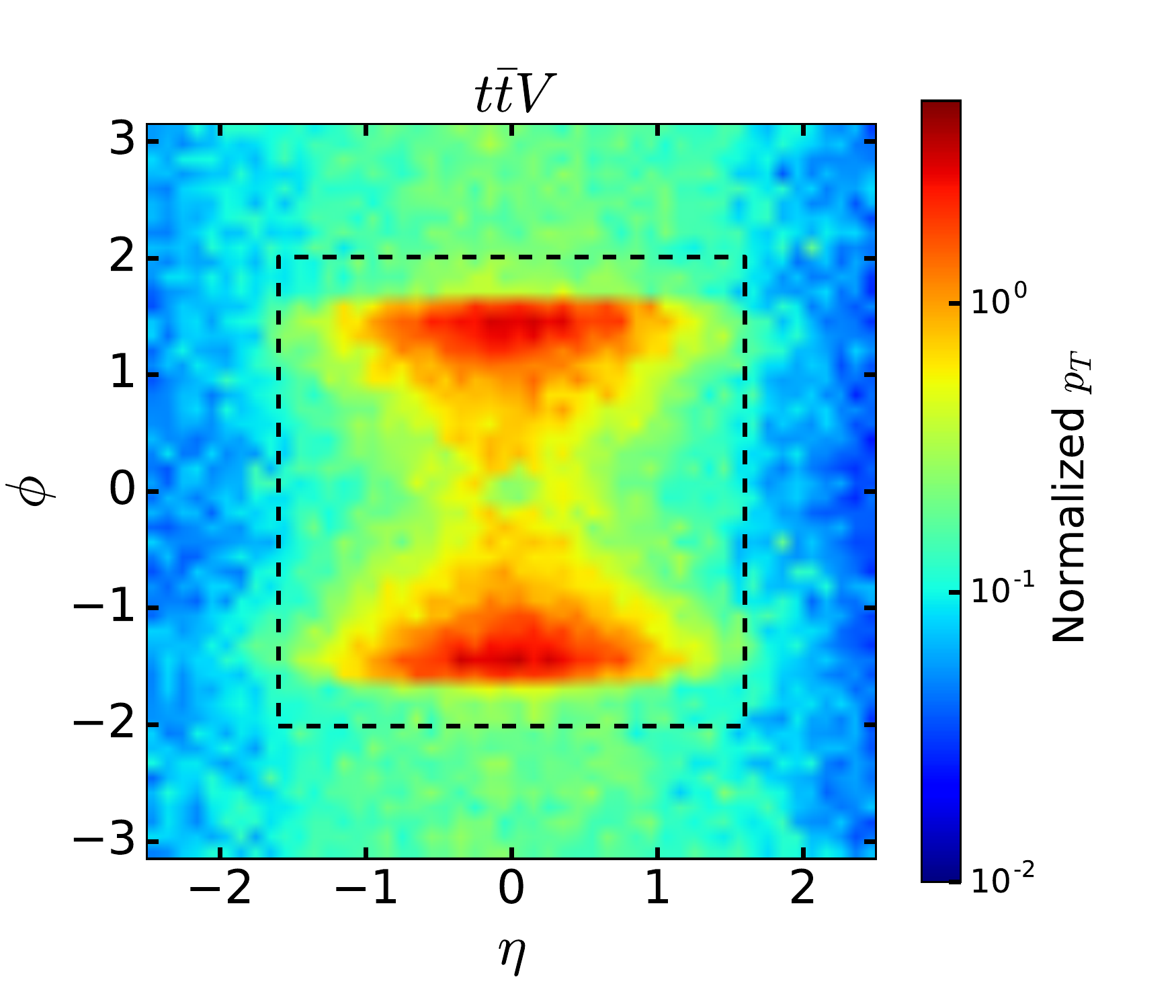} \hspace*{-1.3cm}
\includegraphics[width=4.67cm]{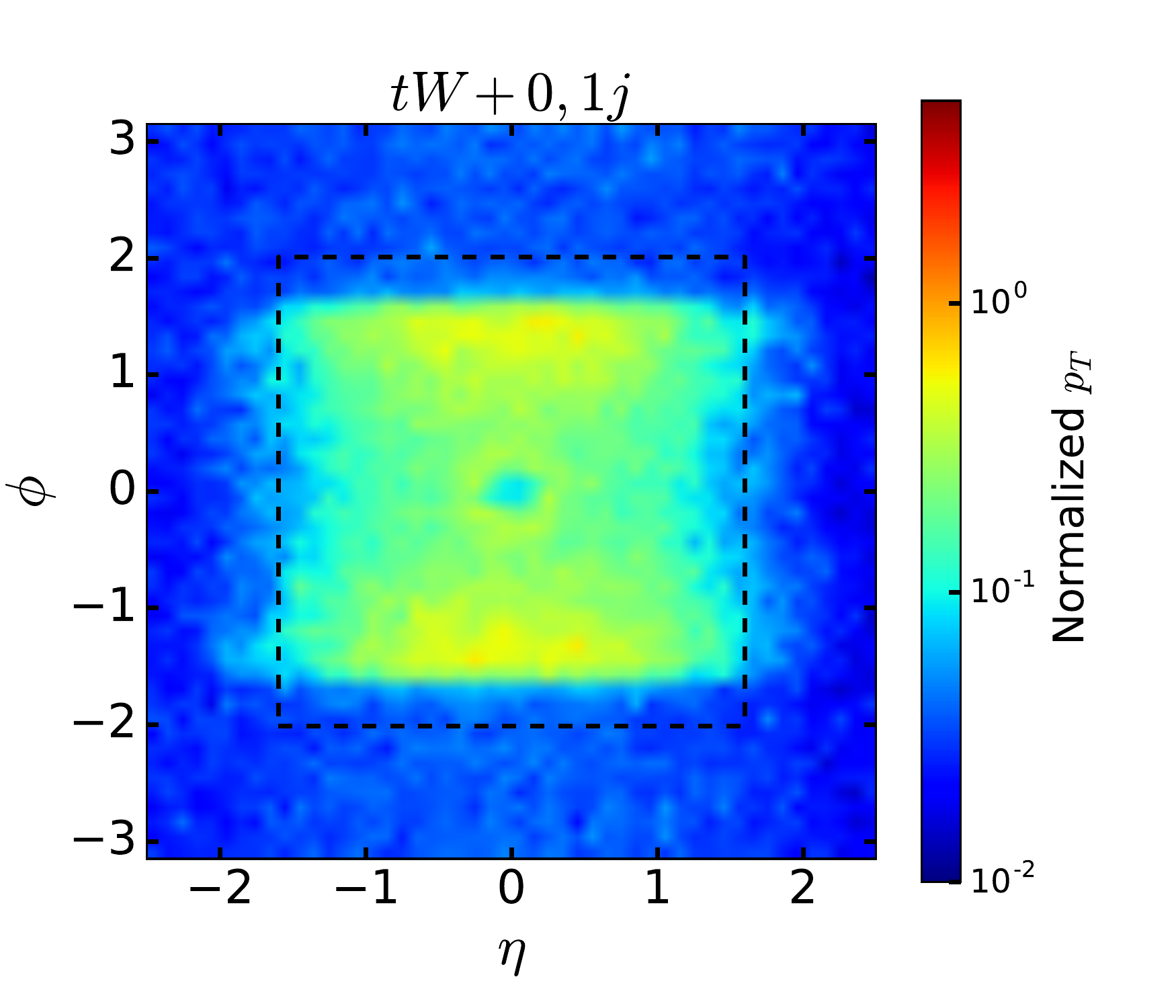} \hspace*{-1.3cm}
\includegraphics[width=4.67cm]{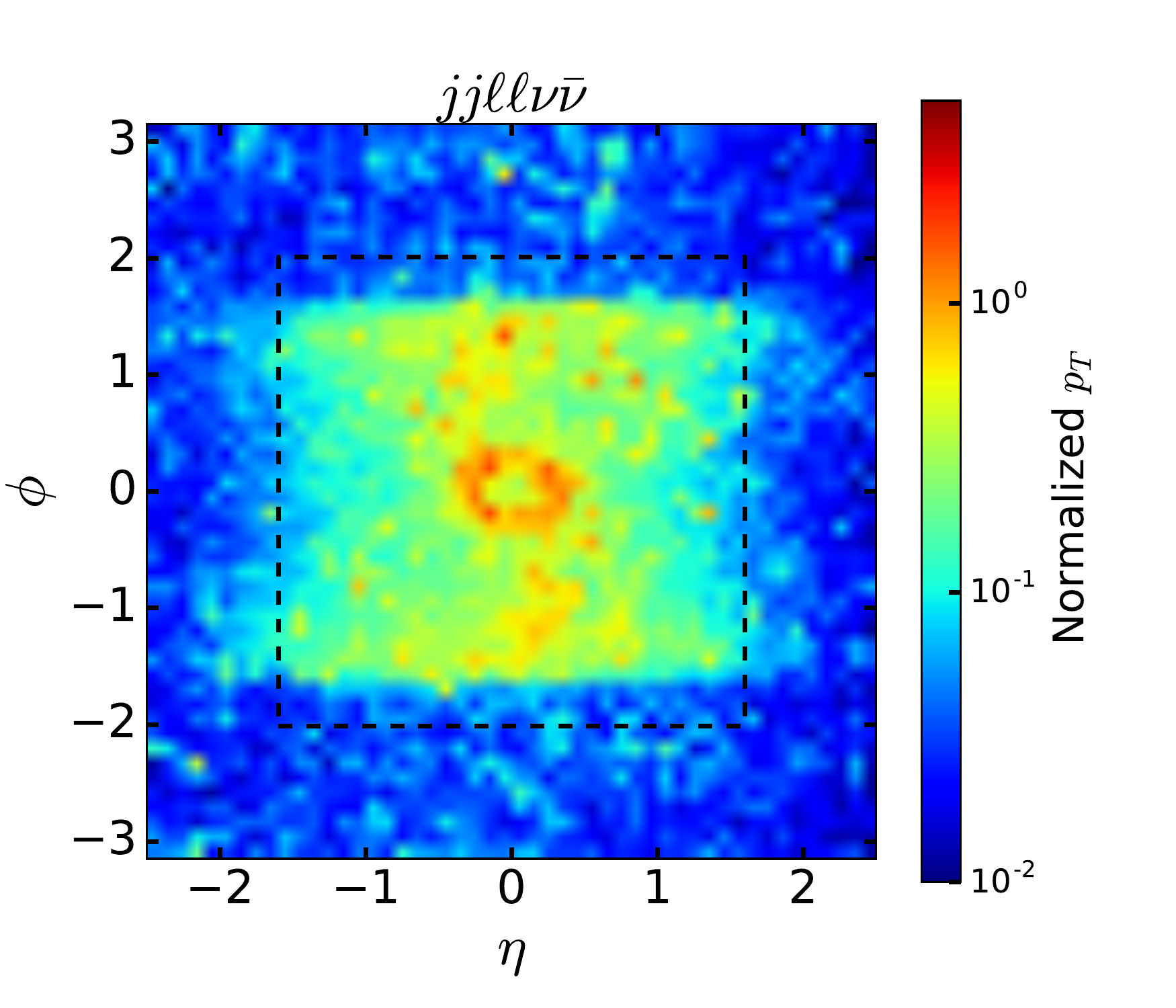}
\vspace*{-0.7cm}
\caption{\label{fig:jetimagebeforecuts} 
The cumulative average of the jet images for the signal and the different background processes before the baseline cuts
(basic cuts at the event generation stage were still imposed). 
The origin of the ($\eta, \phi$) plane is taken to be the center of the $b$ quark pair and the color scheme indicates the total $p_T$ in each pixel.
The black dotted line delineates the region $\-1.6\le \eta\le1.6$ and $-2.01\le\phi\le2.01$ used in the analysis.
}
\end{figure}
\begin{figure}[t]
\centering
\includegraphics[width=4.67cm]{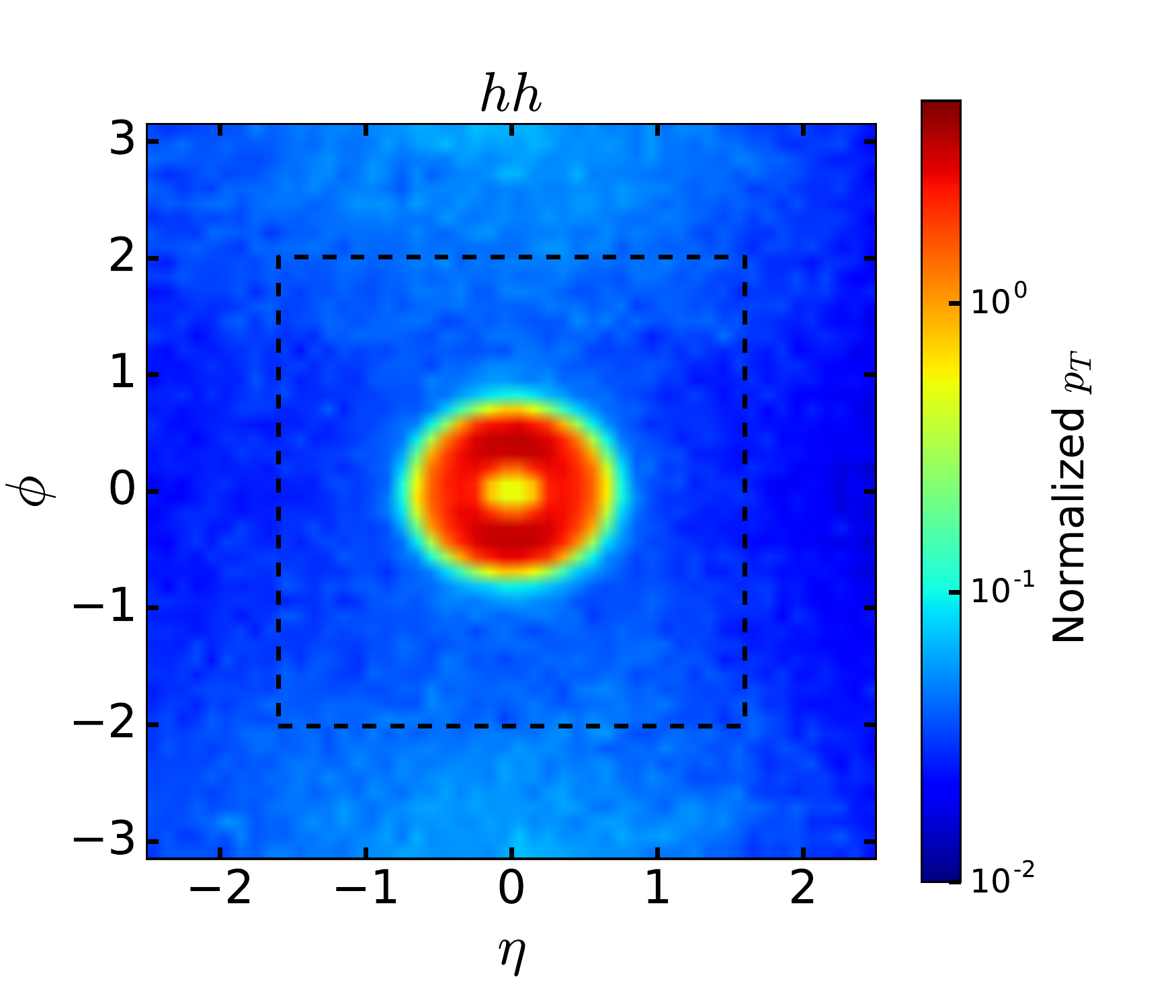} \hspace*{-1.3cm}
\includegraphics[width=4.67cm]{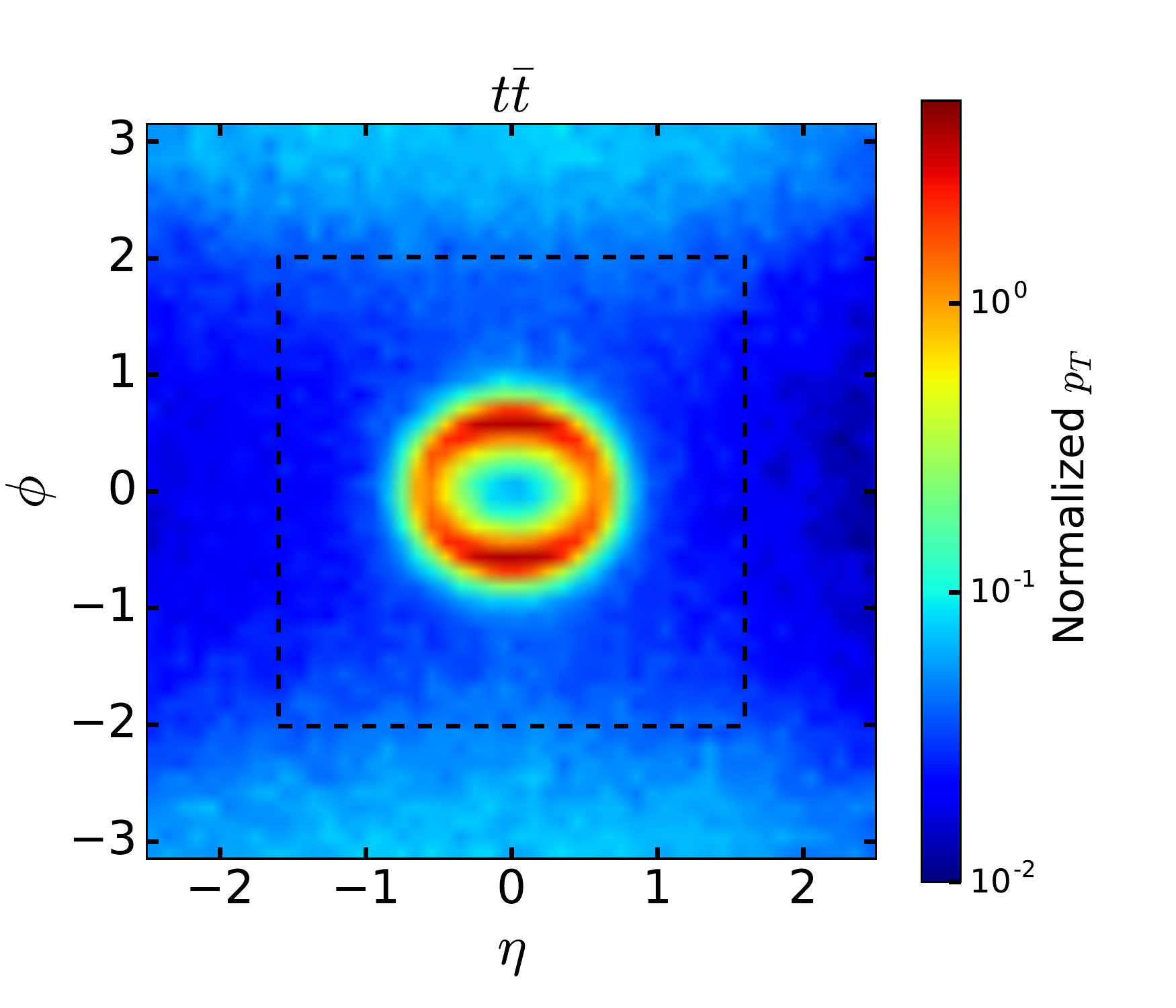} \hspace*{-1.3cm}
\includegraphics[width=4.67cm]{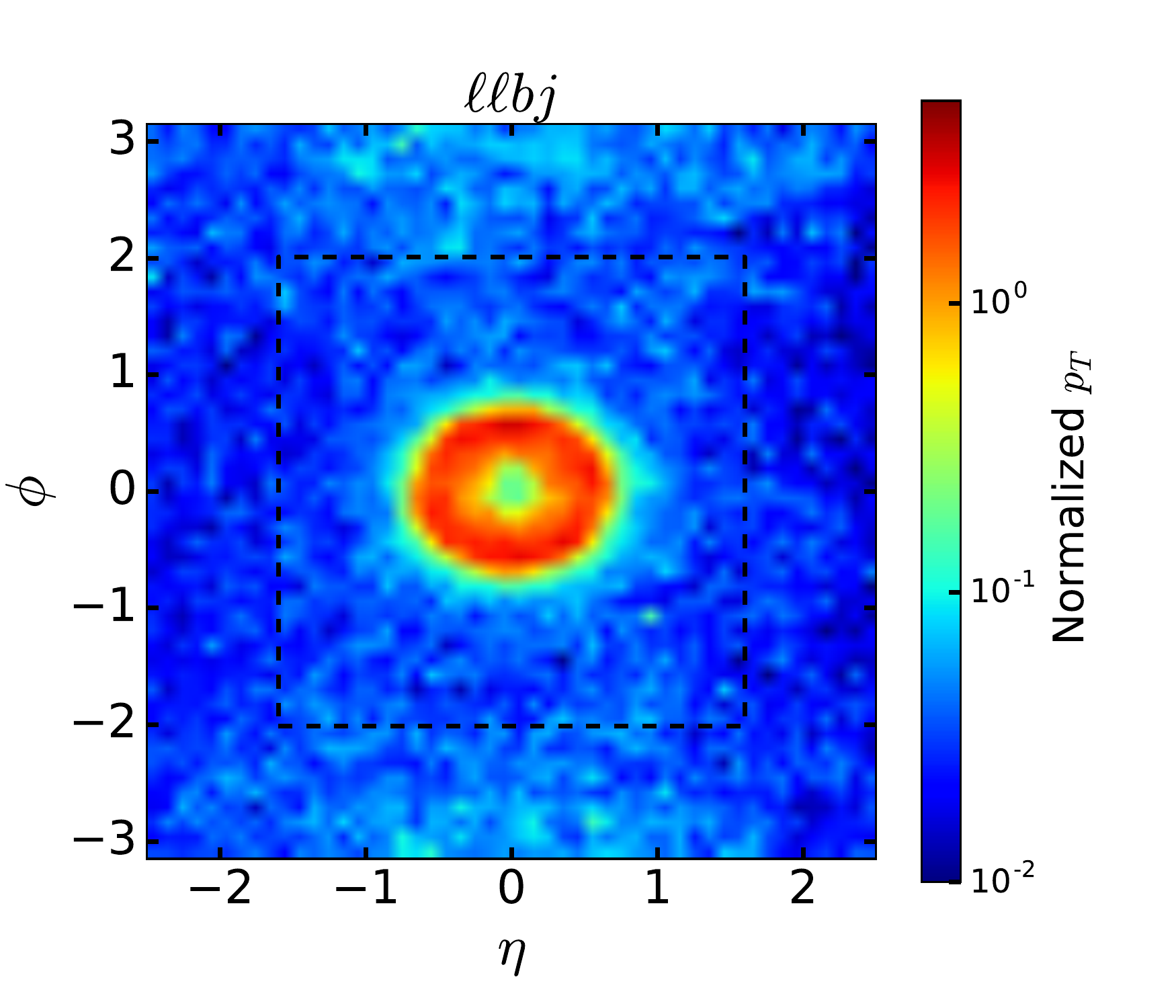} \hspace*{-1.3cm}
\includegraphics[width=4.67cm]{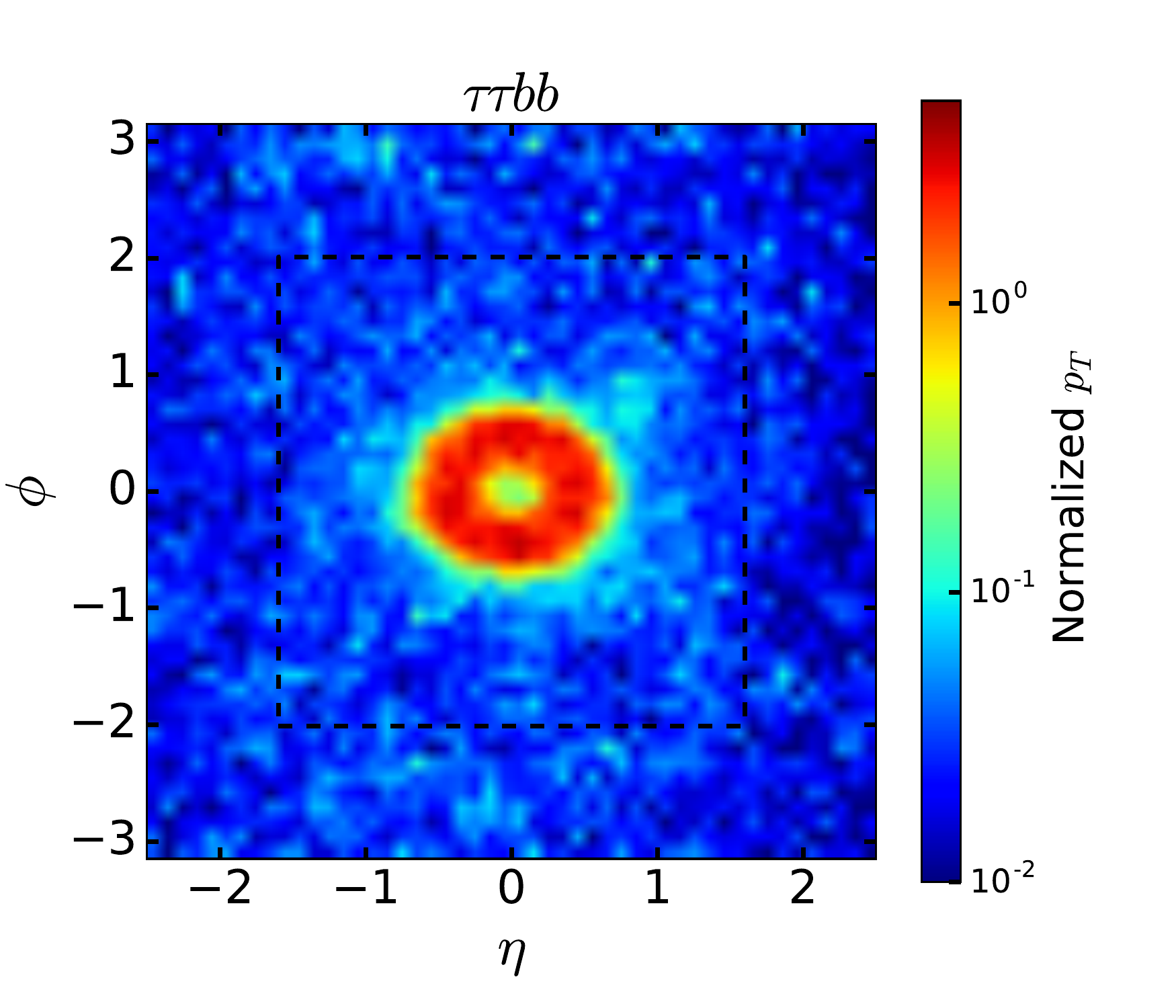}  \\
\includegraphics[width=4.67cm]{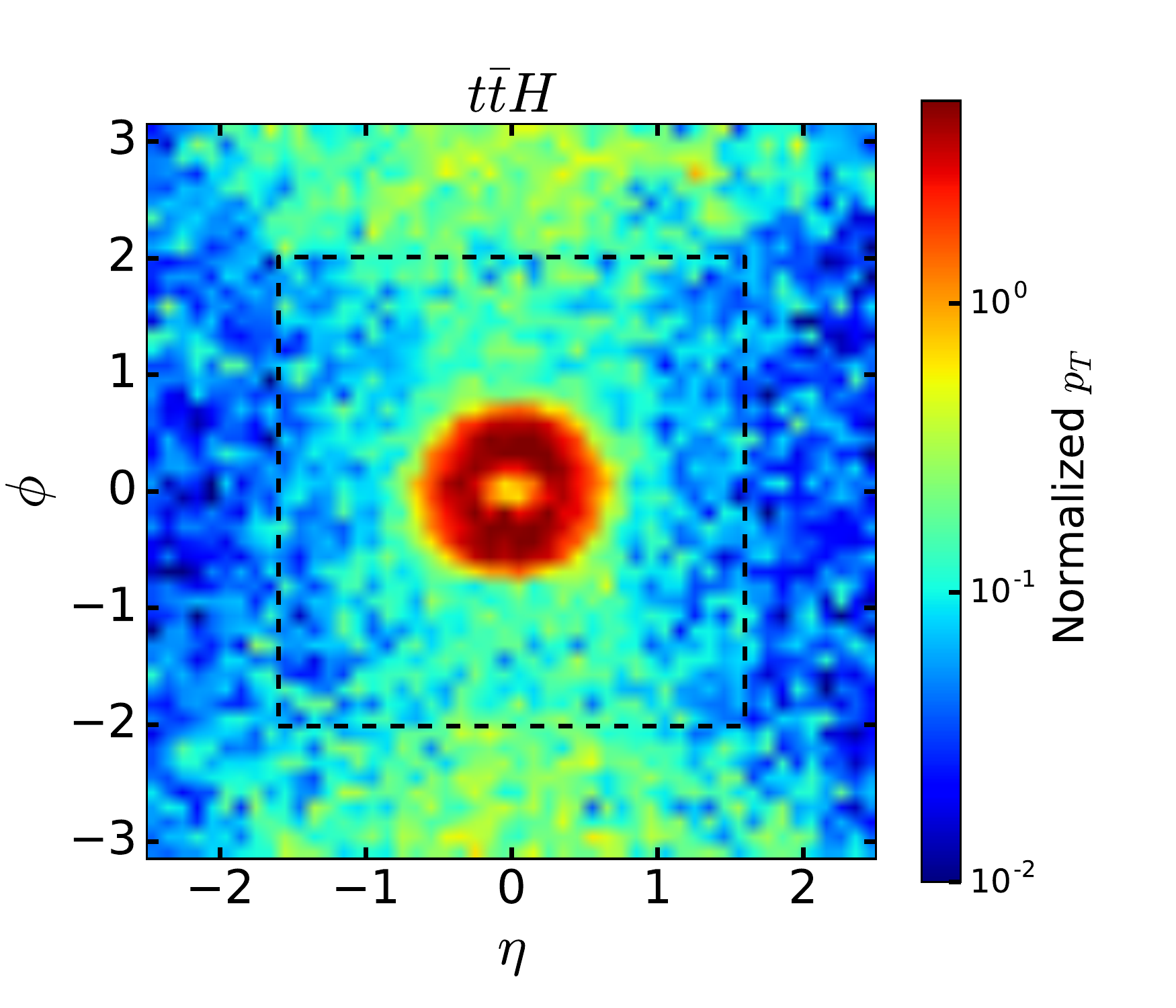} \hspace*{-1.3cm}
\includegraphics[width=4.67cm]{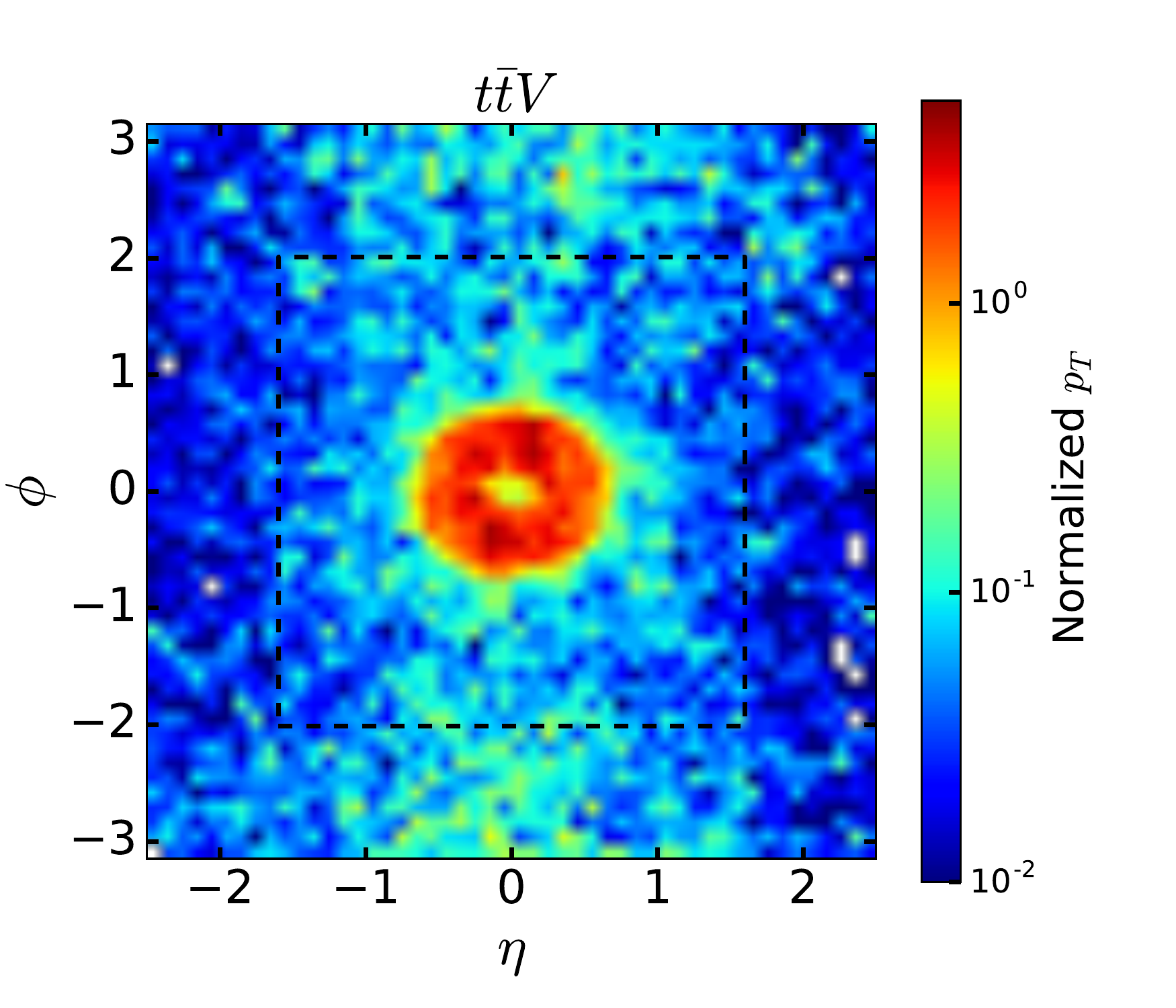} \hspace*{-1.3cm}
\includegraphics[width=4.67cm]{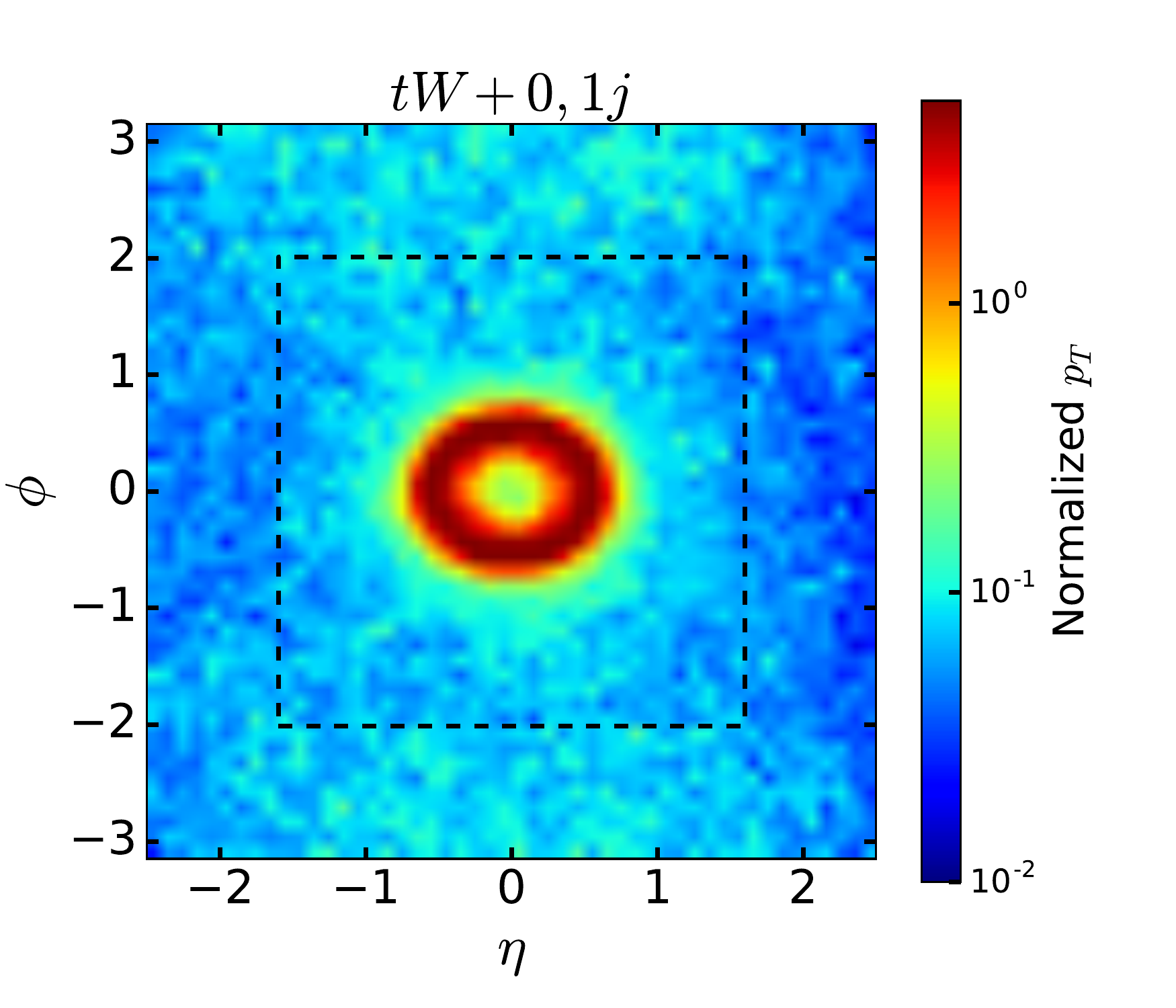} \hspace*{-1.3cm}
\includegraphics[width=4.67cm]{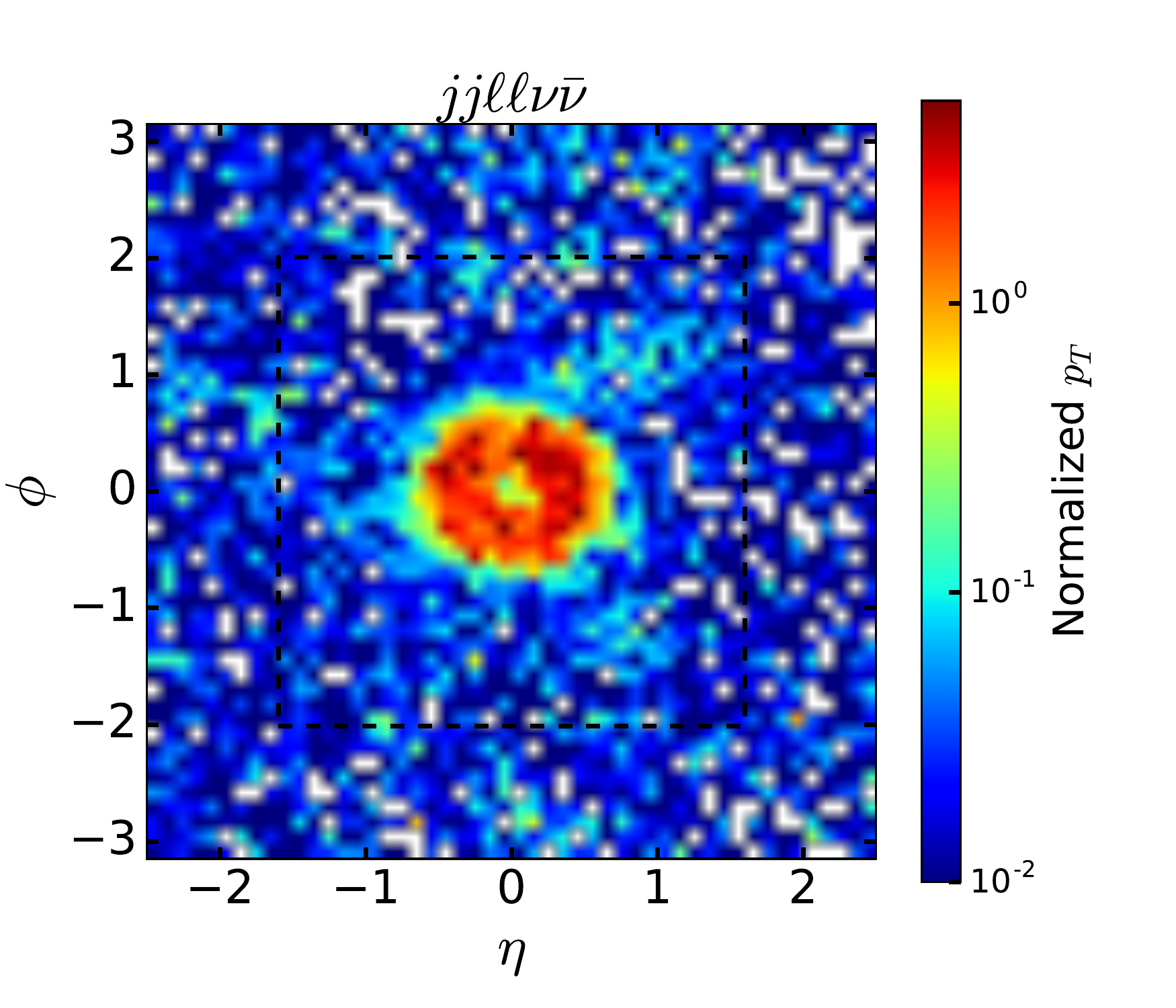}
\vspace*{-0.7cm}
\caption{\label{fig:jetimage} 
The same as Fig.~\ref{fig:jetimagebeforecuts}, but after imposing the baseline cuts introduced in Section \ref{sec:analysis}.
}
\end{figure}

As an alternative to Fig.~\ref{fig:jetimage1}, in Figs.~\ref{fig:jetimagebeforecuts} and \ref{fig:jetimage}, we illustrate the 
effects of color-connection by showing the average of the jet images for 
the signal and the different background processes before and after the baseline cuts, respectively
(some basic generation-level cuts were imposed on the events in Fig.~\ref{fig:jetimagebeforecuts}). 
The origin of the ($\eta, \phi$) plane plane is taken to be the center of the $b$ quark pair and the color scheme indicates the total $p_T$ in each pixel.
The black dotted line delineates the region $\-1.6\le \eta\le1.6$ and $-2.01\le\phi\le2.01$ used in the analysis. 
One can observe a striking difference in density between signal and background events in Fig.~\ref{fig:jetimagebeforecuts} --- the two $b$ quarks tend to be more collimated in the signal
and more spread out in the background.

Unfortunately, after imposing the baseline cuts introduced in Section \ref{sec:analysis}, 
this distinction tends to be washed out and the backgrounds start mimicking the signal:
one can see a similar structure emerging in all panels in Fig.~\ref{fig:jetimage}, albeit with some subtle differences. 
Although one may find it difficult to discriminate signal from backgrounds simply by looking at a particular event, 
the patterns in the average jet images are different, and have been used actively for signal versus background separation \cite{Gallicchio:2010sw,Hook:2011cq,Aaboud:2018ibj}. 
In this paper, instead of quantifying the difference (e.g., with a pull vector \cite{Gallicchio:2010sw}) we will use the images themselves on deep neural networks (DNNs), 
along with the 16 kinematic variables introduced in the previous section.

\section{Analysis using deep learning} \label{sec:dnn}

DNN is known to be very efficient and powerful in image recognition \cite{NIPS2012_4824,0483bd9444a348c8b59d54a190839ec9} and 
the particle physics community has used it for various applications\footnote{The use of neural networks for data analysis in high energy physics can be traced back to the pioneering work by R.~Field and his students in the mid-nineties \cite{Field:1996rw,Field:1996nq,RFtalk}.}.
For instance, one can map the information about the direction and the energy (or transverse momentum) of a particle onto a pixel in an image. 
DNN then provides excellent classification between signal and background in the jet image \cite{Cogan:2014oua, Komiske:2016rsd, Kasieczka:2017nvn,Lin:2018cin}. 
It also shows performance gains in multrivariate analyses over traditional cut-and-count analyses or BDTs \cite{Baldi:2014kfa,Baldi:2016fql}.
In this section, we describe how we organize our analysis in a DNN framework. In the following three subsections, we
address the issues of data pre-processing, DNN architecture and training of the NN.

\subsection{Data pre-processing}

In order to achieve the improved DNN learning performance and to minimize the error, it is important to properly process signal and backgrounds events 
before feeding them into a DNN framework. For each event passing the baseline cuts, the jet images are processed as follows.
\begin{enumerate}
\item {\it Input data:} we use the particle flow for our input data \cite{CMS-PAS-PFT-09-001}. 
\item {\it Particle classification:} we divide the particle flow into two groups: neutral particles and charged particles.  
Neutral particles include photons and neutral hadrons, while charged particles include charged hadrons.
\item {\it Lepton removal:} if there is a lepton, we remove it.
\item {\it Shift:} we shift all particle coordinates in the ($\eta, \phi$) plane with respect to the center of the reconstructed $b$-quark pair, i.e., 
we set $(\frac{\eta_b+\eta_{\bar{b}}}{2},\frac{\phi_b+\phi_{\bar{b}}}{2})$ as the new origin, (0,0).
\item {\it Pixelization:} we discretize the rectangular region in the ($\eta, \phi$) plane defined by
$-2.5 \le \eta \le 2.5$ and $-\pi \le \phi \le \pi$ into a grid of $50 \times 50$ pixels for each particle classification (charged particle set and neutral particle set). 
In each pixel, we record the total transverse momentum as the pixel's intensity (in case of more than one particle, we add the transverse momenta and record the total sum). 
We refer to this $50 \times 50$ discrete image as the jet image \cite{Cogan:2014oua}.
\item {\it Normalization:} we rescale each jet image intensity as $I^{ij}\rightarrow I^{ij}/I^\textrm{max}$, where $i,j=1,2,\ldots,50$,  and $I^{ij}$ represents the 
intensity value in the $(i, j)$ pixel. $I^\textrm{max}$ is defined to be the largest value of pixel intensity found in the two $50 \times 50$ pixel images. 
\item {\it Cropping:} we crop the jet image to $32 \times 32$ pixels, by further restricting to the ($\eta, \phi$) rectangular range of $-1.6\le \eta \le 1.6$ and $-2.01 \le \phi \le 2.01$.
\end{enumerate}
The final jet image has dimension $2 \times 32 \times 32$ and is comprised of one charged particle channel 
with dimension $1 \times 32 \times 32$ and a neutral particle channel with dimension $1 \times 32 \times 32$. 
This pre-processed jet-image is the input to the DNN. We note that Figs.~\ref{fig:jetimagebeforecuts} and \ref{fig:jetimage} showed the combined 
$1 \times 50 \times 50$ jet-image obtained by adding the neutral and charged particle layers. The black dotted rectangular area in those figures showed 
the restricted $1 \times 32 \times 32$ pixel area.

\subsection{DNN architecture}
\label{sec:dnna}

\begin{figure}[t]
\centering
\includegraphics[width=15.cm]{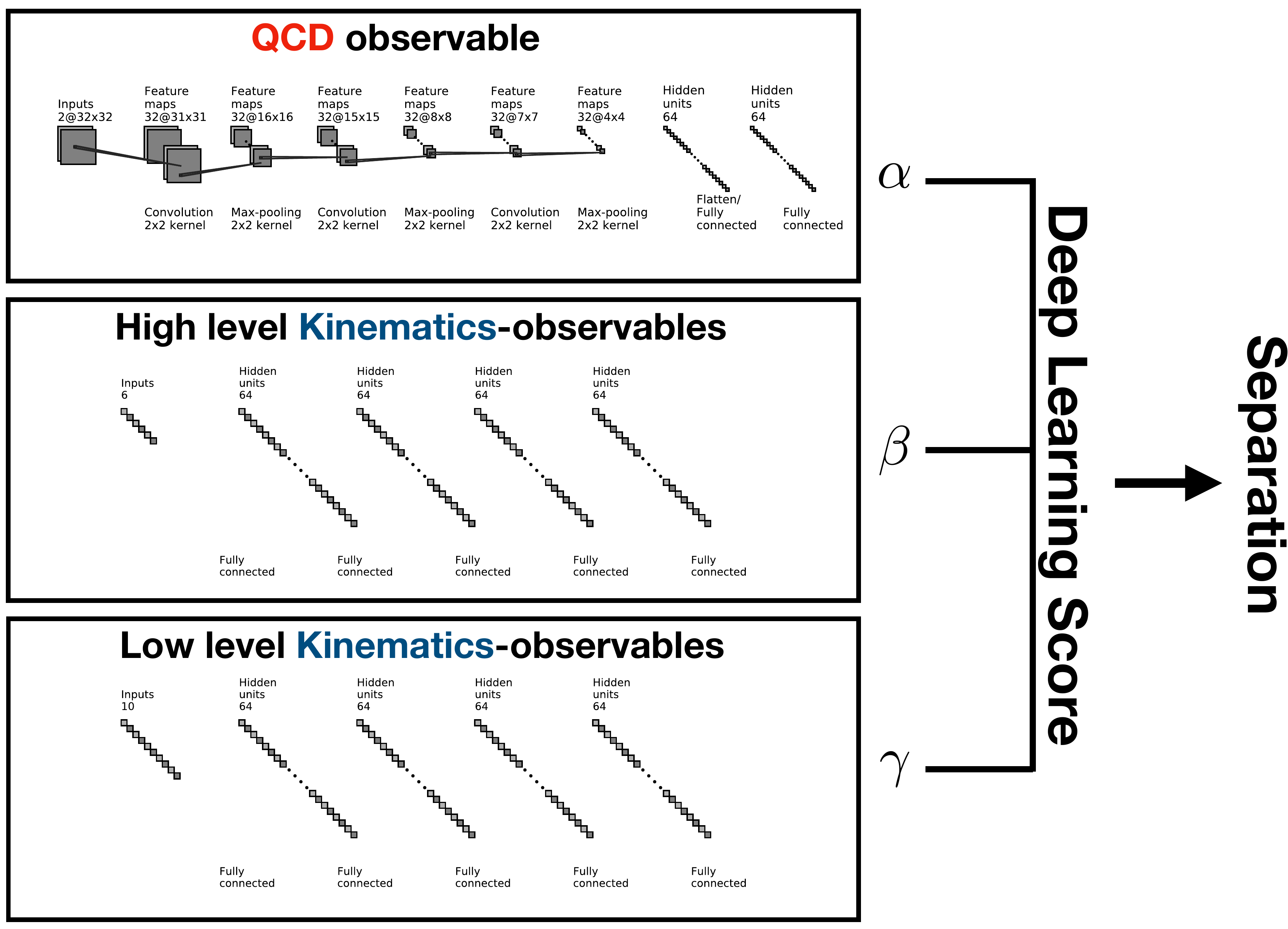}
\caption{\label{fig:dlshape} 
Illustration of the concept of combined Deep Learning\protect\footnotemark[1].}
\end{figure}
\footnotetext[1]{Parts of Fig.~\ref{fig:dlshape} are generated using the Python script in  
\url{https://github.com/gwding/draw_convnet}.}

Our DNN architecture consists of three sub-architectures, which will merge later, as illustrated in Fig.~\ref{fig:dlshape}. 
Combined deep learning (DL) is not yet very common\footnote{Combined DL is similar to ensemble learning \cite{Dietterich:2000:EMM:648054.743935}.}, 
but recently there have been several studies in particle physics \cite{Lin:2018cin},
as well as in other areas \cite{7821017, 7477589}, which showed improved results over simple DL. 
In this subsection, we provide some details of our DNN layer architecture as follows:
\begin{enumerate}
\setcounter{enumi}{-1}
\item {\it Initialization.}
Since DNN has a lot of parameters, it is important to give non-biased initial values for the (weight, bias) before running DNN with all input data. We use the He uniform initialization method as in Ref. \cite{He:2015dtg}, among several other algorithms for parameter initialization \cite{pmlr-v9-glorot10a,DBLP:journals/corr/SaxeMG13,He:2015dtg}.
\item {\it Jet images.} They are represented by the top panel in Fig.~\ref{fig:dlshape}.
\begin{enumerate}
\item {\it Input data:} we use pre-processed jet images as inputs.
\item {\it Convolutional neural networks layers:} we use three layers of 
convolutional neural networks (CNN). Each layer has a $32 \times 2 \times 2$ filter with no stride and no padding. 
We proceed with the batch normalization process after filtering \cite{DBLP:journals/corr/IoffeS15}, 
using the ReLU function as our activation function \cite{pmlr-v15-glorot11a}. 
After activation, we introduce the max pooling layer which has a $2 \times 2$ shape with $2 \times 2$ strides and padding.
\item {\it Dense layers:} we feed the output of the CNN into two fully connected $1 \times 64$ dense layers, using ReLU as the activation function. 
\end{enumerate}
\item {\it The 6 high level variables.} Those are illustrated by the middle panel in Fig.~\ref{fig:dlshape}.
\begin{enumerate}
\item {\it Input data:} $\sqrt{\hat{s}}_{min}^{(bb\ell \ell)}$,  
$\sqrt{\hat{s}}_{min}^{(\ell \ell)}$, $M_{T2}^{(b)}$, $M_{T2}^{(\ell)}$, Higgsness and Topness.
\item {\it Dense layers:} 
we introduce four fully connected $1 \times 64$ dense layers with the ReLU activation function.
All four layers have the batch normalization process before activation. 
\end{enumerate}
\item {\it The 10 low level variables}: Those are illustrated by the bottom panel in Fig.~\ref{fig:dlshape}.
\begin{enumerate}
\item {\it Input data:} 
$p_{T\ell_1}$, $p_{T\ell_2}$, $\mpt$, $m_{\ell\ell}$, $m_{bb}$, $\Delta R_{\ell\ell}$, $\Delta R _{bb}$, $p_{Tbb}$, $p_{T\ell\ell}$, $\Delta\phi_{bb,\ell\ell}$.
\item {\it Dense layers:} we follow the same procedure as in the case with the 6 high level variables above.
\end{enumerate}

\item {\it Combination}
\begin{enumerate}
\item
{\it Merge:} we apply three single ($1 \times 1$) dense layers to the jet image, 
the 6 high level variables and the 10 low level variables. These layers are denoted as $\alpha$, $\beta$, and $\gamma$, respectively, as shown in Fig.~\ref{fig:dlshape}. 
To merge the three sub-architectures, we introduce the final dense layer of dimension $1 \times 3$ without an activation function.
\item {\it Final output:} to distinguish signal from backgrounds, we apply a layer of dimension $1 \times 2$ without an activation function. 
\end{enumerate}
\end{enumerate}

\subsection{DNN training}

We now proceed with deep learning on the DNN architecture described in Sec.~\ref{sec:dnna}, using the pre-processed input data.
We use Microsoft CNTK \cite{cntk} as the main DNN library on GPU with an Nvidia CUDA platform. 
We use the Adam optimizer \cite{DBLP:journals/corr/KingmaB14} with cross entropy with SoftMax loss function and classification error function. 
The sizes of the training data set and the testing data set are about 40k and 17k, respectively. 
The size of the mini-batch is 128 and that of the epoch is 30.

For each event, we prepare the jet images and the 16 variables.  
The dimension of the final output is $1 \times 2$, 
($\mathcal{P}_\textrm{sig}$, $\mathcal{P}_\textrm{bknd}=1-\mathcal{P}_\textrm{sig}$). 
If the deep learning score is equal to 1, i.e., $\mathcal{P}_\textrm{sig} =1$ ($\mathcal{P}_\textrm{bknd}=0$), the corresponding event is taken to be a signal event. 
If $\mathcal{P}_\textrm{sig} =0$ ($\mathcal{P}_\textrm{bknd}=1$), the event is considered to be background.

\section{Results} \label{sec:results}

In this section we present our results.
First we validate our framework by repeating the analysis performed in Ref.~\cite{Kim:2018cxf} under similar 
assumptions.\footnote{Our current analysis has several notable improvements over the one carried out in Ref.~\cite{Kim:2018cxf}.
First, the detector simulation is different --- in the current study, we use {\sc Delphes}, which assumes (on average) 
$\sim 90$\% ($\sim 80$\%) reconstruction efficiency for leptons ($b$-jets), while Ref.~\cite{Kim:2018cxf} assumed 
100\% reconstruction efficiency for both. In addition, the {\sc Delphes} detector resolution itself is slightly different 
from one used in Ref.~\cite{Kim:2018cxf}. In particular we find that the resolution of the missing transverse momentum 
is worse in {\sc Delphes} and hence our current results are more conservative (if not more realistic). Finally, 
as mentioned earlier, we are now including $tW+j$ production, which turned out to be the next dominant background, 
yet was missing from all previous studies. These effects should be kept in mind when comparing our results here
to previous results in the literature.} 
We obtained consistent results for the conventional cut-and-count method with {\sc Delphes} detector simulation. 
When we added deep learning, the signal significance improved slightly by 5-10\%. 

\begin{figure}[t]
\centering
\includegraphics[width=8cm]{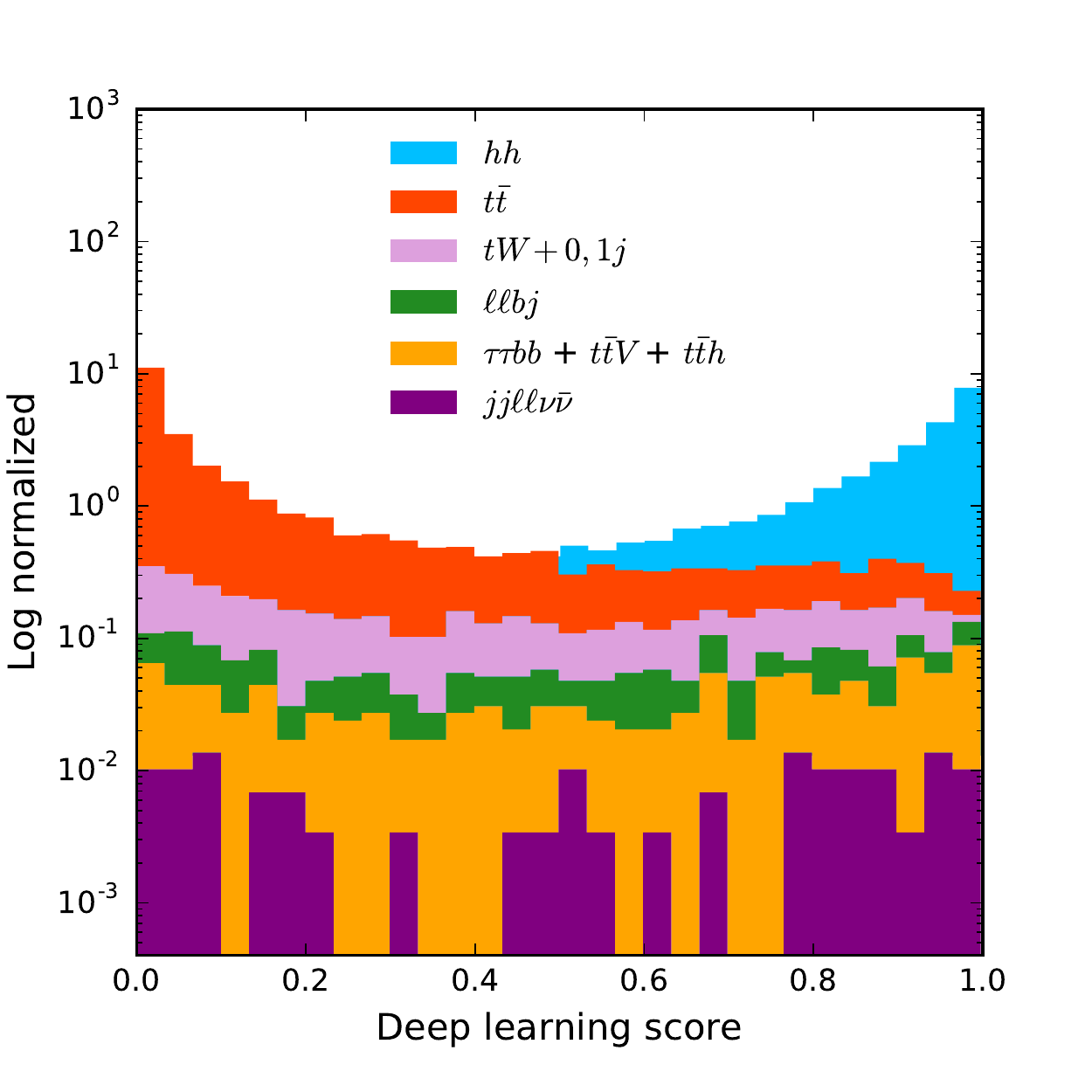}
\caption{\label{fig:score} 
Deep learning score for the signal and the individual backgrounds. 
}
\end{figure}

Now considering {\it all} relevant backgrounds and using {\it all 16 variables and jet images}, 
we show the deep learning score for the signal and the individual background processes in Fig.~\ref{fig:score}. 
The signal should peak near $\mathcal{P}_\textrm{sig}=1$ by construction, and indeed this is what is observed in the figure. 
Note that the $t \bar t$ and $tW$ processes are well separated from the signal and both peak near $\mathcal{P}_\textrm{sig}=0$.
This is direct consequence of the improvements made in our analysis --- introducing the proper kinematic variables and jet images,
which were meant to target the dominant background ($t \bar t$ production), as evidenced in Figs.~\ref{fig:var1}, 
\ref{fig:THness} and \ref{fig:jetimage}. Although the subdominant backgrounds are also reduced in this process, 
they remain rather flat in Fig.~\ref{fig:score}. 

\begin{figure}[t]
\centering
\includegraphics[width=7cm, height=7cm]{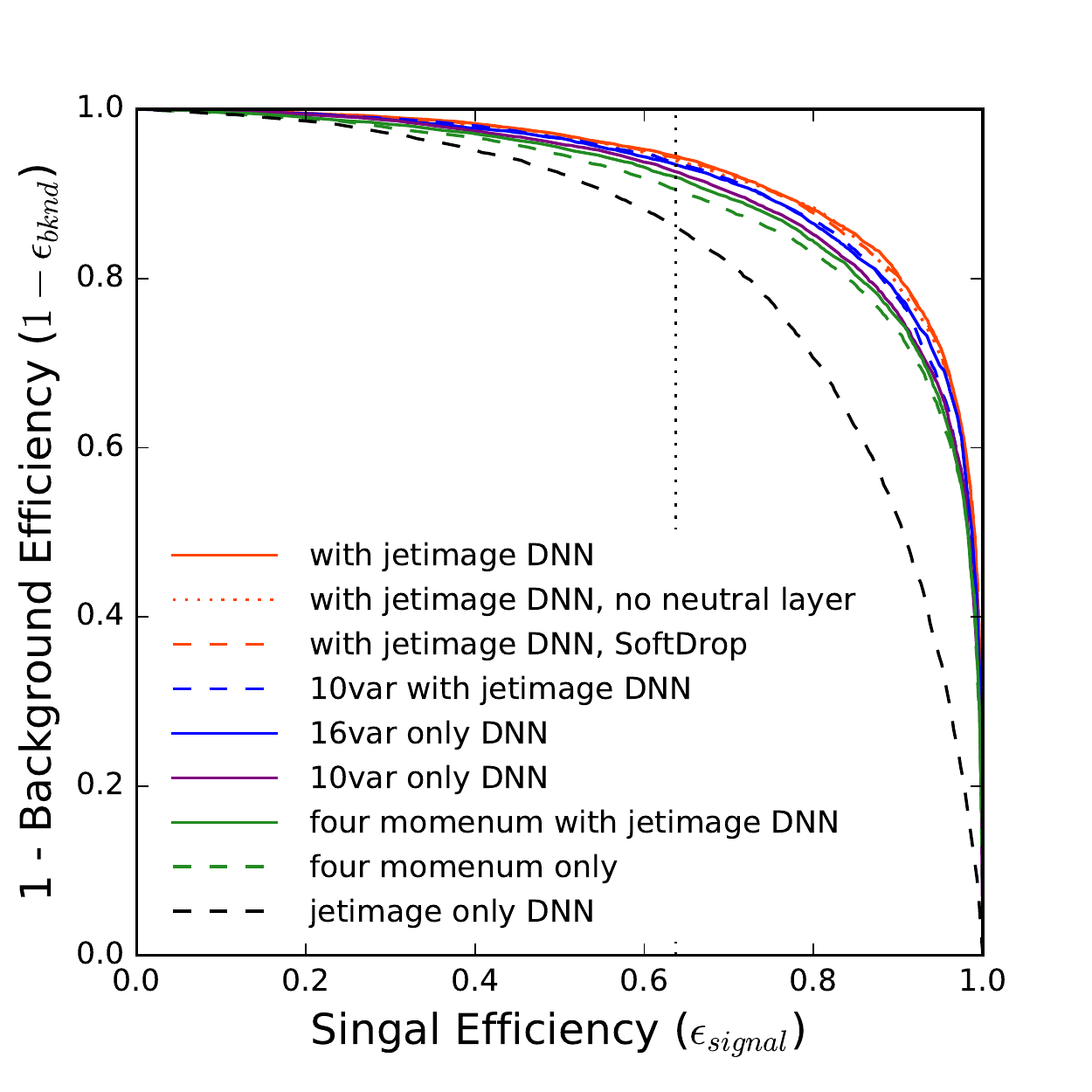} 
\includegraphics[width=7cm, height=7cm]{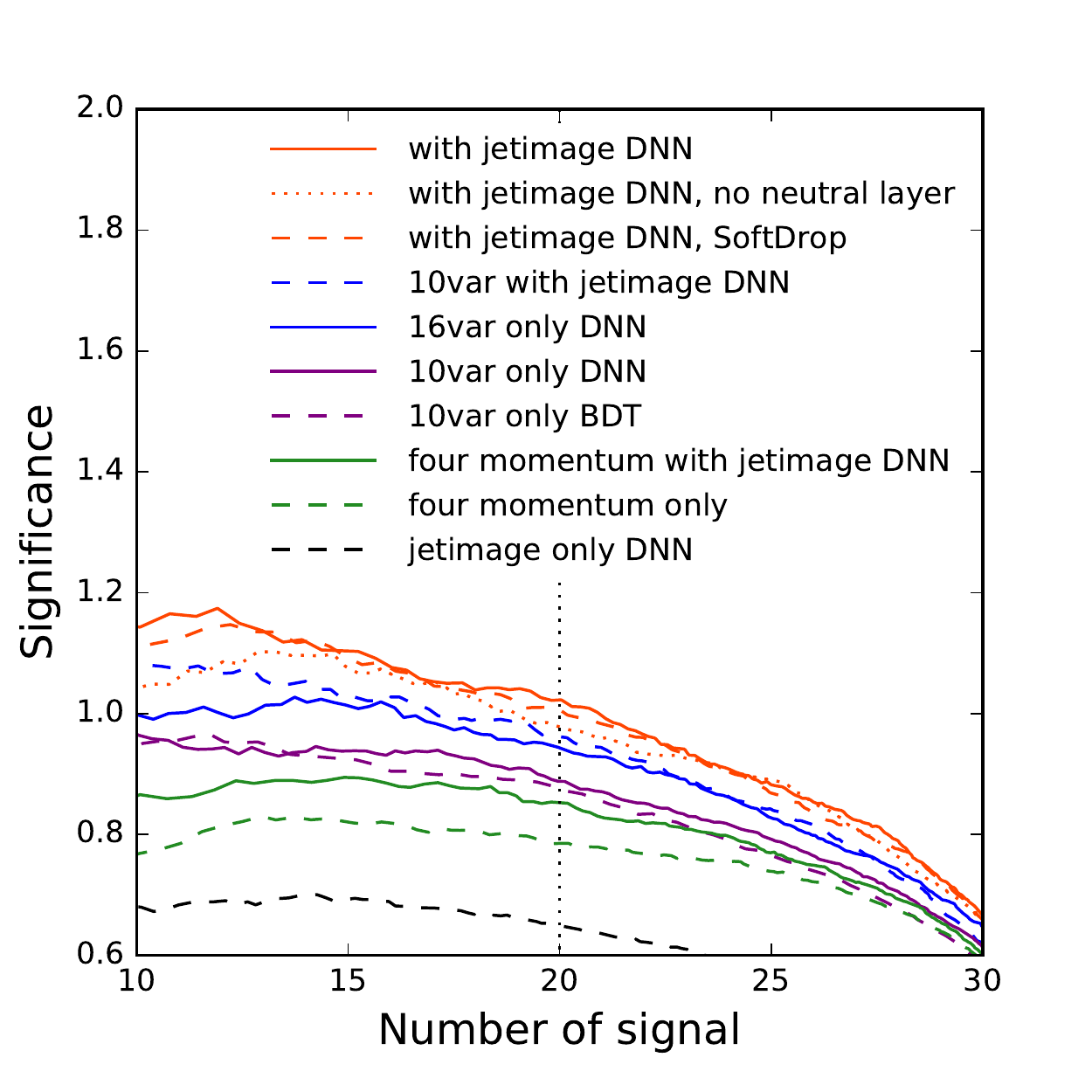} 
\caption{\label{fig:eff} 
A ROC curve (left panel) and signal significance as a function of the number of signal events (right panel). 
The vertical lines mark $N=20$ signal events, which corresponds to $\epsilon_{\rm signal}=0.64$.
}
\end{figure}

The deep learning score shown in in Fig.~\ref{fig:score} can now be used as a signal-to-background discriminator.
By placing a lower cut and counting the number of surviving signal and background events, one 
obtains the efficiency curve (also known as a receiver operating characteristic (ROC)) shown in the left panel of Fig.~\ref{fig:eff}. 
The curve contains several independent runs of deep learning and 
shows the signal efficiency ($\epsilon_{\rm signal}$) versus the fraction of rejected background events, i.e., 
$1 - \epsilon_{\rm bknd}$, where $\epsilon_{\rm bknd}$ is the background efficiency. 
The efficiency corresponding to the results in Fig.~\ref{fig:score} is shown with the red solid curve labeled ``with jetimage DNN''. 
The other two solid lines show the efficiencies which would be obtained if we were to remove the jet images from the analysis:
the purple solid curve (labelled ``10var only DNN'') is obtained with the help of the 10 low-level kinematic variables, while
the blue solid curve (labelled ``16var only DNN'') shows the improvement when we add the 6 high-level variables and 
use the full set of 16 variables from Section~\ref{sec:method1}, but still without jet images. 
The black dotted curve (labeled ``jet image only DNN'') shows the result when we use jet images alone, with no help from any of the 16 kinematic variables.
Finally, the blue dashed line (labelled ``10var with jetimage DNN'') shows the result from an analysis combining jet images with the 10 low-level kinematic variables only.
The corresponding signal significances are shown as a function of the number of events in the right panel of Fig.~\ref{fig:eff}. 
Note that the right panel contains an additional curve (the purple dashed line labeled ``10var only BDT'')
where we use the 10 low-level variables and adopt a BDT algorithm using the {\sc TMVA} tool kit \cite{Hocker:2007ht}.
The comparison of the latter line against the ``10var only DNN'' result (purple solid line) reveals the relative performance of DNN versus BDT.

In order to examine the effects of pile-up, we use several methods as follows. 
In the first method, we use the Soft Drop algorithm \cite{Larkoski:2014wba} to remove soft jet activity which is exacerbated by pile-up. 
We set $\beta = 0$ and $z_\textrm{cut} = 0.1$ with $R = 1.2$ anti-$k_T$ clustered fatjets. 
Then we select the closest fatjet to the $b\bar b$ momentum in the 
$\eta$-$\phi$ plane and replace the particle flow data with the charged and neutral jet constituents of the selected fatjet. 
Soft Drop does not affect the jet images and retains the same shapes as in Fig.~\ref{fig:jetimage}. 
In second method, we remove the neutral jet image layer in the analysis.
Unlike charged particles, which can be cleaned up from pile-up relatively easily by checking the longitudinal vertex information \cite{Bertolini:2014bba}, 
neutral particles cannot be treated the same way and suffer from non-removable pile-up effects. 
The corresponding results with these two pile-up mitigation methods are also shown in Fig.~\ref{fig:eff}
with the red dotted line labelled ``16var with jetimage DNN, SoftDrop'' 
and the red, dashed line labelled ``16var with jetimage DNN, no neutral layer'', respectively.

We also examine the performance of the DNN with four momentum information as input. The corresponding results are shown in Fig.~\ref{fig:eff}, where
the green-dashed (green-solid) curve represents the significance with four momentum information only (four momentum information plus jet images). 
The inputs are 18 real numbers, {\it i.e.,} the four momenta of the two leptons and the two $b$-tagged jets and the missing transverse momentum.
For this exercise, we use a 4 $\times$ 128 dense layer instead of a 4 $\times$ 64 dense layer. 
We notice that the DNN performance with kinematic variables is better. 
This is because, in general, the use of four momenta requires a large training sample in order to be effective, while the
kinematic variables already perform efficiently with a smaller data set.
If the architecture is deep enough with a large amount of data, the DNN performance with four momentum information would be comparable (or better) to that with kinematic variables only. This exercise illustrates the importance of the appropriate use of kinematic variables. 

In summary, Fig.~\ref{fig:eff} demonstrates that jet images (which capture the effects of color flow) can improve performance over the baseline selection cuts.  
At the same time, DL with jet image substructure alone does not show the best performance, and becomes fully effective (and still stable under pile-up)
only when it is combined with the full set of 16 variables, including the high-level ones.

%
\begin{table*}[t]
\begin{center}
\setlength{\tabcolsep}{1.1mm}
\renewcommand{\arraystretch}{1.4}
\scalebox{0.77}{
\begin{tabular}{|c||c|c|c|c|c|c|c|c|c|c|c|}
\hline   
& Signal   
& $t\bar{t}$   & $t\bar{t}h$  & $t\bar{t}V$  & $\ell\ell bj$  
& $\tau\tau b b$ & $tw +j$ & $jj\ell\ell \nu\nu$ 
& $\sigma$  & $S /B$
\\
\hline \hline
\bf{Baseline cuts}:  $\mpt> 20 \gev$, 
& \multirow{4}{*}{$0.01046$} & \multirow{4}{*}{$1.8855$} & \multirow{4}{*}{$0.0269$} 
& \multirow{4}{*}{$0.0179$} & \multirow{4}{*}{$0.0697$} & \multirow{4}{*}{$0.0250$} 
& \multirow{4}{*}{$0.2209$} & \multirow{4}{*}{$0.0113$} 
& \multirow{4}{*}{$0.38$} & \multirow{4}{*}{$0.0046$} 
\\   
$p_{T,\ell}  > 20 \gev$,  $\Delta R_{\ell\ell}<1.0$,    & & & & & & & & & & 
\\
$p_{T,b}  > 30 \gev$,  $\Delta R_{bb}<1.3$,  &   & & & & & & & & & 
\\  
$m_{\ell\ell}<65 \gev$,  $95<m_{b{b}}<140\gev$ & & & & & & & & & & 
\\
\hline
jet-image DL  &  0.00667 & 0.1855 & 0.0147 & 0.00731 & 0.0243 & 0.0128 & 0.0626 & 0.00786 
& 0.65 & 0.021
\\
\hline
10 low-level variables DL  &  0.00668 & 0.0738 & 0.0132 & 0.00529 & 0.0184 & 0.00842 & 0.0424 & 0.00516
& 0.89 & 0.040
\\
\hline
16 variables DL  &  0.00668 & 0.0676 & 0.0109 & 0.00454 & 0.0163 & 0.00689 & 0.0376 & 0.00418 
& 0.94 & 0.045
\\
\hline
10 variables + jet-image DL  &  0.00667 & 0.0630 & 0.00964 & 0.00429 & 0.0194 & 0.00791 & 0.0343 & 0.00393 
& 0.96 & 0.047
\\
\hline
16 variables + jet-image DL  &  0.00668 & 0.0602 & 0.00914 & 0.00252 & 0.0133 & 0.00689 & 0.0299 & 0.00344 
& 1.0 & 0.053
\\
\hline
\end{tabular}}
\end{center}
\vspace*{-0.3cm}
\caption{
Signal and background cross sections in fb after baseline cuts (first row) and at different stages of analysis, using a combination of kinematic variables and jet images while requiring $N = 20$ signal events. 
The significance $\sigma$ is calculated using the log-likelihood ratio for a luminosity of 3~$\rm{ab}^{-1}$ at the 14 TeV LHC. 
} 
\label{tab:Cutflow}
\end{table*}

Table \ref{tab:Cutflow} summarizes the signal and background cross sections in fb at different stages of the analysis for the case of $N = 20$ signal events. The last two columns show the signal significance $\sigma$ and the signal-to-background ratio $S /B$. The significance is calculated using the log-likelihood ratio for a luminosity of 3~$\rm{ab}^{-1}$ at the 14 TeV LHC.

\begin{figure}[t]
\centering
\includegraphics[width=8cm]{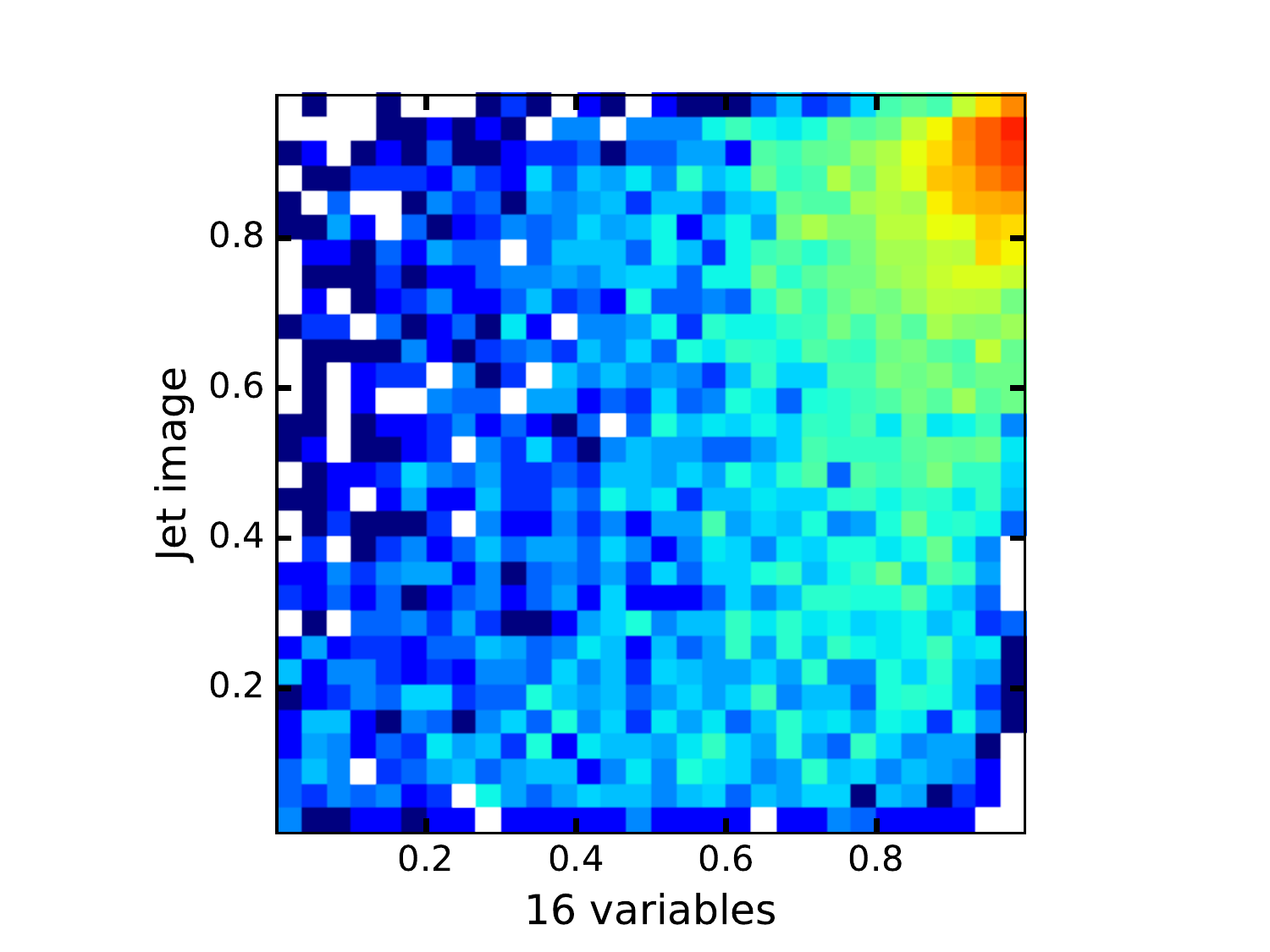} \hspace*{-0.9cm}
\includegraphics[width=8cm]{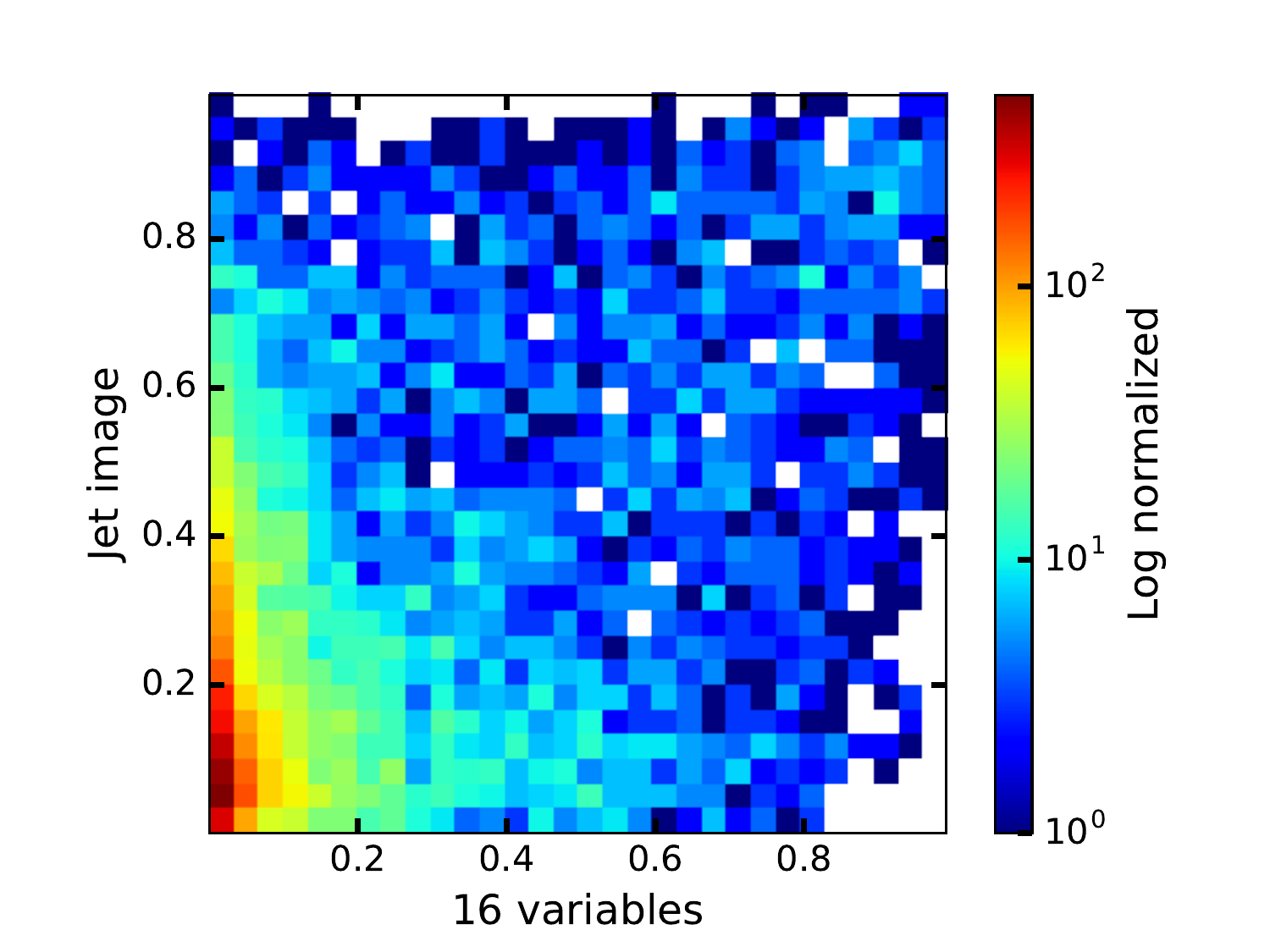}
\caption{\label{fig:correlation} 
Correlation of the deep learning scores obtained in independent DL analyses using the 16 kinematic variables only ($x$-axis)
and jet images only ($y$-axis) for signal (left panel) and background (right panel). 
}
\end{figure}

In order to understand the correlation between jet images and the 16 kinematic variables, we performed two independent runs with 
``jet images only; no kinematic variables'' and ``16 kinematic variables; no jet images''. The corresponding results are shown in Fig.~\ref{fig:correlation}.
Since the two DLs are trained separately, both the $x$-axis and the $y$-axis are normalized to unity. 
As expected, Fig.~\ref{fig:correlation} reveals a degree of correlation between the jet images and the 16 kinematic variables, 
which is somewhat stronger for the signal and less so for the background.

\begin{figure}[t]
\centering
\hspace*{-0.3cm}
\includegraphics[width=16.0cm]{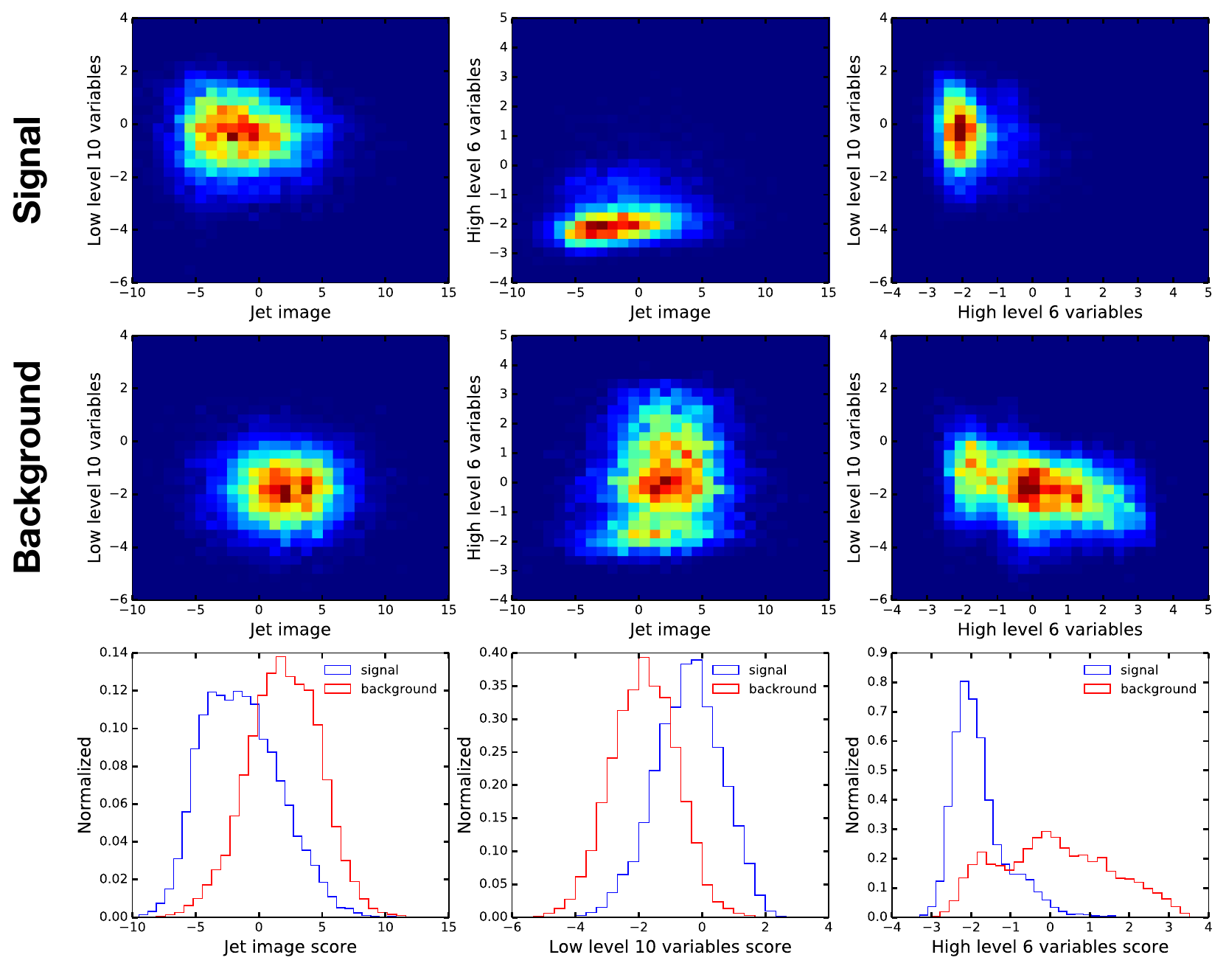} 
\caption{\label{fig:correlation2}Correlations among the intermediate DL substructure scores for jet images, the 6 high level variables and the 10 low level variables. 
The top (middle) row shows the correlations for signal (background) events and the bottom row shows the corresponding distributions for each individual substructure score.
}
\end{figure}

In our main analysis, we performed simultaneous runs as shown in the deep learning architecture in Fig.~\ref{fig:dlshape}. Before calculating our final deep learning score, 
we obtain three intermediate values, $\alpha$, $\beta$, and $\gamma$, which represent the DL scores for the respective substructure 
corresponding to the jet images, the 6 high level variables and the 10 low level variables. 
The first 6 panels in Fig.~\ref{fig:correlation2} show the pair-wise correlations between these three intermediate scores
for the signal (top row) and the background (middle row). 
The bottom three panels in the figure show the one-dimensional distributions of the 
intermediate scores for signal (blue histograms) and background (red histograms).
We observe that the score from jet images ($\alpha$) is relatively uncorrelated to the kinematic variables scores $\beta$ and $\gamma$,
which motivates the simultaneous training on jet images and kinematic variables together.

\begin{figure}[t]
\centering
\includegraphics[width=5.5cm]{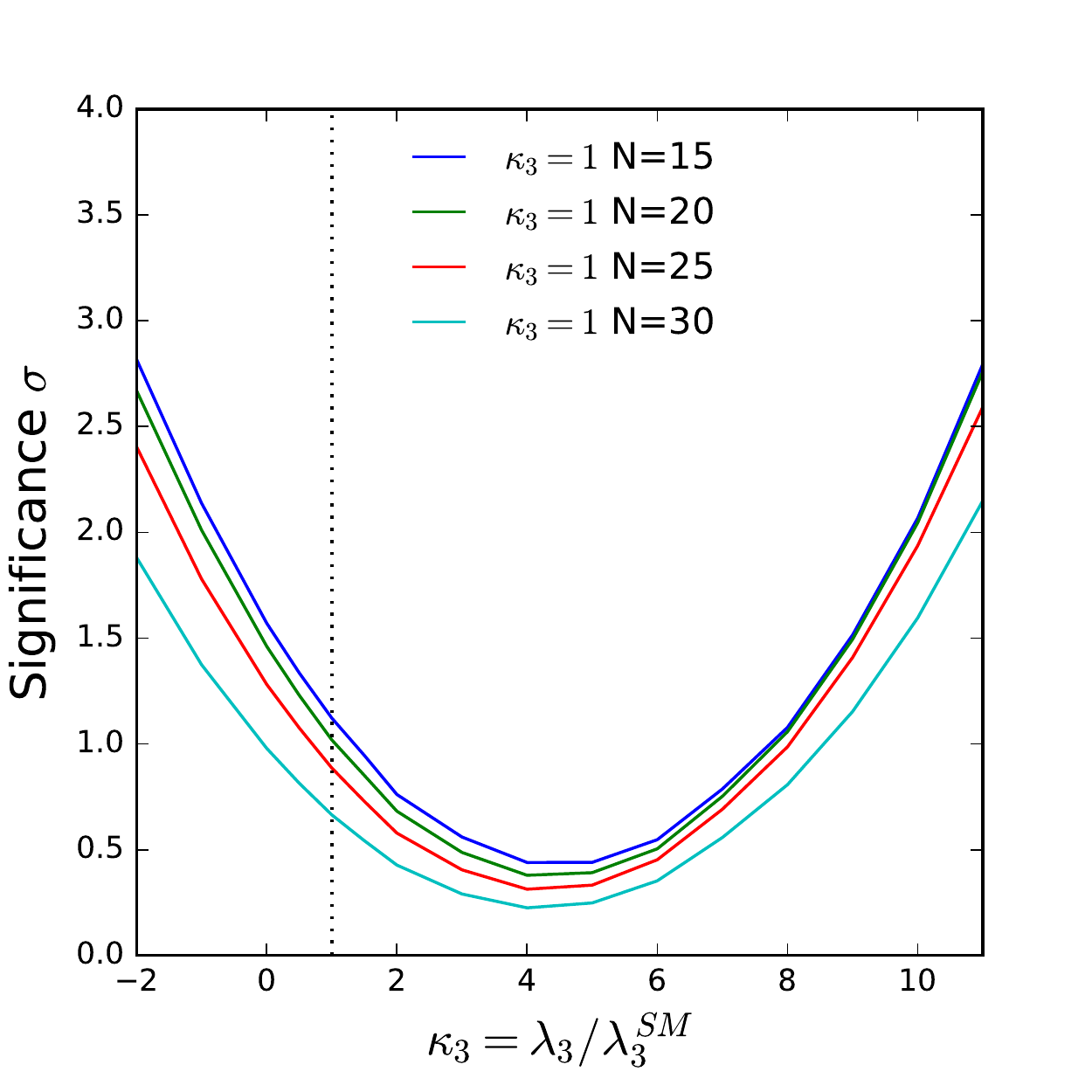} \hspace*{-0.5cm}
\includegraphics[width=5.5cm]{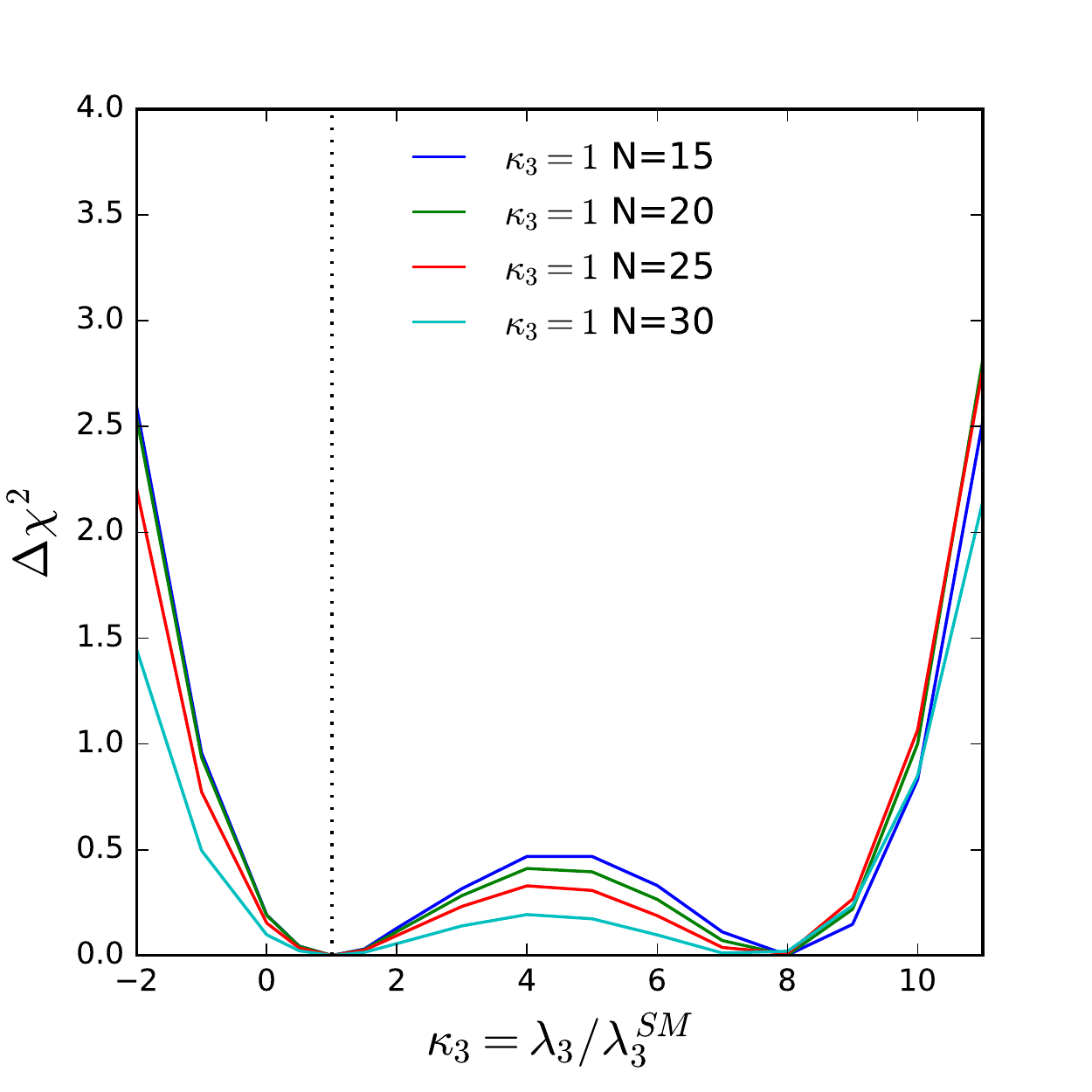} \hspace*{-0.5cm}
\includegraphics[width=5.5cm]{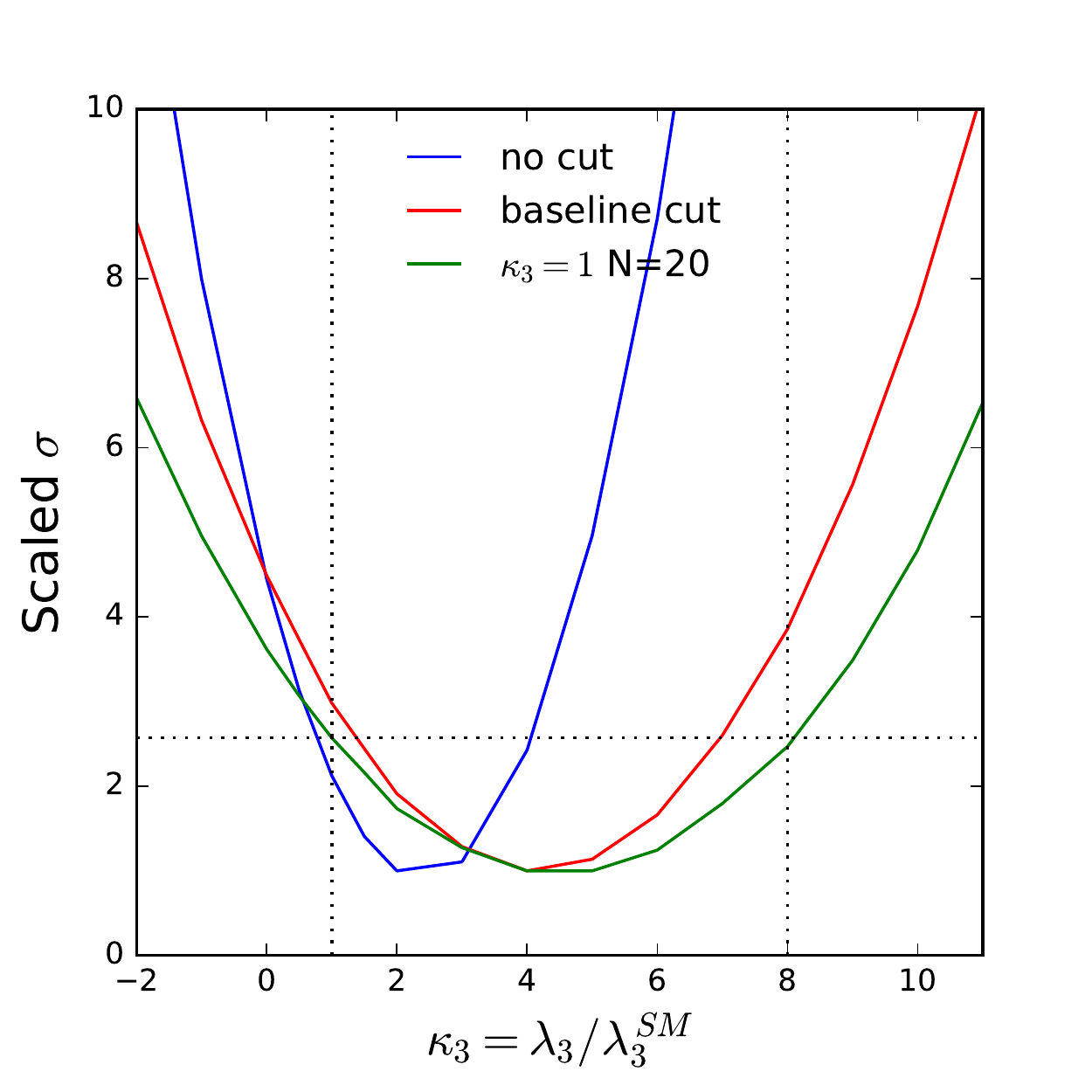}\hspace*{-0.5cm}
\caption{\label{fig:kappa3} 
The discovery significance (left panel) and precision (middle panel) as a function of $\kappa_3$. 
Both the significance and the precision are calculated fixing DL cuts that would give a certain number 
of signal events ($N=15, 20, 25, 30$) for the SM at $\kappa_3=1$ (marked with the dotted vertical line). 
The shape of the significance roughly follows the cross section ratios between the case of $\kappa_3 \neq 1$ to the case of $\kappa_3=1$.
This is illustrated in the right panel, which shows the scaled cross section, computed as ${\sigma(\kappa_3)} / { {\rm min} \big ( \sigma(\kappa_3) \big ) }$ for each curve.
}
\end{figure}

Finally, in Fig.~\ref{fig:kappa3} we scan over different values of the triple Higgs coupling $\kappa_3$ 
and show the discovery significance (left panel) and precision (middle panel) as a function of $\kappa_3$. 
Both the significance $\sigma$ and the precision $\Delta \chi^2$ are calculated fixing DL cuts that would give a certain number 
of signal events ($N=15, 20, 25, 30$) for the SM at $\kappa_3=1$ (marked with the dotted vertical line). 
For the significance, we used the log-likelihood-ratio
\beq
  \sigma_{dis} \equiv
    \sqrt{-2\,\ln\bigg(\frac{L(B | S \!+\!B)}{L( S \!+\!B| S \!+\!B)}\bigg)}
  \;\;\;\;\; \text{with}\;\;\;
  L(x |n) =  \frac{x^{n}}{n !} e^{-x} \,,
\label{Eq:SigDis} \eeq
where $S$ and $B$ are the expected number of signal and background events, respectively. 
We define $\Delta \chi^2$ as 
\begin{equation}
\Delta \chi^2 = \left ( \frac{ \Big ( S(\kappa_3) + B \Big ) -  \Big ( S(\kappa_3=1) + B \Big )    }{ \sqrt{S(\kappa_3=1) + B} } \right )^2 \, .
\end{equation}

The shape of the significance roughly follows the cross section ratios between the case of $\kappa_3 \neq 1$ to the case of $\kappa_3=1$.
This is illustrated in the rightmost panel of Fig.~\ref{fig:kappa3}, which shows the cross section 
scaled as ${\sigma(\kappa_3)} / { {\rm min} \big ( \sigma(\kappa_3) \big ) }$,
i.e., normalized with respect to the minimum cross section for each curve. 
The blue curve represents the double Higgs production cross section before cuts, and in this case we find the 
minimum of the cross section somewhere between $\kappa_3=2$ and $\kappa_3 = 3$. After baseline cuts (the red solid line), 
the minimum shifts to around $\kappa_3\sim 4$,
and after DL cuts (the green solid line), the minimum shifts even further out to around $\kappa_3\sim 5$. 
In the latter case, we observe that the signal cross sections for $\kappa_3=1$ and $\kappa_3=8$ are numerically very close, 
as indicated by the two vertical dotted lines in the right panel.
This provides an explanation for the double dip structure seen in the middle panel of Fig.~\ref{fig:kappa3}. 

As demonstrated in the rightmost panel of Fig.~\ref{fig:kappa3}, the analysis cuts modify the signal cross section so that the
location of its minimum shifts to higher values of $\kappa_3$. This can be understood as follows.
At leading order, the Higgs pair production cross section is given by 
\begin{eqnarray}\label{eq:higgs_xsec}
\sigma_{gg\rightarrow hh}(\hat{s})&=&\frac{\alpha^2_s}{2^{15}v^4\pi^2\hat{s}^2}\int d\hat{t}
(|F_1|^2+|F_2|^2) \approx c_\triangle \, \kappa_3^2+c_{\triangle,\Box} \, \kappa_3+c_\Box \, ,
\end{eqnarray}
before convoluting with the parton distribution functions \cite{Glover:1987nx,Borowka:2016ypz}. 
Here $F_1$ represents a parity-even triangle and box diagram contribution, 
while $F_2$ is a parity-odd box diagram contribution. Now $F_1$ can be rewritten as $F_1=\kappa_3 F_\triangle+F_\Box$, where 
$F_\triangle$ is the triangle diagram contribution and $F_\Box$ is the box diagram contribution.
Therefore the cross section can be parameterized as a quadratic function of $\kappa_3$, 
where the $c$ coefficients are related to contributions from $\triangle$ and $\Box$ diagrams. 

The observation that the baseline cuts and the DL cut shift the minimum cross section to a larger $\kappa_3$ value
implies that the effects of the cuts are stronger on $c_\triangle$ than $c_{\triangle,\Box}$. In other words, 
our cuts are more likely to affect the triangle diagram which contains the triple Higgs coupling. 
Unlike the box diagram, the triangle diagram includes an off-shell Higgs in the $s$-channel. 
Since it is harder to produce a Higgs pair from an $s$-channel off-shell Higgs, 
the Higgs pair generated from the triangle diagram is not as energetic as the one coming from the box diagram,
and will therefore tend to have lower transverse momentum. As discussed in Section~\ref{sec:method2}, several of
the cuts on our kinematic variables, namely, $\Delta R_{bb}$, $\Delta R_{\ell\ell}$, $\Delta \phi_{bb, \ell\ell}$, $p_{Tbb}$ and $p_{T\ell\ell}$,
rely on the fact that the Higgs bosons are produced with a significant boost.
Consequently, the effect of the cuts will be to suppress the $c_\triangle$ term and enhance the box diagram contribution, 
which in turn shifts the location of the minimum to a larger value of $\kappa_3$.

\begin{figure}[t]
\centering
\includegraphics[width=5.75cm]{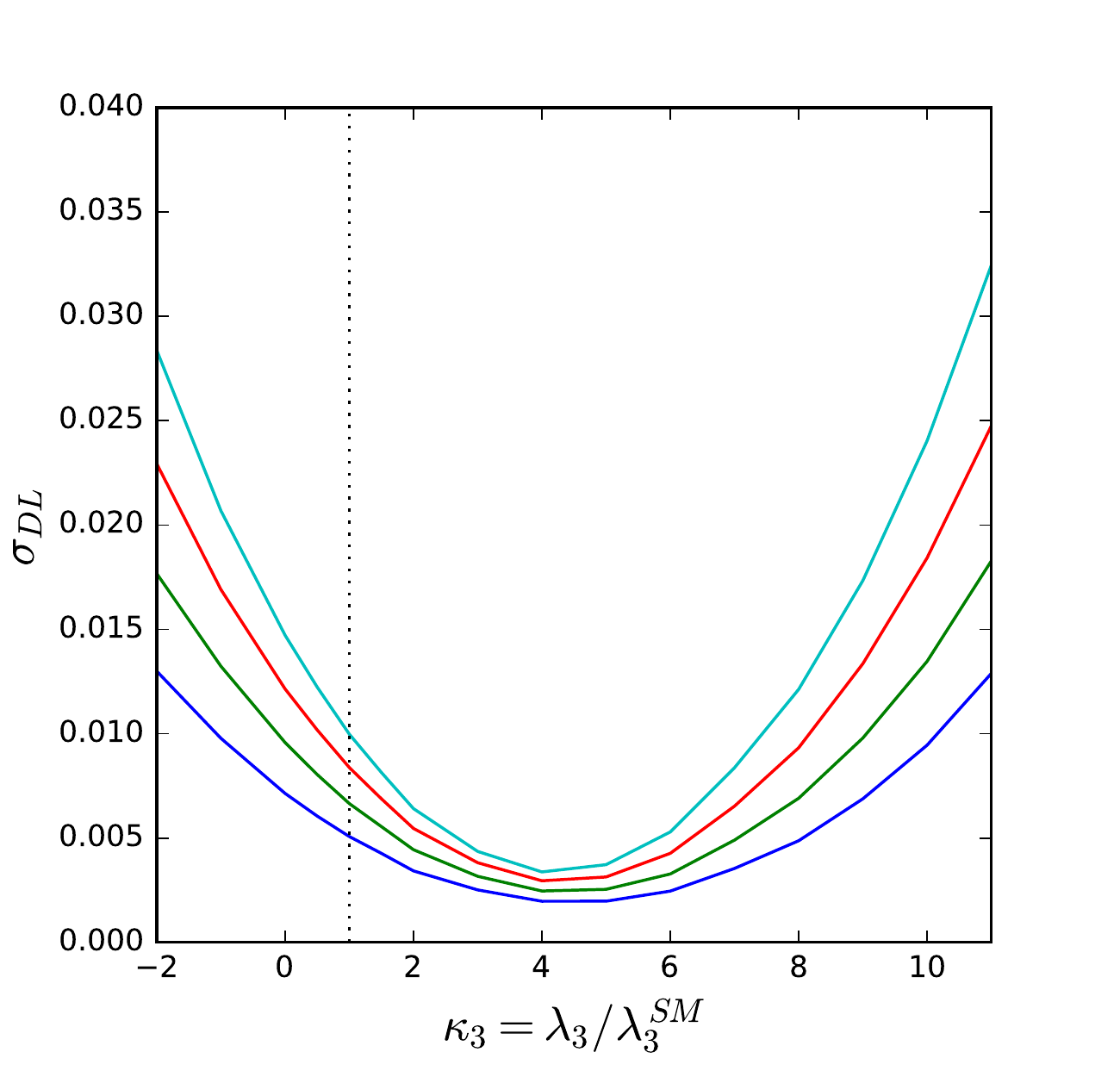}  
\includegraphics[width=5.5cm]{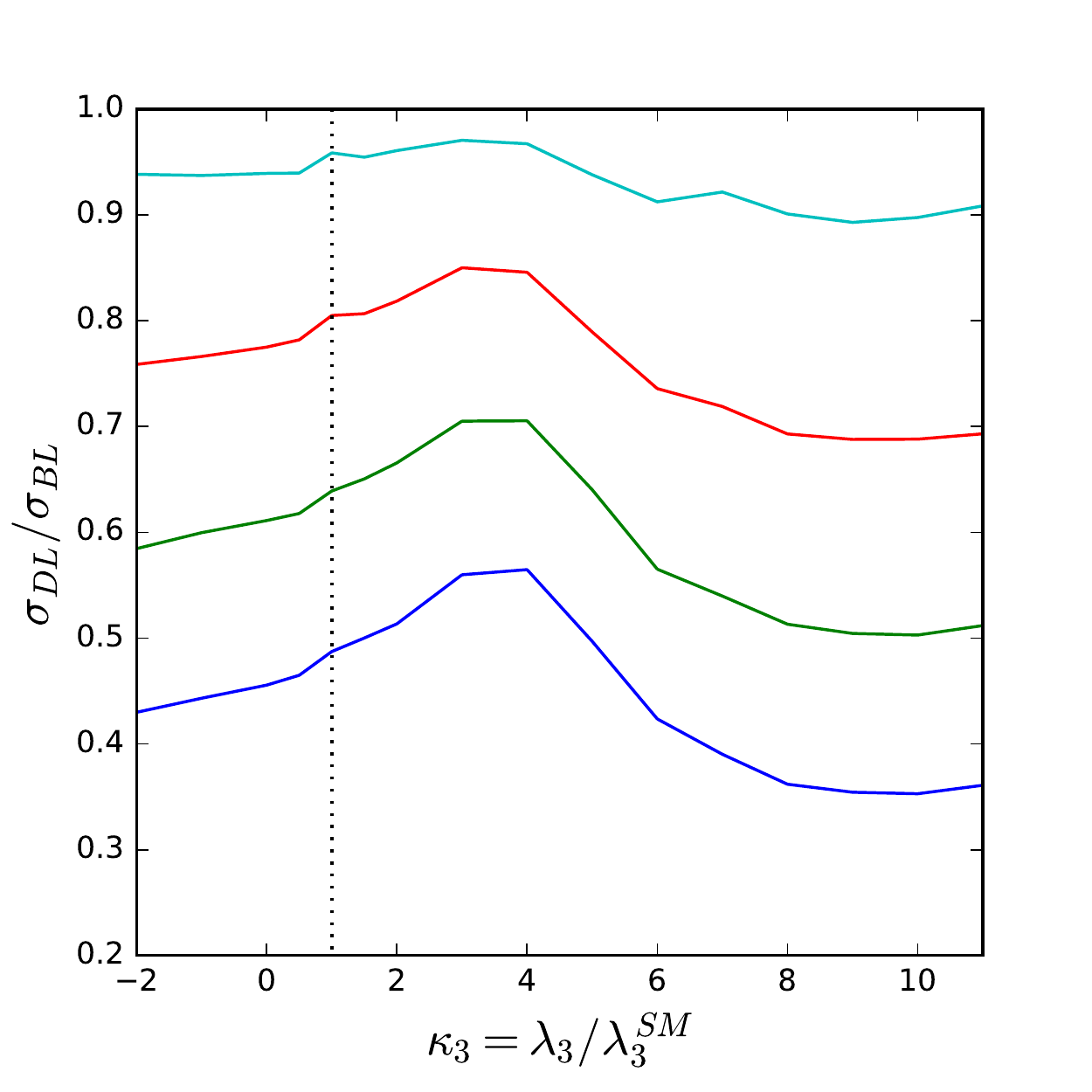} 
\includegraphics[width=11cm]{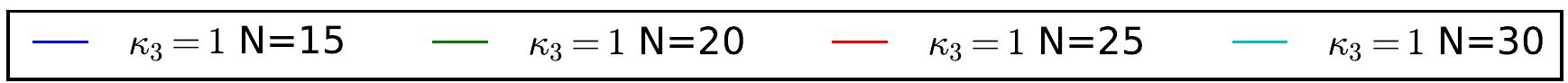} 
\caption{\label{fig:xsec_comparison} 
Cross section $\sigma_{DL}( p p \to h h \to bb\ell^+\ell^-\nu\bar \nu)$ in pb after cutting on the DL score (left panel) 
and the ratio $\sigma_{DL} / \sigma_{\rm baseline}$ between the cross section $\sigma_{DL}$ after the DL cut 
and the cross section $\sigma_{\rm baseline}$ after baseline cuts (right panel). 
}
\end{figure}

Note that the results for the significance and the precision in Fig.~\ref{fig:kappa3} 
do not change dramatically when we require a different number of signal events at the SM point.
This means that the dependence on the DL cut is relatively mild, since the kinematics remains similar when we vary $\kappa_3$, so that
the dependence on the cross section is more important. 
This is illustrated in Fig.~\ref{fig:xsec_comparison}, which shows the cross section (in pb) after cutting on the DL score, $\sigma_{DL}( p p \to h h \to bb\ell^+\ell^-\nu\bar \nu)$, (left panel) and the ratio $\sigma_{DL} / \sigma_{\rm baseline}$ between the cross section $\sigma_{DL}$ after the DL cut and the cross section $\sigma_{\rm baseline}$ after baseline cuts (right panel).

\section{Discussion} \label{sec:discussion}

In this paper, we investigated double Higgs production in the $hh \to bbWW^* \to b b \ell \ell + \mptvec$ final state.
It is known to be one of the difficult channels due to the large backgrounds, ${\sigma_{\rm bknd}}/{\sigma_{\rm hh}}\sim 10^5$.
We performed a detailed analysis by adopting a deep learning framework and successfully combining new kinematic variables and jet image information.
As a result, we obtained a sizable increase in signal sensitivity and an improved signal-to-background ratio compared to the existing analyses.

Our results showed that the dominant $t\bar t$ background can be brought down to the level of the other remaining backgrounds, 
without sacrificing too much in the signal rate. This is mostly due to the use of Higgsness, Topness and the subsystem variable $M_{T2}^{(b)}$. 
Other backgrounds like $b b\tau\tau$ can be reduced further by the use of $M_{T2}^{(\ell)}$. 
Finally, additional improvements are possible with the use of jet images. 
After all cuts, we find that all backgrounds contribute at similar levels. 

We find from recent CMS and ATLAS analyses with 36 fb$^{-1}$ of LHC data at 13 TeV that 
the 95\% confidence level observed (expected) upper limit on the production cross section is
22.2 (12.8) times the standard model value \cite{Sirunyan:2018two} for CMS and 6.7 (10.4) times the predicted Standard Model cross-section \cite{ATLAS:2018otd} for ATLAS. 
The leading channel in CMS is $bb\gamma\gamma$ followed by $bb\tau\tau$, while the leading channels in ATLAS are $bb\tau\tau$ and $bbbb$, followed by $bb\gamma\gamma$.  
The main difference arises due to the superior $b$-tagging efficiency for the ATLAS detector \cite{Aaboud:2018knk}. 
In both studies, the $bbWW^*$ channel was largely overlooked due to the expected poor significance. 
However, our study suggests that double Higgs production may be probed in the dilepton $bbWW^*$ channel as well,
and would contribute to the combined analysis on par with the other final states, increasing the overall significance. 
For example, in Ref. \cite{ATLAS:2018combi}, the ATLAS collaboration showed that the combined significance of $ hh\to bbbb$, $hh \to bb\tau\tau$, and $hh \to bb\gamma\gamma$ is 3.5 (3.0) without (with) systematic uncertainties at the 14 TeV LHC with 3 ab$^{-1}$. Their individual significance is 1.4, 2.5 and 2.1 (0.61, 2.1 and 2.0), respectively without (with) systematics. 
They did not combine with the $hh \to bbWW^*$ channel  but a naive estimate shows that when including our channel, the combined significance would be about 3.7.

We urge the experimental collaborations to consider the ideas presented in this paper and test them in the LHC data.
We would also like to mention that the proposed method can be easily generalized to the semi-leptonic channel from 
$hh \to bbWW^*$ production, as well as to other processes with similar final states.

\section*{Acknowledgments}
This work is supported in part by United States Department of Energy (DE-SC0010296, DE-SC0017988, DE-SC0017965, DE-SC0019474) and Korea NRF-2018R1C1B6006572. We thank KIAS for providing computing resources. 
We thank Georgios Anagnostou for valuable comments and Anja Butter, Tilman Plehn and Chris Rogan for general discussion on machine learning. 
KK is grateful to the Mainz Institute for Theoretical Physics, which is part of the DFG Cluster of Excellence PRIMA$^+$ (Project ID 39083149), for its hospitality and its partial support during the completion of this work.

 \appendix

\section{Deep Neural Network} \label{sec:appendix}

The artificial neural network (ANN) is one of the most popular approaches to pattern recognition in machine learning algorithms. 
The structure of an ANN is defined by a succession of non-linear and linear transformations between nodes or artificial neurons, which are located on input, output or hidden layers. 
A hidden layer which uses an ordinary one-dimensional layer is called a dense layer or a fully connected layer.

The linear operation consists of weights and bias:
\begin{equation}
O[i]=\sum_{j=0}^{n_{I}-1}(\mathcal{W}[i,j]I[j]+\mathcal{B}[i]) \, , ~~~ i = 0,\cdots,n_{O}-1 \, ,
\end{equation}
where $I[j]$ is the value of the $j$-th neuron (input) in the prior layer, $O[i]$ the value of the $i$-th neuron (output) in the subsequent layer, 
$\mathcal{W}[i, j]$ are the weights, and $\mathcal{B}[i]$ the bias. The index $i$ ($j$) takes the values $0,\cdots,n_{O}-1$ ($0,\cdots,n_{I}-1$) and $n_{O}$ ($n_{I}$) is the dimension of the output (input). 
The input initially can be given in more than one dimension. For example, if the input results from a convolution and has dimension $n\times n$, it may be rearranged as follows:
\begin{equation}
I[j]=(I[1,1]\cdots I[1,n]\cdots I[n,1]\cdots I[n,n]) \, ,
\end{equation}
where the corresponding dimension of the input would be $n_I = n^2$.

The non-linear transformation is often called activation function, which imitates the action potential of biological neurons. 
Similar to how each neuron adjusts how much signal it needs to deliver to the next neuron using an electric action potential, 
the activation function determines the output of a particular neuron for a set of given inputs from neurons on the previous layer, 
and the output is then used as input for the next artificial neuron. 
The commonly used activation functions are
\begin{eqnarray}
 \textrm{ReLU}(x[i]) &=& \textrm{max}(0,x[i]) \, ,\\ [2mm]
 \textrm{Sigmoid}(x[i]) &=&\frac{1}{1+e^{-x[i]}}\, ,\\ [2mm]
 \textrm{SoftMax}(x[i])&=&\frac{e^{x[i]}}{\sum_i e^{x[i]}} \, ,
 \end{eqnarray}
where $x[i]$ represents the value of the $i$-th neuron.

If the neural network has sufficiently many hidden layers, the network is called deep neural network (DNN). 
DNN can learn from the input data to obtain the desirable output by adjusting the parameters in the hidden layers. 
We note that the proper normalization of the input data helps improve convergence during training.
The goal of the training is to determine the parameters (weights and biases) by minimizing the loss, which represents the difference between the target output and the actual DNN output.
There are various algorithms for optimization of the parameters \cite{DBLP:journals/corr/KingmaB14,Duchi:2011:ASM:1953048.2021068, rmsprop}. 
Some well known loss functions are
\begin{eqnarray}
\textrm{Mean Square Error} &=& \frac{1}{n}\sum_{i=1}^{n}(x[i]-t[i])^2\, ,\\
\textrm{Cross Entropy}  &=&  -\sum_{i=1}^{n} t[i] \textrm{log}(x[i])\, , \\
\textrm{Cross Entropy with SoftMax} &=&  -\sum_{i=1}^{n} t[i] \textrm{log}(\textrm{SoftMax}(x[i]))\, ,
\end{eqnarray}
where $\{t[i]\}$ is the true answer (either 1 or 0 in our current study), $\{x[i]\}$ is the DNN final output, and $n$ is the number of neurons in the output layer. 

Instead of feeding the entire data into the DNN all at once, one splits the input data into several subsets with random selection and takes one subset, called mini-batch, for a given iteration, 
which helps avoid the over-fitting problem \cite{pmlr-v40-Ge15,DBLP:journals/corr/abs-1804-07612}. When the full training set is used, the cycle is called epoch,
and one uses several epochs to obtain a well-trained DNN. 
When training DNN with a mini-batch, the corresponding loss is defined by the sum of all losses over the mini-batch or by their average.

Once the training is over, for testing one uses a different data set from the one used in the DNN training, in order to avoid the over-fitting problem. 
In order to test the trained DNN model one can use either the loss function or the classification error function. 
If the number of test events is $n$, the classification error function is defined by
\begin{equation}
\textrm{Classification Error} = \frac{1}{n}\sum_{j=1}^{n} \delta[\textrm{ArgMax}(\{t[i]\}_j),
\textrm{ArgMax}(\{x[i]\}_j)],
\end{equation}
where ArgMax($\{y[i]\}$) gives the position $i_\textrm{max}$ where the value of $\{y[i]\}$ is maximized. 
$j$ represents the $j$-th test event, the $\delta$ is Kronecker delta function.

Often one takes additional steps such as dropout for reducing over-fitting in neural networks \cite{Srivastava:2014:DSW:2627435.2670313} 
and batch normalization for improving the performance and stability of artificial neural networks 
\cite{DBLP:journals/corr/IoffeS15}. Dropout makes a random drop of units (both hidden and visible) in a neural network and is considered 
an efficient way of performing model averaging. The batch normalization procedure normalizes the input layer by adjusting and scaling the activations:
 \begin{eqnarray}
 O[i]       &=& \gamma\hat{I}[i]+\beta \, , \\
\hat{I}[i] &=& \frac{I[i]-\mu[i]}{\sqrt{(\sigma[i])^2+\epsilon}} \, , \\
\mu[i]	 &=& \frac{1}{n}\sum_{\alpha=1}^nI[i]_\alpha \, , \\
\Big (\sigma[i] \Big )^2 &=& \frac{1}{n}\sum_{\alpha=1}^n\Big (I[i]_\alpha-\mu[i] \Big )^2 \, ,
\end{eqnarray}
where $\alpha$ represents the $\alpha$-th input in a mini-batch and $n$ is the size of the mini-batch.
The dimensions of input and output are the same. 
Note that ($\gamma$, $\beta$) are the learned parameters during the training and $\epsilon$ is a parameter added to avoid a divergence in the denominator.
The batch normalization allows each layer of a network to learn by itself independently of the other layers.

A convolutional neural network (CNN) is a class of DNN, most commonly used to analyze images.
CNN utilizes filters made of a set of neurons with a fixed size. 
The value of parameters in each filter is learned during the training process.
By varying the position of the filters on the input and learning the values of different filters, 
CNN can find local features of the input data. This process is called convolution and a hidden layer which uses convolution is called a convolutional layer. 
With $n_f'$ filters whose size is $(n_{fs}\times n_{fs})$, the convolution is defined as follows
\begin{equation}
\label{convolution}
O[i,j,k]=\sum^{n_f-1}_{\gamma=0}\sum_{\alpha,\beta}
\big(\mathcal{W}[\alpha,\beta,\gamma,k]I[\alpha,\beta,\gamma]+ \mathcal{B}[k]\big) \, ,
\end{equation}
where the dimension of the input is $n_f\times(n\times n)$ and the dimension of output is $n_f'\times(n'\times n')$. The corresponding ranges of the parameters are 
$k=\{0,\cdots,n_f'-1$\}, 
$\alpha = \{i,\cdots,i+n_{f_s}\}$, 
$\beta = \{ j,\cdots,j+n_{f_s}\}$, 
$i,j = \{0,n_s,2n_s,\cdots,n'\}$, 
$n'=n/n_s-n_{fs}+n_s$, and $n_s$ is called the stride.

Since each filter has a finite size, the output size decreases, after applying the convolution (\ref{convolution})
on the input or on the output from a previous layer. In order to prevent the size reduction, CNN incorporates the padding process:
\begin{equation}\label{padding}
I=
\begin{pmatrix}
\alpha&\beta\\\gamma&\delta
\end{pmatrix}
\rightarrow
O=
\begin{pmatrix}
0&0&0&0\\0&\alpha&\beta&0\\0&\gamma&\delta&0\\0&0&0&0
\end{pmatrix} \, ,
\end{equation}
which increases the size of the original input by adding zeros around it.
Usually the padding is used before applying convolution or pooling. 

CNN may include local or global pooling layers (often called sub-sampling), which combine the output of several neurons at one layer to a single neuron in the next layer. 
For example, max (average) pooling takes the maximum (average) value from a set of neurons at the previous layer and passes it to next layer. 
For a pooling dimension $n_p$, the relation between the output with dimension $n_f\times(n'\times n')$ and the input with dimension $n_f\times(n\times n)$ is given by
\begin{equation}
O[i,j,k]= \textrm{Max (Average)}(\{I[\alpha,\beta,k]\}) \, ,
\end{equation}
where $\alpha = \{i,\cdots,i+n_p\}$, $\beta = \{j,\cdots,j+n_p\}$, $k=\{0,\cdots,n_f-1\}$, $i,j=\{0,n_s,2n_s,
\cdots,n'\}$, $n'=n/n_s-n_p+n_s$, and $n_s$ is the stride. 

Another beneficial feature of a CNN is the reduction of the number of parameters via convolution and pooling, which effectively increases the learning speed in deep neutral networks.  
A typical DNN architecture consists of a combination of convolutional layers and dense layers, 
which provides better performance compared to a NN with only one type of layers \cite{Krizhevsky:2012:ICD:2999134.2999257}.


\bibliographystyle{JHEP}
\bibliography{draft}

\end{document}